\documentclass{article}  

\let\ssection=\section
\renewcommand{\section}{\setcounter{equation}{0}\ssection}
\usepackage{graphicx}
\usepackage{psfrag}
\usepackage{amsmath,amssymb,amsfonts}
\newtheorem{theorem}{Theorem}[section]
\newtheorem{lemma}[theorem]{Lemma}
\newtheorem{definition}[theorem]{Definition}

\title{Smoothness at null infinity and the structure of initial data}
\author{Helmut Friedrich\\ 
Max-Planck-Institut f\"ur Gravitationsphysik\\
Am M\"uhlenberg 1\\
14476 Golm\\
Germany}

\begin{document}
\maketitle

\begin{abstract}
We describe our present understanding of the relations between the
behaviour of asymptotically flat Cauchy data for Einstein's vacuum field 
equations near space-like infinity and the asymptotic behaviour of their 
evolution in time at null infinity. 
\end{abstract}


\newpage

\section{Introduction}

There are no doubts any longer that the idea of {\it gravitational
radiation} refers to a real physical phenomenon. Framing, however, a
precise underlying mathematical concept still poses problems. The
work on gravitational radiation by Pirani, \cite{pirani}, Trautman
\cite{trautman}, Sachs \cite{sachs:waves VI}, \cite{sachs:waves VIII} 
Bondi \cite{bondi:et.al}, Newman and Penrose \cite{newman:penrose} and
others, which was brought in a sense to a conclusion by Penrose
\cite{penrose:scri:let}, \cite{penrose:scri}, is based on the idealization
of an {\it isolated self-gravitating system}. It requires information on
the long time evolution of gravitational fields which at the time could only
be guessed. Ten years before these developments Y. Choquet-Bruhat had
achieved a breakthrough in the mathematical analysis of the local Cauchy
problem for Einstein's field equations \cite{choquet-bruhat}. However, the
technical means to derive the fall-off behaviour of gravitational fields at
far distances and late times from `basic principles' were not available in
the 1960's. In the meantime there has been a considerable progress in
controlling the asymptotic structure of solutions to Einstein's field
equations but it is still not quite clear which `basic principles' to assume
here.

In the following we shall report on work which aims at closing
various gaps in the study of gravitational radiation, the analysis of the
Einstein equations, and the calculation of wave forms. 
Sections \ref{conffieldequ} - \ref{Rfivp} present a fairly detailed
discussion of the underlying analytical structures and of the recent results
which led to the author's present understanding of the situation. To maintain
the flow of the arguments, the reader is referred for derivations to the
original literature. In sections \ref{confextstaticvacuum} and
\ref{stavacatI}  will be given new results and detailed arguments.

\vspace{.3cm}

Penrose's proposal to characterize far fields of {\it isolated
systems} in terms of their conformal structure 
(\cite{penrose:scri:let}, \cite{penrose:scri}) has been criticized over the
years on several grounds; various variations, alternatives,
etc. have been proposed (cf. \cite{bartnik:norton}, \cite{christ},
\cite{christ:klain}, \cite{cms}, \cite{ellis:1984}, \cite{schutz:2002},
\cite{valiente kroon:2001}, \cite{winicour}, and references given therein).
Some authors consider the smoothness requirements on the conformal boundary
as too restrictive and suggest generalizations (cf. \cite{cms},
\cite{valiente kroon:2001},
\cite{winicour}). Doubts have been raised as to whether non-trivial
asymptotically simple solutions to the vacuum field equations exist at all
(\cite{christ:klain}) and it has been argued that the smoothness of the
conformal boundary required in \cite{penrose:scri:let} excludes interesting
physics (\cite{christ}). The wide range of opinions on the subject is
illustrated by the curious contrast between this emphasis on 
subtleties of the asymptotic smoothness and claims that `null infinity is too
far away for modelling real physics' (cf. \cite{ellis:1984},
\cite{schutz:2002}).

In \cite{ellis:1984} even the {\it asymptotically flat model} is abandoned
and replaced by a {\it time-like cut model}. The latter introduces a
spatially compact time-like hypersurface ${\cal T}$ which is chosen in an
ad hoc fashion to cut off `the system of interest' from the rest of the
ambient universe. The idea then is to study the system which has thus been
`isolated' as an object of its own.

The usefulness of any such suggestion can only be demonstrated by analysing
its mathematical feasibility. This becomes clear when one tries to calculate
wave forms numerically. Such calculations cannot be based on hand waving or
physical intuition. The design of an effective numerical computer code
requires a precise mathematical formulation.  

The analysis of the time-like cut model reduces to a study of the initial
boundary value problem for Einstein's field equations in which boundary data
are prescribed on ${\cal T}$ and Cauchy data are given on a space-like
hypersurface ${\cal S}$ which intersects ${\cal T}$ in the space-like
surface $\partial{\cal S}$.  In \cite{friedrich:nagy} has been given a
fairly complete analysis of this problem for Einstein's vacuum field
equations.  This study is only local in time, but it provides insights into
the basic problem. So far, the time-like cut model raises many more
questions than it appears able to answer.

How is ${\cal T}$ to be chosen ? Physical considerations may lead to
suggestions when the system of interest is `sufficiently far' away from other
systems. However, there is in general no preferred physical or geometrical
choice for ${\cal T}$. (It is instructive to compare this with the anti-de
Sitter-type solutions, where the time-like boundary ${\cal J}$ at space-like
and null infinity is determined geometrically and the boundary data can be
prescribed in covariant form (cf.
\cite{friedrich:AdS}).)

The boundary must be characterized by some implicit or explicit geometrical
condition. A natural choice is to prescribe its mean extrinsic curvature. 
Its evolution in time is then defined implicitly by a quasi-linear wave
equation which itself depends in a non-local way on the data given on 
${\cal S}$ and ${\cal T}$  (cf. \cite{friedrich:nagy}). Long time
calculations thus require an extra effort to control the regularity of the
boundary. 

The gauge is related on the time-like boundary ${\cal T}$ directly to the
evolution process. It depends on the (implicit) choice of a time-like unit
vector field tangent to ${\cal T}$. While the data which are prescribed on
the space-like hypersurface ${\cal S}$ allow one to analyse the local
geometry near ${\cal S}$ at any desired order, the data which can be
prescribed on the boundary ${\cal T}$ provide very little information on the
local geometry near ${\cal T}$. All this makes it particularly difficult to
show that the gauge and the constraints are preserved under the evolution in
time.

These properties imply in general a non-covariance of the boundary
conditions and data. Moreover, due to the fact that no causal direction is
distinguished on ${\cal T}$ there does not seem to exist a natural `no
incoming radiation condition' and, in particular, no natural concept of
`outgoing radiation'. In fact, it appears difficult to associate with the
initial boundary value problem any `simple' quantities which characterize
the system and its dynamics and which can be related to observational data.

While the discussion in \cite{friedrich:nagy} singles out data which are
mathematically admissible, it is far from clear what should be prescribed on
${\cal T}$ from the physical point of view. The `correct' data induced by
the ambient universe will never be known. The information fed into `the
system' by the data prescribed on ${\cal T}$ can hardly be assessed. In long
time calculations it may alter the character of the system drastically.

Because of these difficulties the time-like cut model appears not very
promising. Nevertheless, it is of interest  because of its similarity to the
standard approach to numerical relativity, where an artificial time-like
boundary is introduced to render the computational grid finite. It is
expected here that the assumption of asymptotic flatness together with a
judicious choice of the boundary will alleviate some of the difficulties
pointed out above. 

At present the only satisfactory solution to the gravitational radiation
problem is based on the assumption of asymptotical flatness and the most
elegant and geometrically natural definition  of the latter is provided by
the idea of the {\it conformal boundary at null infinity} introduced in
\cite{penrose:scri:let}. While useful physical concepts can be associated
with a conformal boundary which is sufficiently smooth (cf. \cite{ashtekar},
\cite{geroch}, \cite{penrose:rindler:I} and the references given there),
the possible degree of differentiability, which encodes the fall-off
behaviour of the gravitational field, still poses questions. This article
deals with this particular issue and tries to disentangle its various
aspects and difficulties. 

\vspace{.4cm}

Einstein's field equations admit certain conformal representations which
in the following will be referred to as {\it conformal field equations}.
These equations are `regular' in the sense that they imply in a suitable
gauge equations which are hyperbolic even at points of null infinity
(\cite{friedrich:1981a}, \cite{friedrich:1981b}). This fact has been used to
show that the smoothness of the conformal boundary is preserved if it is
guaranteed on the initial slice ${\cal S}$ of an hyperboloidal initial value
problem (\cite{friedrich:hypivp}, \cite{friedrich:n-geod}, cf. also
\cite{friedrich:tueb}). The subsequent analysis of hyperboloidal initial
data (\cite{friedrich:ACF}, \cite{andersson:chrusciel:as},
\cite{andersson:chrusciel:ph}) showed the existence of a large class of
smooth hyperboloidal data for the conformal field equations. The
construction of  such data requires the `free data' to satisfy a finite
number of conditions at the space-like boundary $\partial {\cal S}$ at which
the hyperboloidal slice ${\cal S}$ intersects future null infinity ${\cal
J}^+$.

However, the work referred to above also shows the existence of a large class
of hyperboloidal data which are smooth on ${\cal S} \setminus \partial {\cal
S}$ but possess a non-trivial {\it polyhomogeneous expansion} at $\partial
{\cal S}$, i.e. an asymptotic expansion in terms of $x^k\,\log^j x$ where
$x$ is a defining function of the boundary $\partial {\cal S}$, which
vanishes on $\partial {\cal S}$.  Logarithmic terms can occur as a
consequence of the constraint equations  even if the free data extend
smoothly to $\partial {\cal S}$. Recently, it has been shown that certain
hyperboloidal data which are polyhomogeneous at
$\partial {\cal S}$ evolve into solutions to the conformal field equations
which possess {\it generalized conformal boundaries} near the initial slice
(\cite{chrusciel:lengard}, \cite{lengard}). While the precise behaviour of
these solutions near that boundary still needs to be analysed, the result
shows that the use of the conformal field equations and the characterization
of the edge of space-time in terms of its conformal structure are not
restricted to asymptotically regular situations.

We conclude from these results that in the standard Cauchy problem the 
field equations decide on the degree of smoothness of the conformal
boundary at  null infinity in arbitrarily small neighbourhoods of
space-like infinity.

There are other reasons to study the region near space-like infinity.
The hyperboloidal initial value problem is intrinsically time-asymmetric.
To analyse in the same picture incoming radiation, a
non-linear scattering process, and outgoing raditation, one needs to include
space-like infinity (as pointed out already in \cite{penrose:scri}). Also, if
the hyperboloidal data are not distinguished by special features as, for
instance, the presence of a trapped surface, it is not clear which part of
the imagined space-time is covered by their evolution. They
could represent a hypersurface close to time-like infinity or close
to a Cauchy hypersurface (a difficulty shared with the characteristic
initial value problem and the initial boundary value problem).

This should not obscure the fact that numerical calculations of space-times
from hyperboloidal data allow one to determine wave forms for many
`realistic' physical processes. So far the only semi-global calculations of
space-times, including their radiation fields at null infinity, are based on
hyperboloidal and characteristic initial value problems (cf.
\cite{frauendiener:tueb}, \cite{huebner:2001}, \cite{husa:tueb}, and the
article by L. Lehner and O. Reula, this volume).

We are thus left with the following task: (i) characterize the data which
evolve near space-like infinity into solutions of prescribed smoothness
at null infinity, (ii) analyse for which of these data  physical concepts and
requirements (linear and angular momentum at  space-like and null infinity,
reduction of the asymptotic gauge (BMS) group to a Poincar\'e group, $\dots$)
can be meaningfully introduced and a satisfactory physical picture can be
established.

The first step is technically the most difficult one. It requires us to 
control under fairly general assumptions the effect of the quasi-linear,
gauge hyperbolic field equations over infinite regions of space-time.
Moreover, the asymptotic behaviour of the solutions has to be determined
with a precision which excludes any further refinement.

Once this step has been taken, many considerations of the second step will
reduce to straightforward, though possibly quite lengthy, calculations. 
Nevertheless, the second step is of crucial importance. At this stage one has
to observe that the notion of asymptotic flatness is not part of the
general theory; it is an idealization which chosen to serve a purpose.  While
it is suggested to us by important solutions such as those of the Kerr
family, it is far from being determined by the equations alone. There
remains a large freedom to decide on the asymptotic behaviour of the fields. 

To make one's choice, one needs to know the mathematical options and 
has to decide on the physical questions to be answered. A theorem which
characterizes the most general Cauchy data on $\tilde{{\cal S}} =
\mathbb{R}^3$ for which the maximal globally hyperbolic Einstein development
is null geodesically complete and for which the Riemann tensor goes to zero
at (null) infinity would be mathematically  quite an achievement but, by
itself, insufficient from the point of view of physics.

We are not interested here in discussions of asymptotically flat 
solutions with `observations' referring to the roughness of the
asymptotic structure as, for instance, in \cite{valiente kroon:2001}. 
We rather wish to understand whether (i) the solution models a `system of
physical interest' and (ii) its far field and asymptotic structure
allow one to extract information on the system which characterizes its
physical nature and can be related to observational data.

This task is neither easy nor well defined.  The studies of the last $40$
years provide some understanding of the situations one may expect to observe
(collapse to a black hole, mergers of black holes, $\ldots$). 
By exploring, however, the questions above in a general setting,
new phenomena may be encountered (cf. \cite{choptuik}, \cite{gundlach} for an
example). But given that the interior is understood to some degree,
what do we do about (ii) ?

Recent results on the constraint equations exhibit possibilities to modify
asymptotically flat vacuum data `far out' without affecting the interior.
The data can be made to agree near space-like infinity with exact
Schwarzschild or Kerr data (\cite{corvino}, \cite{corvino:schoen}), with
even more general static resp. stationary data, or with data which are
only {\it asymptotically static resp. stationary}
(\cite{chrusciel:delay:2003}) (cf. also the discussion in the article by R.
Bartnik and J. Isenberg, this volume, for other techniques of modifying
or extending solutions to the constraints).

These results have been used to settle a question
which has been open for a long time. Since data which are static or
stationary near space-like infinity evolve into solutions which possses a
smooth conformal boundary at null infinity  (cf. \cite{dain:stationary}),
these solutions contain smooth hyperboloidal
hypersurfaces. Recently P. Chru\'sciel and E. Delay have shown the
existence of families of Cauchy data on $\mathbb{R}^3$ which are static
outside a fixed radius and have members of arbitrarily small ADM-mass. The
corresponding solutions contain hyperboloidal hypersurfaces to which the
results of \cite{friedrich:n-geod} apply. This demonstrates the existence of
non-trivial asymptotically simple solutions to Einstein's vacuum field
equations with prescribed smoothness of the asymptotic structure
(\cite{chrusciel:delay:2002}). 

More recently S. Klainerman and F. Nicol\`o revisited their work in
\cite{klainerman:nicolo} and showed (\cite{klainerman:nicolo:II}) that their
solutions will have the {\it Sachs peeling property} (\cite{sachs:waves VI},
\cite{sachs:waves VIII}) if the data are subject to certain asymptotic
conditions. However, the class of data which meet these
requirements still needs further analysis.

The new flexibility in constructing asymptotically flat initial
data also allows one to illustrate some difficulties of the asymptotically
flat space-time  model. Let $({\cal S},d)$ denote the initial data where
${\cal S}$ is the hypersurface considered in our discussion of the
time-like cut model and $d$ indicates the fields induced on ${\cal S}$ by
the cosmological model. Suppose that $({\cal S}', d')$ is an
asymptotically flat initial data set for which there exists an embedding
$\phi: {\cal S} \rightarrow {\cal S}'$ such that the push forward
of $d$ by $\phi$ is in a suitable sense `close' to $d'$ on
$\phi({\cal S})$. The evolution in time of the data $({\cal S}', d')$ can
then be considered in some neighbourhood of $\phi({\cal S})$ as a good
approximation of the evolution of $d$ in the cosmological model.

If the set ${\cal S}$ is chosen large enough and close to the region where
the system is undergoing a wave generation process, the main part of the
wave signal will reach null infinity at a finite retarded time. The fact
that for very late times the data on ${\cal S}' \setminus \phi({\cal S})$
will create a deviation of our solution from the cosmological one is
likely to be irrelevant in many interesting situations. From a pragmatical
point of view it may be considered the main purpose of the asymptotically
flat space-time extension beyond the domain of dependence of 
$\phi({\cal S})$ to allow perturbations of the gravitational field
generated near $\phi({\cal S})$ to unfold into a clean wave signal which
can be read off at null infinity. 

Since changes near space-like infinity affect the field, however weakly,
at all later times, they may have an important effect in the case of black
hole solutions. One may envisage the collapse of pure gravitational
radiation to a black hole as being modelled by
vacuum solutions which arise from smooth asymptotically
flat data on $\mathbb{R}^3$, admit smooth, complete (cf.
\cite{geroch:horowitz}) conformal boundaries ${\cal J}^{\pm}$, and  possess
future event horizons while all past directed null geodesics require
endpoints on ${\cal J}^-$. At present nothing is known about such solutions
and they may not exist. Is it conceivable then that there exist  solutions
which show all the (suitably generalized) features listed above but have a
rough conformal boundary ? Could such boundaries allow for a `higher
radiation content' ? If that were the case the restriction to smooth
conformal boundaries might preclude the discussion of certain interesting
physical phenomena.

Clearly, the large freedom in constructing asymptotically flat extensions
should neither be used to import irrelevant information into the system nor
to suppress important features. The replacement of an extension by one which
is strictly Kerr (say) near space-like infinity introduces a transition zone
on the initial hypersurface which mediates between the given and the Kerr
data. The resulting `wrinkles' in the solution are recorded in the radiation
field at null infinity. Is this information physically
insignificant or does it indicate that something important has been
ironed out by forming the new extension ? 

This question points again to the need of understanding the detailed
behaviour of the fields near space-like infinity. In the standard conformal
representation space-like infinity with respect to the solution space-time
is represented by a point, usually denoted by $i^0$.  With respect to an
asymptotically flat Cauchy hypersurface $\tilde{{\cal S}}$ space-like
infinity is then also represented by a point, denoted by $i$, which becomes
under conformal compactification a point in the extended 3-manifold ${\cal
S} = \tilde{{\cal S}} \cup \{i\}$.  Unfortunately, if $m_{ADM} \neq 0$ the
conformal initial data are strongly singular at $i$. This is the basic
stumbling block for analysing the field near space-like infinity in terms of
the standard conformal rescaling. 

The constraint equations on the Cauchy hypersurface $\tilde{{\cal S}}$
impose only weak restrictions on the asymptotic behaviour of the data near
space-like infinity. It is easy to construct data which at higher orders
will become quite `rough' near $i$ and which can be expected to affect the
smoothness of the fields near null infinity in a physically meaningless 
way. Thus one will have to make a reasonable choice and find a class of data 
which allows one to perform a sufficiently detailed analysis of their
evolution in time while still being sufficiently general to recognize basic
features of the asymptotic behaviour at null infinity.
 
In the following it is assumed that the data on $\tilde{{\cal S}}$ represent 
a space-like slice of time reflection symmetry, so that the second
fundamental form vanishes, and that their conformal structure, represented
by a conformal $3$-metric $h$ on ${\cal S}$, extends smoothly to the point
$i$. Only these conditions will be used in the following discussion, no a
priori assumptions on the evolution in time will be made.  We note that the
assumptions are made to simplify the calculations, there exists a large space
for generalizations. 

Somewhat unexpectedly, the attempt to analyse for data as described above the
evolution near space-like infinity $i$ in the context of the conformal field
equations led to a {\it finite regularization of the singularity at
space-like infinity} (\cite{friedrich:i-null}).

In a certain conformal scaling of the conformal initial data the choice of
the frame and the coordinates is  combined with a blow-up of the point $i$ to
a sphere ${\cal I}^0$ such that the initial data and the gauge of the
evolution system become smoothly extendable to ${\cal I}^0$ {\it in a
different conformal scaling}.  Moreover, the {\it general conformal field
equations} imply in that scaling a system of hyperbolic reduced
equations which also extends smoothly to ${\cal I}^0$ (section
\ref{sjetsnearcriticalsets}). This allows one to define a {\it regular
finite initial value problem at space-like infinity}. 

Under the evolution defined by the extended reduced system the set
${\cal I}^0$ evolves into a set ${\cal I} = \,]-1 , 1[ \times {\cal I}^0$,
which represents a boundary of the physical space-time. This {\it
cylinder at space-like infinity} is defined solely in terms of 
conformal geometry and the general conformal field equations. 

In the given coordinates, the sets ${\cal J}^{\pm}$ which represent
near space-like infinity the conformal boundary at null infinity are at a
finite location. They `touch' the set  ${\cal I}$ at certain {\it critical
sets} ${\cal I}^{\pm} = \{\pm 1\} \times {\cal I}^0$, which can be regarded
as the two components of the boundary of ${\cal I}$ and, simultaneously, as
boundaries of ${\cal J}^{\pm}$.  Due to the peculiarities of the
construction the solution is determined on the closure $\bar{{\cal I}}
\equiv {\cal I}  \cup {\cal I}^-
\cup {\cal I}^+ \simeq [-1 , 1] \times {\cal I}^0$ of ${\cal I}$ uniquely by
the data on $\tilde{{\cal S}}$ and there is no freedom to prescribe boundary
data on $\bar{{\cal I}}$.

This setting discloses the structure which decides on the asymptotic 
smoothness of the fields. At the critical sets ${\cal I}^{\pm}$ occurs a
{\it break-down of the hyperbolicity  of the reduced equations}. 
As explained in section \ref{specificsofrfivpsli},
a subtle interplay of this degeneracy with the structure of
the initial data near ${\cal I}^0$, which is mediated by certain {\it
transport equations} on ${\cal I}$, turns out critical for the smoothness of
the conformal structure at null infinity.  This peculiar situation is not
suggested by general PDE theory, it is a specific feature of Einstein's
theory, the geometric nature of the field equations, and general properties
of conformal structures.

The transport equations, which are linear hyperbolic equations interior to 
${\cal I}$, allow one to calculate the coefficients $u^p$ of the Taylor
series of the solution at ${\cal I}$ from data implied on ${\cal I}^0$ by
the Cauchy data on ${\cal S}$ (cf. \ref{ITaylorexp}). Near ${\cal I}^{\pm}$
this series can be interpreted as an asymptotic expansion. The coefficients
$u^p$ are smooth functions on ${\cal I}$ which can be calculated order by
order by following an algorithmic procedure. 

It turns out that the coefficients $u^p$ develop in general logarithmic
singularities at the critical sets ${\cal I}^{\pm}$. This behaviour
foreshadows a possible non-smoothness of the conformal structure at null
infinity. In the linearized setting it follows that the logarithmic
singularities are transported along the generators of null infinity
(\cite{friedrich:spin-2}). In the non-linear case their effect on the
conformal structure at null infinity is not under control yet, however, the
solutions are unlikely to be better behaved than in the linear case.

The evidence obtained so far suggests cases which
range from conformal structures of high differentiability
to ones with low smoothness at null infinity. The non-smoothness may 
take the form of polylogarithmic expansions in terms of
expressions $c\,(1 - \tau)^k\,\log^j \,(1 - \tau)$. Here
$\tau$ is a coordinate with $\tau \rightarrow 1$ from below on ${\cal
J}^+$, the coefficients $c = c(\rho, t)$ are smooth functions of a
coordinate $\rho$ along the null generators and suitable angular
coordinates $t$ on the set of null generators of ${\cal J}^+$, and $k$, $j$
are non-negative integers. If $k$ is small enough Sachs peeling fails and
Penrose compactification results in weak differentiability.

\vspace{.4cm}

Can one `loose physics' if one insists on extensions which are smooth at 
null infinity ? This certainly would be true if the coefficients $c$ would
encode important physical information. The discussion of the regularity
conditions in section \ref{sjetsnearcriticalsets} suggests
that at low orders the coefficients are determined by the data in an
arbitrarily  small neighbourhood of space-like infinity. By the results on
the constraints referred to above these data seem to be rather arbitrary,
only weakly related to `the system' characterized by the data on ${\cal S}$,
and thus of little relevance. As described in the following the situation is
more complicated at higher orders, depends then in a more subtle way on the
non-linearity of the equations, and requires further study. 

Since the coefficients $u^p$ can be calculated at arbitrary orders,
we expect that this analysis will also allow us to describe in detail the
relations between physical concepts defined at space-like infinity and
concepts defined on null infinity (ADM resp. Bondi linear and angular
momentum, etc.). The behaviour of the fields at the sets ${\cal I}^{\pm}$
is also critical for the possibility to reduce the BMS group to the
Poincar\'e group (cf. \cite{friedrich:kannar1}). Thus, the precise
understanding of the  behaviour of the fields near the critical sets should
provide us with a rather complete physical picture. 

Eventually one would like to make statements on the smoothness of the 
solution space-time at null infinity in terms of properties of the initial
data. Thus one needs to control how the behaviour of the asymptotic
expansion at the critical sets depends on the structure of the initial data
and to derive {\it regularity conditions} on the initial data
which are necessary and sufficient for the smoothness of the coefficients
$u^p$ at the critical sets. 

The information on the coefficients which is available so far has been used
to derive conditions which are {\it necessary} for the regularity
(\cite{friedrich:i-null}). The derivation of the complete condition is still
difficult because of the algebraic complexities of the calculations
involved. 

Recently J. A. Valiente Kroon obtained with the help of an algebraic
computer program complete and explicit expressions at higher order which are
pointing at the possibility that {\it asymptotic staticity} (or, if the  time
reflection symmetry is dropped, {\it asymptotic stationarity}) may play a
decisive role in deriving sufficient regularity conditions (\cite{valiente
kroon:2003}, \cite{valiente kroon:2003B}).

We are thus led to revisit the static vacuum solutions (sections
\ref{conformallystatic}, \ref{confextstaticvacuum} and \ref{stavacatI}).
Because of the loss of hyperbolicity of the conformal field equations at
${\cal I}^{\pm}$, it is not clear whether the smoothness of the conformal
structure at null infinity observed for static and stationary vacuum
solution can be understood as resulting from the possible regularity of the
extended solutions at the critical sets. We show that for static solutions
our setting is smooth, in fact real analytic, in a neighbourhood of the set
${\cal I}^- \cup {\cal I} \cup {\cal I}^+$. 

This narrows down the range in which the final regularity condition is to be
found. We know that asymptotic staticity is sufficient and that the
conditions found in \cite{friedrich:i-null}, which are implied by asymptotic
staticity, are necessary for the regularity of the asymptotic expansion on 
$\bar{{\cal I}}$. There is still the possibility that the final
condition ` fizzles out' and depends on the specific data but we expect to
arrive at the end at a definite, geometric condition.

To assess how restrictive such conditions would be, it is instructive to
consider the results by Chru\'sciel and Delay in \cite{chrusciel:delay:2003}.
They allow us to conclude that there exist large classes of solutions to the
constraints, which are essentially arbitrary on given compact subsets of the
initial hypersurface $\tilde{{\cal S}}$, whose evolutions in time admit
asymptotic expansions at ${\cal I}^{\pm}$ with coefficients which extend
smoothly to the critical sets ${\cal I}^{\pm}$ up to a given or at all
orders. 

\vspace{.4cm}

So far we ignored a question which is of central importance for gravitational
wave astronomy: can the replacement of an asymptotically flat extension by
another one result in a drastic change of the wave signal at null infinity ?
If that were the case, it would be hard to see how specific physical
processes could be identified in the wave forms calculated at ${\cal J}^+$. 

There should be characteristics of wave signals which are specific to `the
system' and which are stable under changes of the asymptotically flat
extension if these extensions are restricted to `reasonable' classes. 
This problem should be amenable to analytical and numerical investigations
and we expect our analysis to contribute to its solution. In fact, it
appears that with the regularity conditions mentioned above the field
equations themselves hint at a first `reasonable' class of asymptotically
flat extensions.

\section{Conformal field equations}
\label{conffieldequ}

Our analysis of the gravitational field near space-like and null
infinity relies on a certain conformal representation of Einstein's
vacuum field equations referred to as the {\it general conformal field
equations}. We give a short introduction to these equations and point out 
various facts about the equations and the underlying mathematical
structures which will be important in the following.
For derivations, detailed arguments, and further background material such as
the theory of {\it normal conformal Cartan connections}, which is the
natural home of many of the concepts used in the following, we refer the 
reader to the original article \cite{friedrich:AdS} and the survey
article \cite{friedrich:tueb}.

The aim is to discuss a solution $(\tilde{M}, \tilde{g})$ to Einstein«s
vacuum field equation  
\begin{equation}
\label{vaceinstequ}
R_{\mu \nu}[\tilde{g}] = 0,
\end{equation}
in terms of a suitably chosen {\it conformal factor} $\Theta$ and the {\it
conformal metric}
$g = \Theta^2\,\tilde{g}$. Denoting by $\nabla$ the Levi-Civita connection
of $g$, the transformation law of the Ricci tensor under the conformal
rescaling above takes in four dimensions the form
\begin{equation}
\label{RiccimGtraf}
R_{\nu \rho}[g] = R_{\nu \rho}[\tilde{g}] 
-\frac{2}{\Theta}\,\nabla_{\nu}\,\nabla_{\rho}\,\Theta - 
g_{\nu \rho}\,g^{\lambda \delta}
\left(\frac{1}{\Theta}\,\nabla_{\lambda}\,\nabla_{\delta}\,\Theta
- \frac{3}{\Theta^2}\,\nabla_{\lambda}\,\Theta\,\nabla_{\delta}\,\Theta
\right).
\end{equation}
If $\Theta$ is considered here as being given, equation
(\ref{vaceinstequ}) implies with (\ref{RiccimGtraf}) an equation for $g$
with a similar  principal part as (\ref{vaceinstequ}).  

As explained in the next section, we will mainly be interested in
the behaviour of the field in space-time domains where $\Theta
\rightarrow 0$. Because the right hand side of (\ref{RiccimGtraf}) is
formally singular in this limit, an abstract discussion of the
solutions near the sets $\{\Theta = 0\}$ becomes very delicate. It will be
seen, however, that under suitable assumptions on the initial data for the
field and with an appropriate behaviour of the conformal factor the right
hand side of (\ref{RiccimGtraf}) can attain smooth limits. This result is
obtained by a more sophisticated use of the behaviour of the
fields and the equations under transformations which preserve the conformal
structure.

\subsection{The general conformal field equations}
\label{genconffieldequ}

In \cite{friedrich:1981a}, \cite{friedrich:1981b} has been obtained a
system of equations which is regular in the sense that there occur no factors
$\Theta^{-1}$ on the right hand sides or factors $\Theta$ in the principal
part of the equations. Its unknowns are $\Theta$, $g$, and fields derived
from them such as the {\it rescaled conformal Weyl tensor}
$W^i\,_{jkl} \equiv \Theta^{-1}\,C^i\,_{jkl}$.
These have been used to derive various results about the asymptotic behaviour
of solutions to the Einstein equations. The specific behaviour of the
conformal fields near space-like infinity discussed in the next sections
asks, however, for a particularly careful analysis of the equations and the
gauge conditions. It turns out that this is considerably simplified by
making use of the full freedom offered by conformal structures. 

\vspace{.3cm}

A {\it Weyl connection}\index{Weyl connection} for the conformal structure
defined by $g$ is a torsion free connection $\hat{\nabla}$ which satisfies 
\begin{equation}
\label{wdg}
\hat{\nabla}_{\rho}\,g_{\mu \nu} = - 2\,f_{\rho}\,g_{\mu \nu},
\end{equation}
with some 1-form $f_{\rho}$. It is distinguished by the fact that it
preserves the conformal structure of $g$ (and thus of $\tilde{g}$). If a
frame $\{e_k\}_{k = 0, 1, 2, 3}$ is conformal at a given point $p$ in the
sense that it satisfies there $g(e_j, e_k) = \Lambda^2\,\eta_{jk}$ with
some
$\Lambda > 0$ and $\eta_{jk} = diag\,(1, -1, -1, -1)$, then it satisfies
such a relation with a point dependent function $\Lambda$ along a given
curve $\gamma$ through $p$ if it is parallely transported along $\gamma$
with respect to the connection $\hat{\nabla}$. In particular, if the 1-form
$f_{\rho}$ is closed the connection
$\hat{\nabla}$ is locally the Levi-Civita connection of a
metric in the conformal class.

Assuming under $\tilde{g} \rightarrow g = \Theta^2\,\tilde{g}$ the
transformation law
\[
\tilde{f}_{\rho} \rightarrow 
f_{\rho} = \tilde{f}_{\rho} - \Theta^{-1}\,\tilde{\nabla}_{\rho}\,\Theta,
\]
the defining property (\ref{wdg}) is expressed in terms of the metric
$\tilde{g}$ equivalently by $\hat{\nabla}_{\rho}\,\tilde{g}_{\mu
\nu}  = - 2\,\tilde{f}_{\rho}\,\tilde{g}_{\mu \nu}$ where $\hat{\nabla}$
denotes the Levi-Civita connection of $\tilde{g}$.
It follows from (\ref{wdg}) that the connection $\hat{\nabla}$ defines with 
the connection $\nabla$ the difference tensor $\hat{\nabla} - \nabla = S(f)$
given by the specific expression
\begin{equation}
\label{WGtraf}
 S(f)_{\mu}\,^{\rho}\,_{\nu} \equiv
\delta^{\rho}\,_{\mu}\,f_{\nu} 
+ \delta^{\rho}\,_{\nu}\,f_{\mu} 
- g_{\mu \nu}\,g^{\rho \lambda}\,f_{\lambda}. 
\end{equation}
This, in turn, can be used to specify $\hat{\nabla}$ in terms of $\nabla$
and the 1-form $f_{\rho}$. The three connections are related by
\begin{equation}
\label{threeconnrel}
\hat{\nabla} - \tilde{\nabla} = S(\tilde{f}),\,\,\,\,\,\,\,
\nabla - \tilde{\nabla} = S(\Theta^{-1}\,d\,\Theta),\,\,\,\,\,\,\,
\hat{\nabla} - \nabla = S(f).
\end{equation}
Important for us will also be the 1-form 
\begin{equation}
\label{d=Ttfdef}
d_{\mu} \equiv \Theta\,\tilde{f}_{\mu} =
\Theta\,f_{\mu} + \nabla_{\mu} \Theta. 
\end{equation}

The decomposition
\begin{equation}
\label{curvedecomp}
R^{\mu}\,_{\nu \lambda \rho} =
2\,\{g^{\mu}\,_{[\lambda}\,L_{\rho]\nu}  -
\,g_{\nu[\lambda}\,L_{\rho]}\,^{\mu}\} 
+ C^{\mu}\,_{\nu \lambda \rho}, 
\end{equation}
of the curvature tensor of $\nabla$ in terms of the
trace free conformal Weyl tensor $C^{\mu}\,_{\nu \lambda \rho}$ and the
Schouten tensor
\begin{equation}
\label{mLtens}
L_{\mu \nu} = \frac{1}{2}\,R_{\mu \nu}
- \frac{1}{12}\,R\,g_{\mu \nu}, 
\end{equation}
which carries the information about the Ricci tensor
$R_{\mu \nu} = R^{\rho}\,_{\mu  \rho \nu}$, has an analogue for
$\hat{\nabla}$ which takes the form 
\begin{equation}
\label{Wcurvedecomp}
\hat{R}^{\mu}\,_{\nu \lambda \rho} =
2\,\{g^{\mu}\,_{[\lambda}\,\hat{L}_{\rho]\nu}  -
g^{\mu}\,_{\nu}\,\hat{L}_{[\lambda \rho]}  -
\,g_{\nu[\lambda}\,\hat{L}_{\rho]}\,^{\mu}\} 
+ C^{\mu}\,_{\nu \lambda \rho}, 
\end{equation}
where 
\begin{equation}
\label{WLtens}
\hat{L}_{\mu \nu} = \frac{1}{2}\,\hat{R}_{(\mu \nu)}
- \frac{1}{4}\,\hat{R}_{[\mu \nu]}
- \frac{1}{12}\,\hat{R}\,g_{\mu \nu},  
\end{equation}
contains the information about the
Ricci tensor $\hat{R}_{\mu \nu} = \hat{R}^{\rho}\,_{\mu  \rho \nu}$
and the Ricci scalar $\hat{R} = g^{\mu \nu}\,\hat{R}_{\mu \nu}$.
The tensors (\ref{mLtens}) and (\ref{WLtens}) are related by
\begin{equation}
\label{WLtensrel}
\nabla_{\mu}\,f_{\nu} - f_{\mu}\,f_{\nu}
+ \frac{1}{2}\,g_{\mu \nu}\,f_{\lambda}\,f^{\lambda}
= L_{\mu \nu} - \hat{L}_{\mu \nu}.
\end{equation}

\vspace{.3cm}

To take care of the specific direction dependence of the various fields near
space-like infinity it is convenient to express the conformal field
equations in terms of a suitably chosen orthonormal frame field.
Let $\{e_k\}_{k = 0, 1, 2, 3}$ be a frame field satisfying 
$g_{ik} \equiv g(e_i, e_k) = \eta_{ik}$, denote by $\nabla_k$ and
$\hat{\nabla}_k$ the covariant derivative with respect to $\nabla$ and
$\hat{\nabla}$ in the direction of $e_k$, and define the connection
coefficients $\Gamma_i\,^j\,_k$ and $\hat{\Gamma}_i\,^j\,_k$ of $\nabla$ and
$\hat{\nabla}$ in this frame field by
$\nabla_i e_k = \Gamma_i\,^j\,_k\,e_k$ and
$\hat{\nabla}_i e_k = \hat{\Gamma}_i\,^j\,_k\,e_k$ respectively. Then 
$\hat{\Gamma}_i\,^j\,_k = \Gamma_i\,^j\,_k +
\delta^j\,_i\,f_k + \delta^j\,_k\,f_i 
- g_{ik}\,g^{jl}\,f_l$ with $f_k = f_{\mu}\,e^{\mu}\,_k$, where 
$e^{\mu}\,_k = \,<d\,x^{\mu},\,e_k>$ denote the frame coefficients with
respect to an as yet unspecified coordinate system $x^{\mu}$, 
$\mu = 0, 1, 2, 3$. We note that 
$f_i = \frac{1}{4}\,\hat{\Gamma}_i\,^k\,_k$ because 
$\Gamma_i\,^j\,_k\,g_{jl}  + \Gamma_i\,^j\,_l\,g_{jk} = 0$.

If all tensor fields (except the $e_k$) are expressed in terms of the
frame field and the corresponding connection coefficients, the
{\it conformal field equations}\index{conformal field equations} used in the
following are written as equations for the unknown
\begin{equation}
\label{gcfeunknown}
u = (e^{\mu}\,_k,\,\,\,\,\,\hat{\Gamma}_i\,^j\,_k,\,\,\,\,\,\hat{L}_{jk},
\,\,\,\,\,W^{i}\,_{jkl}),  
\end{equation}
and are given by 
\begin{equation}
\label{gentor}
[e_{p},e_{q}] =
(\hat{\Gamma}_{p}\,^{l}\,_{q} - \hat{\Gamma}_{q}\,^{l}\,_{p})\,e_{l},
\end{equation}
\begin{equation}
\label{gencorv}
e_{p}(\hat{\Gamma}_{q}\,^{i}\,_{j}) - 
e_{q}(\hat{\Gamma}_{p}\,^{i}\,_{j}) 
- \hat{\Gamma}_{k}\,^{i}\,_{j}(\hat{\Gamma}_{p}\,^{k}\,_{q} -
\hat{\Gamma}_{q}\,^{k}\,_{p}) +
\hat{\Gamma}_{p}\,^{i}\,_{k} \hat{\Gamma}_{q}\,^{k}\,_{j} 
- \hat{\Gamma}_{q}\,^{i}\,_{k} \hat{\Gamma}_{p}\,^{k}\,_{j}  
\end{equation}
\[
= 2\,\{g^{i}\,_{[p}\,\hat{L}_{q]j} 
- g^{i}\,_{j}\,\hat{L}_{[pq]} 
- \,g_{j[p}\,\hat{L}_{q]}\,^i\}
+ \Theta \, W^{i}\,_{jpq}, 
\]
\begin{equation}
\label{gencbian}
\hat{\nabla}_{p}\,\hat{L}_{qj} - \hat{\nabla}_{q}\,\hat{L}_{pj} 
= d_i \, W^{i}\,_{jpq},
\end{equation}
\begin{equation}
\label{genbian}
\nabla_{i} \,W^{i}\,_{jkl} = 0.
\end{equation}

The square brackets in the first equation denote the commutator of
vector fields. The connection $\nabla$, which appears in the last equation,
can be expressed by the relations given above in terms of $\hat{\nabla}$ and
$f_k$. The last equation, referred to in the following as the {\it Bianchi
equation}, is in a sense the core of the system. It is obtained from the
contracted vacuum Bianchi identity $\tilde{\nabla}_{\mu}\,C^{\mu}\,_{\nu
\lambda \rho} = 0$ by using the specific conformal identity 
$\Omega^{-1}\,\tilde{\nabla}_{\mu}\,C^{\mu}\,_{\nu \lambda \rho} =
\nabla_{\mu}\,(\Omega^{-1}\,C^{\mu}\,_{\nu \lambda \rho})$.  The
first three equations are then essentially the {\it structural equations} of
the theory of normal conformal Cartan connections.

\vspace{.5cm}

No equations are given so far for the fields $\Theta$ and $d_k = \Theta\,f_k
+ \nabla_k \Theta$. They reflect the {\it conformal gauge freedom}
artificially introduced here into Einstein's field equations. 
These fields cannot be prescribed quite arbitrarily. For solution for
which the limit $\Theta \rightarrow 0$ is meaningful the latter should
imply $d_k \rightarrow \nabla_k \Theta$.

The theory of normal conformal Cartan connections associates with each
conformal structure a distinguished class of curves which provides a useful
way of dealing with the gauge freedom. 
A {\it conformal geodesic}\index{conformal geodesics} for $(\tilde{M},
\tilde{g})$ is a curve
$x(\tau)$ in $\tilde{M}$ which solves, together with a 1-form 
$\tilde{f} = \tilde{f}(\tau)$ along it, the system of ODE's
\begin{equation}
\label{acgxequ}
(\tilde{\nabla}_{\dot{x}}\dot{x})^{\mu}
+ S(\tilde{f})_{\lambda}\,^{\mu}\,_{\rho}\,
\dot{x}^{\lambda}\,\dot{x}^{\rho}
= 0, 
\end{equation}
\begin{equation}
\label{bcgbequ}
(\tilde{\nabla}_{\dot{x}}\tilde{f})_{\nu} - \frac{1}{2}\, 
\tilde{f}_{\mu}\,S(\tilde{f})_{\lambda}\,^{\mu}\,_{\nu}\,\dot{x}^{\lambda} 
= \tilde{L}_{\lambda \nu}\,\dot{x}^{\lambda},
\end{equation}
where $S(\tilde{f})$ and $\tilde{L}$ are given by (\ref{WGtraf}) and
(\ref{mLtens}) with $g$ replaced by $\tilde{g}$. For any given metric in
the conformal class there are more conformal geodesics than metric
geodesics  because for given initial data  
$x_* \in M$, $\dot{x}_* \in T_{x_*} M$, $\tilde{f}_* \in T^*_{x_*} M$
there exists a unique solution $x(\tau)$,
$\tilde{f}(\tau)$ to (\ref{acgxequ}), (\ref{bcgbequ}) near $x_*$ satisfying
for given $\tau_* \in \mathbb{R}$
\begin{equation}
\label{cgindat}
x(\tau_*) = x_*,\,\,\,\,\,\,\dot{x}(\tau_*) =
\dot{x}_*,\,\,\,\,\,\,\tilde{f}(\tau_*) = \tilde{f}_*.
\end{equation}
The sign of $\tilde{g}(\dot{x},\dot{x})$ is preserved near $x_*$
but not its modulus. 

Conformal geodesics admit, unlike metric geodesics, general fractional
linear maps as parameter transformations. They are {\it conformal
invariants}. Denote by
$b$ a smooth 1-form field. Then, if $x(\tau)$,
$\tilde{f}(\tau)$ solve the conformal geodesics equations (\ref{acgxequ}),
(\ref{bcgbequ}), the pair $x(\tau)$,
$\tilde{f}(\tau) - b|_{x(\tau)}$ solves the same equations with
$\tilde{\nabla}$ replaced by the connection $\hat{\nabla} =
\tilde{\nabla} + S(b)$ and $L$ by $\hat{L}$, i.e. the curve $x(\tau)$, and
in particular its parameter $\tau$, are independent of the Weyl connection
in the conformal class which is used to write the equations
(cf. \cite{friedrich:cg on vac}).

Let there be given a smooth congruence of conformal geodesics which
covers an open set $U$ of $\tilde{M}$ such that the associated 1-forms
$\tilde{f}$ define a smooth field on $U$. Denote by $\hat{\nabla}$ the
torsion free connection on $U$ which has with the connection 
$\tilde{\nabla}$ difference tensor 
$\hat{\nabla} - \tilde{\nabla} = S(\tilde{f})$ and denote by $\hat{L}$ the
tensor (\ref{WLtens}) derived from $\hat{\nabla}$. Comparing with 
(\ref{WLtensrel}), we find that equations (\ref{acgxequ}), (\ref{bcgbequ})
can be written 
\begin{equation}
\label{Wacgxequ}
\hat{\nabla}_{\dot{x}}\,\dot{x} = 0,\,\,\,\,\,\, 
\hat{L}_{\lambda \nu}\,\dot{x}^{\lambda} = 0.
\end{equation}
Let $e_k$ be a frame field satisfying along the congruence
\begin{equation}
\label{Wccgxequ}
\hat{\nabla}_{\dot{x}}\,e_k = 
\tilde{\nabla}_{\dot{x}}\,e_k + <\tilde{f}, e_k>\dot{x} 
+ <\tilde{f}, \dot{x}>e_k 
- \tilde{g}(\dot{x}, e_k)\tilde{g}^{\sharp}(\tilde{f},\,.\, ) 
= 0.
\end{equation}
Suppose that $\tilde{{\cal S}}$ is a hypersurface which is transverse
to the congruence, meets each of the curves exactly once,
and on which $\tilde{g}(e_i, e_k) = \Theta^2_*\eta_{ik}$ with some
function $\Theta_* > 0$. It follows that  
$\tilde{g}(e_i, e_k) = \Theta^2\,\eta_{ik}$ on $U$ with a function
$\Theta$ which satisfies
\begin{equation}
\label{Wdcgxequ}
\hat{\nabla}_{\dot{x}}\,\Theta = \Theta <\dot{x}, \tilde{f}>,
\,\,\,\,\,\,
\Theta|_{\tilde{S}} = \Theta_*.
\end{equation}

\vspace{.5cm} 

The observations above allow us to construct a special gauge for the
conformal equations. Let $\tilde{{\cal S}}$ be a space-like hypersurface in
the given vacuum solution $(\tilde{M}, \tilde{g})$. We choose on
$\tilde{{\cal S}}$ a positive `conformal factor' $\Theta_*$, a frame field
$e_{k*}$, and a 1-form
$\tilde{f}_*$ such that $\Theta^2_*\,\tilde{g}(e_{i*}, e_{k*}) =
\eta_{ik}$  and $e_{0*}$ is orthogonal to $\tilde{{\cal S}}$. 
Then there exists through each point $x_* \in \tilde{{\cal S}}$ a unique
conformal geodesic $(x(\tau), \tilde{f}(\tau))$ with $\tau = 0$ on
$\tilde{S}$ which satisfies there the initial conditions $\dot{x} =
e_{0*}$, $\tilde{f} = \tilde{f}_*$. 

If all data are smooth these curves define in some neighbourhood $U$ of
$\tilde{{\cal S}}$ a smooth caustic free congruence which covers $U$.
Furthermore, $\tilde{f}$ defines a smooth 1-form  on $U$ which supplies a
Weyl connection $\hat{\nabla}$ as described above. A smooth frame field $e_k$
and the related conformal factor $\Theta$ are then obtained on $U$ by
solving (\ref{Wccgxequ}), (\ref{Wdcgxequ}) for given initial data
$e_{k} = e_{k*}$, $\Theta = \Theta_*$ on $\tilde{S}$. The frame field
is orthonormal for the metric $g = \Theta^2\,\tilde{g}$. Dragging along
local coordinates
$x^{\alpha}$, $\alpha = 1, 2, 3$, on $\tilde{S}$ with the congruence and
setting $x^0 = \tau$ we obtain a coordinate system.
This gauge is characterized on $U$ by the explicit gauge conditions
\begin{equation}
\label{wgaugecond}
\dot{x} = e_0 = \partial_{\tau},\quad
\hat{\Gamma}_0\,^j\,_k = 0,\quad
\hat{L}_{0k} = 0.
\end{equation}
Coordinates, a frame field, and a conformal factor as above are said to
define a {\it conformal Gauss gauge}\index{conformal Gauss gauge}. Since
metric Gauss systems are well known to quickly develop caustics, it may be
mentioned that conformal Gauss systems can cover large space-time domains 
in a regular fashion (cf. \cite{friedrich:cg on vac}).

\vspace{.3cm}

To obtain a closed system for all the fields entering equations
(\ref{gentor}) - (\ref{genbian}), we could now supplement the latter 
by equations which are implied for the fields $\Theta$ and $d_k$ in a
conformal Gauss gauge. It turns out that such a gauge implies
quite simple ordinary differential equations along the conformal
geodesics defining the gauge, it holds in fact
\[
\ddot{d}_0 = 0,\,\,\,\,\,\,\,\,
\dot{\Theta} = d_0,\,\,\,\,\,\,\,\,   
\dot{d}_a = 0,\,\,\,\,\,a = 1, 2, 3,
\]
where the dot denotes differentiation with respect to the parameter $\tau$.

Thus, the fields $\Theta$ and $d_k$ given by a conformal
Gauss   can be determined in our
situation explicitly (\cite{friedrich:AdS}): {\it  If $\tilde{g}$ is a
solution to Einstein's vacuum equations (\ref{vaceinstequ}), the fields 
$\Theta$ and  $d_k$ are given by the explicit expressions
\begin{equation}
\label{Thetaexpl}
\Theta = \Theta_*\,\left(1 + \tau\,<\tilde{f}, \dot{x}>_*
+ \frac{\tau^2}{4}\,\Theta_*^2\,
\left((\tilde{g}(\dot{x}, \dot{x}))^2\,
\tilde{g}^{\sharp}(\tilde{f}, \tilde{f})\right)_*\right)
\end{equation}
\[
= \Theta_*\,\left(1 + \tau\,<\tilde{f}, \dot{x}>_*
+ \frac{\tau^2}{4}\,\left(g^{\sharp}(\tilde{f}, \tilde{f})\right)_*\right),
\]
\begin{equation}
\label{bexpl}
d_0 = \dot{\Theta},\,\,\,\,\,\,d_a = \Theta_*<\tilde{f}_*,\,e_{a*}>,
\,\,\,a = 1, 2, 3, 
\end{equation}
where $g^{\sharp}$ denotes the contravariant version of $g$ and the
quantities with a subscript star are considered as constant along the
conformal geodesics and given by their values on $\tilde{{\cal S}}$}. 

Assuming for $\Theta$ and $d_k$ the expressions (\ref{Thetaexpl}) and
(\ref{bexpl}), equations (\ref{gentor}) - (\ref{genbian}) provide a
complete system for $u$. In spite of the fact that we use a special gauge,
we refer to this system as the 
{\it general conformal field equations}\index{conformal field
equations, general} to indicate that they employ the full gauge freedom
preserving conformal structures.

Equally important for us are the facts that the expression (\ref{Thetaexpl})
offers the possibility to control in a conformal Gauss gauge the location
of the set where $\Theta \rightarrow 0$ and that (\ref{Thetaexpl}),
(\ref{bexpl}) imply with the relation $d_k = \Theta\,f_k + \nabla_k \Theta$
in sufficiently regular situations that 
\begin{equation}
\label{scricaus}
\nabla_k\,\Theta\,\nabla^k\,\Theta \rightarrow 0
\quad\mbox{as}\quad \Theta \rightarrow 0.
\end{equation}

\subsection{Spinor version of the general conformal field equations}
\label{spincfequ}

Writing the conformal equations in the spin frame formalism leads to
various algebraic simplifications. We introduce here only the basic notions
of this formalism and refer the reader to \cite{penrose:rindler:I} for a
comprehensive introduction. It should be noted, however, that our
notation does not completely agree with that of \cite{penrose:rindler:I}.   
In particular, if a specific frame field is used it will always be
pointed out in the text but not be indicated by gothic indices. 

Starting with the orthonormal frame introduced above we define
null frame vector fields $e_{AA'} = \sigma^k\,_{AA'}\,e_k$ with constant van
der Waerden symbols $\sigma^k\,_{AA'}$ (here and in the following 
indices $A, B, \ldots$, $A', B', \ldots$ take values $0$ and $1$ and the
summation rule is assumed) such that  
\[
e_{00'} = \frac{1}{\sqrt{2}}\,(e_0 + e_3),\,\,\,
e_{11'} = \frac{1}{\sqrt{2}}\,(e_0 - e_3),
\]
\[
e_{01'} = \frac{1}{\sqrt{2}}\,(e_1 - i\,e_2),\,\,\,
e_{10'} = \frac{1}{\sqrt{2}}\,(e_1 + i\,e_2).
\]
Then $e_{00'}$, $e_{11'}$ are real and $e_{01'}$, $e_{10'}$ are complex
(conjugate) null vector fields and their scalar products are given by 
$g(e_{AA'},e_{CC'}) = \epsilon_{AC}\,\epsilon_{A'C'}$ 
where $\epsilon_{AC}$, $\epsilon_{A'C'}$, $\epsilon^{AC}$,
$\epsilon^{A'C'}$ denote the anti-symmetric spinor fields with 
$\epsilon_{01} = \epsilon_{0'1'} = \epsilon^{01} = \epsilon^{0'1'} =1$.
The latter are also used to raise and lower spinor indices according to the
rules $X^A = \epsilon^{AB}\,X_B$ and $X_B = X^A\,\epsilon_{AB}$
so that $\epsilon_A\,^B = \epsilon_{AC}\,\epsilon^{BC}$ denotes the
Kronecker spinor (similar rules hold for primed indices) .

If connection coefficients $\Gamma_{AA'}\,^{BB'}\,_{CC'}$ are defined by
writing 
$\nabla_{e_{AA'}}\,e_{CC'} = \Gamma_{AA'}\,^{BB'}\,_{CC'}\,e_{BB'}$, the 
spinor connection coefficients are given by 
$\Gamma_{AA'}\,^B\,_C = \frac{1}{2}\,\Gamma_{AA'}\,^{BE'}\,_{CE'}$
and one has 
\[
\Gamma_{AA'}\,^{BB'}\,_{CC'} = 
\Gamma_{AA'}\,^B\,_C\,\epsilon_{C'}\,^{B'}
+ \bar{\Gamma}_{AA'}\,^{B'}\,_{C'}\,\epsilon_C\,^B.
\]
Here it is observed, as usual, that the relative order of primed and
unprimed indices is irrelevant and that under complex conjugation primed
indices are converted into unprimed indices and vice versa.
Covariant derivatives of spinor fields are now given by similar
rules as in the case of the standard frame formalism. Writing
$\nabla_{AA'}$ for $\nabla_{e_{AA'}}$ we have e.g. for a spinor field
$X^A\,_B\,^{C'}$ 
\[
\nabla_{DD'}\,X^A\,_B\,^{C'} = e_{DD'}(X^A\,_B\,^{C'})
+ \Gamma_{DD'}\,^A\,_F\,X^F\,_B\,^{C'}
\]
\[
- \Gamma_{DD'}\,^F\,_B\,X^A\,_F\,^{C'}
+ \bar{\Gamma}_{DD'}\,^{C'}\,_{F'}\,X^A\,_B\,^{F'}.
\]
We have similar rules for the connection $\hat{\nabla}$ and its associated
connection coefficients $\hat{\Gamma}_{AA'BC}$ and it holds
\begin{equation}
\label{ccff}
\Gamma_{CC'AB} = \Gamma_{CC'BA},\,\,\,\,\,\,
\hat{\Gamma}_{CC'AB} = \Gamma_{CC'AB} - \epsilon_{AC}\,f_{BC'},  
\end{equation}
so that $\hat{\Gamma}_{CC'}\,^F\,_F = f_{CC'}$ gives the 1-form relating
the connection $\hat{\nabla}$ to $\nabla$.

The general conformal field equations are now written as equations for the
unknowns
\begin{equation}
\label{unknowns}
e_{AA'},\,\,\,\,\,
\hat{\Gamma}_{AA'BC},\,\,\,\,\,
\Theta_{AA'BB'},\,\,\,\,\,
\phi_{ABCD}.
\end{equation}
Here $\Theta_{AA'BB'}$ is the spinor representation of the
tensor field $\hat{L}_{kj}$. It admits a decomposition of the form
\[
\Theta_{AA'BB'} = \Phi_{AA'BB'} -
\frac{1}{24}\,R\,\,\epsilon_{AB}\,\epsilon_{A'B'}  +
\Phi_{AB}\,\epsilon_{A'B'} + \bar{\Phi}_{A'B'}\,\epsilon_{AB},
\]
where $\Phi_{AA'BB'} = \Phi_{BB'AA'} = \bar{\Phi}_{AA'BB'}$ represents the
trace-free part of the tensor $\frac{1}{2}\,\hat{R}_{(jk)}$ provided by
$\hat{\nabla}$ while $R = g^{jk}\,\hat{R}_{jk}$ is the Ricci scalar and the last
two terms, with $\Phi_{ab} = \Phi_{(ab)}$, represent the anti-symmetric tensor
$\frac{1}{4}\,\hat{R}_{[jk]}$. The symmetric spinor field 
$\phi_{ABCD} = \phi_{(ABCD)}$ represents the rescaled conformal Weyl
tensor and is related to the latter by
\[
W_{AA'BB'CC'DD'} = - \phi_{ABCD}\,\epsilon_{A'B'}\,\epsilon_{C'D'}
- \bar{\phi}_{A'B'C'D'}\,\epsilon_{AB}\,\epsilon_{CD}.
\] 
The general conformal field equations in the order (\ref{gentor}),
(\ref{gencorv}), (\ref{gencbian}),  (\ref{genbian}) now take the form
\begin{equation}
\label{s1e}
[e_{BB'}, e_{CC'}] = 
(\Gamma_{BB'}\,^{AA'}\,_{CC'} - \Gamma_{CC'}\,^{AA'}\,_{BB'})\,e_{AA'},
\end{equation}
\begin{equation}
\label{d2e}
e_{CC'}(\hat{\Gamma}_{DD'}\,^A\,_B) -
e_{DD'}(\hat{\Gamma}_{CC'}\,^A\,_B) 
\end{equation}
\[
- \hat{\Gamma}_{CC'}\,^{F}\,_{D}\,\hat{\Gamma}_{FD'}\,^{A}\,_{B}
+ \hat{\Gamma}_{DD'}\,^{F}\,_{C}\,\hat{\Gamma}_{FC'}\,^{A}\,_{B}
- \overline{\hat{\Gamma}}_{CC'}\,^{F'}\,_{D'}\,
\hat{\Gamma}_{DF'}\,^{A}\,_{B}
\]
\[
+ \overline{\hat{\Gamma}}_{DD'}\,^{F'}\,_{C'}
\,\hat{\Gamma}_{CF'}\,^{A}\,_{B}
+ \hat{\Gamma}_{CC'}\,^{A}\,_{F}\,\hat{\Gamma}_{DD'}\,^{F}\,_{B} -
\hat{\Gamma}_{DD'}\,^{A}\,_{F}\,\hat{\Gamma}_{CC'}\,^{F}\,_{B}
\]
\[
= - \Theta_{BD'CC'} \,\epsilon_D\,^A
+ \Theta_{BC'DD'} \,\epsilon_C\,^A
+ \Theta\,\phi^A\,_{BCD}\,\epsilon_{C'D'},
\]
\begin{equation}
\label{d1e}
\hat{\nabla}_{BB'} \Theta_{CC'AA'} - \hat{\nabla}_{AA'} \Theta_{CC'BB'} 
= d^{EE'}(\phi_{EABC} \epsilon_{E'C'} \epsilon_{A'B'}
+ \bar{\phi}_{E'A'B'C'} \epsilon_{EC} \epsilon_{AB}),
\end{equation}
\begin{equation}
\label{bieq}
\nabla^F\,_{A'} \phi_{ABCF} = 0, 
\end{equation}
with the fields $\Theta$, $d_{AA'}$ as given above. The simple form 
(\ref{bieq}) of the spinor version of the Bianchi equation will be useful for
us.

\subsection{The reduced conformal field equations}
\label{redconffe}

The conformal Gauss gauge is not only distinguished by the fact
that it is provided by the conformal structure itself and supplies 
explicit information on $\Theta$ and $d_k$, but also by a remarkable
simplicity of the resulting evolution equations. Setting $p = 0$ 
in (\ref{gentor}) - (\ref{genbian}) and observing the gauge conditions 
(\ref{wgaugecond}) we obtain 

\begin{equation}
\label{redtor}
\partial_{\tau}\,e^{\mu}\,_{q} =
- \hat{\Gamma}_{q}\,^{l}\,_{0}\,e^{\mu}\,_{l},
\end{equation}
\begin{equation}
\label{redcorv}
\partial_{\tau}\,\hat{\Gamma}_{q}\,^{i}\,_{j} = 
- \hat{\Gamma}_{k}\,^{i}\,_{j}\,\hat{\Gamma}_{q}\,^{k}\,_{0} 
+ g^{i}\,_{0}\,\hat{L}_{qj} 
+ g^{i}\,_{j}\,\hat{L}_{q0} 
- \,g_{j0}\,\hat{L}_{q}\,^i
+ \Theta \, W^{i}\,_{j0q}, 
\end{equation}
\begin{equation}
\label{redcbian}
\partial_{\tau}\,\hat{L}_{qj} 
+ \hat{\Gamma}_{q}\,^{k}\,_{0}\,\hat{L}_{kj} 
= d_{i} \, W^{i}\,_{j0q},
\end{equation}
\[
\nabla_{i} W^{i}\,_{jkl} = 0.
\]

While the first three equations are then ordinary differential equations
along the conformal geodesics, we still have to deduce a suitable evolution
system from the last equation. The Bianchi equation represents an
overdetermined system of 16 equations for the 10 independent components of
the rescaled conformal Weyl tensor. It implies a system of wave equations
for $W^{i}\,_{jkl}$ which could be used as the evolution system. For the
application studied in this article it turns out important, however, to
use the first order system. 

There are various ways of extracting from the Bianchi equations 
symmetric hyperbolic evolution systems but these are most easily found in
the spin frame formalism. With the spinor field  
$\tau^{AA'} = \epsilon_0\,^A\,\epsilon_{0'}\,^{A'}
+ \epsilon_1\,^A\,\epsilon_{1'}\,^{A'}$ the gauge conditions 
(\ref{wgaugecond}) can be written 
\begin{equation}
\label{spinwgaugecond}
\tau^{AA'}\,e_{AA'} = \sqrt{2}\,\partial_{\tau},\quad
\tau^{AA'}\,\hat{\Gamma}_{AA'}\,^B\,_C = 0,\quad
\tau^{BB'}\,\Theta_{AA'BB'} = 0.
\end{equation}
Observing $\tau_{AA'}\,\tau^{BA'} = \epsilon_A\,^B$ and its complex
conjugate version, one obtains from (\ref{ccff}) and (\ref{spinwgaugecond})
the relation $\tau^{CC'}\,\Gamma_{CC'AB} = - \tau_A\,^{C'}\,f_{BC'}$ and
thus
\begin{equation}
\label{hatGambyGam}
\hat{\Gamma}_{CC'AB} = \Gamma_{CC'AB} - \epsilon_{AC}\,
\tau^{DD'}\,\Gamma_{DD'EB}\,\tau^E\,_{C'},
\end{equation}
which shows with (\ref{ccff}) that $\hat{\Gamma}_{CC'AB}$ can be expressed
in our gauge in terms of $\Gamma_{CC'AB}$ and vice versa.

Transvecting equations (\ref{s1e}), (\ref{d2e}), (\ref{d1e})
suitably with $\tau^{EE'}$ thus gives the system of ODE's 
\begin{equation}
\label{reds1e}
\sqrt{2}\,\partial_{\tau}\,e^{\mu}\,_{CC'} = 
- \Gamma_{CC'}\,^{AA'}\,_{BB'}\,\tau^{BB'}\,e^{\mu}\,_{AA'},
\end{equation}
\begin{equation}
\label{redd2e}
\sqrt{2}\,\partial_{\tau}\,\hat{\Gamma}_{DD'}\,^A\,_B 
+ (\hat{\Gamma}_{DD'}\,^{F}\,_{C}\,\hat{\Gamma}_{FC'}\,^{A}\,_{B}
+ \overline{\hat{\Gamma}}_{DD'}\,^{F'}\,_{C'}
\,\hat{\Gamma}_{CF'}\,^{A}\,_{B})\,\tau^{CC'}
\end{equation}
\[
= \Theta_{BC'DD'} \,\tau^{AC'}
+ \Theta\,\phi^A\,_{BCD}\,\tau^{C}\,_{D'},
\]
\begin{equation}
\label{redd1e}
\sqrt{2}\,\partial_{\tau}\,\Theta_{CC'AA'} 
+ (\hat{\Gamma}_{AA'}\,^F\,_B\,\Theta_{CC'FB'} 
+ \overline{\hat{\Gamma}}_{AA'}\,^{F'}\,_{B'}\,\Theta_{CC'BF'})\tau^{BB'}
\end{equation}
\[
= - d^{EE'}(\phi_{EABC} \epsilon_{E'C'}\tau^{B}\,_{A'}
+ \bar{\phi}_{E'A'B'C'} \epsilon_{EC}\tau_A\,^{B'}).
\]

We set now $\Lambda_{ABCA'} \equiv \nabla^F\,_{A'} \phi_{ABCF}$.
Equation (\ref{bieq}) is then equivalent to 
$0 = \Lambda_{ABCD} \equiv \Lambda_{ABCA'}\,\tau_D\,^{A'}$.
On the other hand we have the decomposition
$\Lambda_{ABCD} = \Lambda_{(ABCD)} 
- \frac{3}{4}\,\epsilon_{D(C}\,\Lambda_{AB)F}\,^F$ with irreducible
parts
\begin{equation}
\label{spinpropconstr}
\Lambda_{(ABCD)} = - \frac{1}{2}\,\left(P\,\phi_{ABCD} 
- 2\,{\cal D}_{(D}\,^F\,\phi_{ABC)F}\right)
,\,\,\,\,\,\,\,\,\,
\Lambda_{ABF}\,^F = {\cal D}^{EF}\,\phi_{ABEF},
\end{equation}
where $P = \tau^{AA'}\,\nabla_{AA'} = \sqrt{2}\,\nabla_{e_0}$
and ${\cal D}_{AB} = \tau_{(A}\,^{A'}\,\nabla_{B)A'}$ denote covariant 
directional derivative operators such that
${\cal D}_{00} = - \nabla_{01'}$, ${\cal D}_{11} = \nabla_{10'}$, and
${\cal D}_{01} = {\cal D}_{10} = \frac{1}{\sqrt{2}}\,\nabla_{e_3}$,
(cf. \cite{friedrich:global}, \cite{friedrich:AdS} for more details of
the underlying space-spinor formalism). 

In a Cauchy problem one will in general assume $e_0$ to be the future
directed normal to the initial hypersurface $\tilde{{\cal S}}$. The
operators ${\cal D}_{AB}$ then involve only differentiation in directions
tangent to $\tilde{{\cal S}}$ and the equations $\Lambda_{ABF}\,^F = 0$ are
interior equations on $\tilde{{\cal S}}$. They represent the six real
constraint equations implied on $\tilde{{\cal S}}$ by the Bianchi equation.

For a symmetric spinor field $\psi_{A_1 \ldots A_k}$ we define
its (independent) {\it essential components} by $\psi_j = \psi_{(A_1 \ldots
A_k)_j}$, where $0 \le j \le k$ and the brackets with subscript $j$ indicate
that $j$ of the indices in the brackets are set equal to $1$ while the others
are set equal to $0$. 

The five equations $\Lambda_{(ABCD)} = 0$ for the  
components of $\phi_{ABCD}$ contain the operator $P$. Multiplying by
suitable binomial coefficients (and considering the frame and connection
coefficients as given), we find that the system
\begin{equation}
\label{sysyhy}
- \binom{4}{A+B+C+D}\,\Lambda_{(ABCD)} = 0,
\end{equation}
has the following properties. If $\phi$ denotes the transpose of the
$\mathbb{C}^5$-valued `vector' $(\phi_0, \phi_1, \phi_2, \phi_3, \phi_4)$, 
it takes the form
\[
A^{\mu}\,\partial_{\mu}\,\phi = H(x,\phi),
\]
with a $\mathbb{C}^5$-valued
function $H(x, \phi)$ and $5 \times 5$-matrix-valued functions
$A^{\mu}$ which are hermitian, i.e. $^T\bar{A}^{\mu} = A^{\mu}$, and for
which there exists at each point a covector $\xi_{\mu}$ such that
$A^{\mu}\xi_{\mu}$ is positive definite. The system 
(\ref{sysyhy}) is thus {\it symmetric hyperbolic}\index{symmetric
hyperbolic} (\cite{friedrichs}, see also \cite{friedrich:rendall} and the
references given there).

While the constraints implied on a given space-like hypersurface are
determined uniquely, there is a large freedom to select useful evolution
systems. In fact, any system of the form
\[
0 = 2\,a\,\Lambda_{0001'},
\]
\[
0 = (c - d)\,\Lambda_{0011'} - 2\,a\,\Lambda_{0000'},
\]
\begin{equation}
\label{redsymhypbianchi}
0 = (c + d)\,\Lambda_{0111'} - (c - d)\,\Lambda_{0010'},
\end{equation}
\[
0 = 2\,e\,\Lambda_{1111'} - (c + d)\,\Lambda_{0110'},
\]
\[
0 = - 2\,e\,\Lambda_{1110'},
\]
with $a, c, e > 0$ and $ - (2\,e + c) < d < 2\,a + c$, is symmetric
hyperbolic (the system (\ref{sysyhy}) occurs here as a special case). We
note that only the characteristics of these systems which are null
hypersurfaces are of physical significance.

Equations (\ref{redtor}), (\ref{redcorv}), (\ref{redcbian}), respectively
equations (\ref{reds1e}), (\ref{redd2e}), (\ref{redd1e}), combined with a
choice of (\ref{redsymhypbianchi}), will be referred to in the following as
the (general) {\it reduced conformal field equations}\index{conformal field
equations, reduced}. Solutions to these equations solve in fact the complete
system  (\ref{gcfeunknown}), (\ref{gentor}), (\ref{gencbian}),
(\ref{genbian})  if the solution admits a Cauchy hypersurface on which the
latter equations  hold (\cite{friedrich:AdS}). In other words, propagation
by the reduced field equations preserves the constraints.

\subsection{The conformal constraints}
\label{confconstr}

To analyse solutions to the conformal field equations in the context of a
Cauchy problem one needs to study the conformal constraints 
implied on a space-like initial hypersurface $\tilde{{\cal S}}$. It will
be convenient to discuss the evolution equations in terms
of a conformal factor
$\Theta$ which  differs on $\tilde{{\cal S}}$ from the one used 
to analyse the constraints. We thus assume Einstein's
equations (\ref{vaceinstequ}), a conformal rescaling 
\begin{equation}
\label{gtg}
g = \Omega^2 \tilde{g},
\end{equation}
with a positive conformal factor $\Omega$, and denote again 
the Levi-Civita connection of $g$ by $\nabla$. It is also convenient
to derive the conformal constraints from the {\it metric conformal field
equations}\index{conformal field equations, metric}. The latter are written
in terms of the unknown fields 
$g$, $\Omega$, $S \equiv \frac{1}{4}\nabla_{\mu}\nabla^{\mu}\Omega +
\frac{1}{24}\,R\,\Omega$, $L_{\mu \nu}$ as in (\ref{mLtens}), and 
$W^{\mu}\,_{\nu \lambda \rho} = \Omega^{-1} C^{\mu}\,_{\nu \lambda \rho}$ and
are given by equation (\ref{curvedecomp}), with 
$ C^{\mu}\,_{\nu \lambda \rho} = \Omega\,W^{\mu}\,_{\nu \lambda \rho}$, and
the equations
\begin{equation}
\label{boce5}
2\, \Omega\,S - \nabla_{\mu}\Omega\,\nabla^{\mu}\Omega = 0, 
\end{equation}
\begin{equation} 
\label{boce1}
\nabla_{\mu}\, \nabla_{\nu} \Omega = - \Omega \, L_{\mu \nu} 
+ S\, g_{\mu \nu}, 
\end{equation}
which are obtained by rewriting the trace and the trace free part of 
(\ref{RiccimGtraf}),
\begin{equation}
\label{boce2}
\nabla_{\mu}\,S = - L_{\mu \nu} \nabla^{\nu} \Omega, 
\end{equation}
\begin{equation}
\label{boce3}
\nabla_{\lambda} L_{\rho \nu}
- \nabla_{\rho} L_{\lambda \nu}
= \nabla_{\mu} \Omega \, W^{\mu}\,_{\nu \lambda \rho},
\end{equation}
which both can be obtained as integrability conditions of (\ref{boce1}),
and  
\begin{equation}
\label{boce4}
\nabla_{\mu} W^{\mu}\,_{\nu \lambda \rho} = 0.
\end{equation}
In these equations the Ricci scalar $R$ is considered as the {\it conformal
gauge source function}. Its choice, which is completely arbitrary in
local studies, controls together with the initial data $\Omega$ and 
$d \Omega$ on $\tilde{{\cal S}}$ the evolution of the conformal scaling. 

\vspace{.3cm}

To derive the constraints induced by these equations on $\tilde{{\cal S}}$ 
we choose a $g$-orthonormal frame field $\{e_k\}_{k = 0, 1, 2, 3}$ near 
$\tilde{{\cal S}}$ such that $n \equiv e_0$ is $g$-normal to 
$\tilde{{\cal S}}$, write $\nabla_k \equiv \nabla_{e_k}$, $\nabla_k\,e_j =
\Gamma_k\,^l\,_j\,e_l$, and express all fields (except the $e_k$) and
equations in terms of this frame. We assume that indices $a, b, c,..$
from the beginning of the alphabet take values $1,2,3$ and that the
summation convention also holds for these indices. The inner metric $h$
induced by $g$ on $\tilde{{\cal S}}$ is then given by 
$h_{ab} = g(e_a, e_b) = - \delta_{ab}$ and the second fundamental form
by 
\[
\chi_{ab} = g(\nabla_{e_a} n, e_b) = \Gamma_a\,^j\,_0\,g_{jb}
= - \Gamma_a\,^0\,_b.
\]
We set $\Sigma = \nabla_0\,\Omega$ and
$W^*_{\mu \nu \lambda \rho} = \frac{1}{2}\,
W_{\mu \nu \alpha \beta} \epsilon^{\alpha \beta}\,_{\lambda \rho}$.
If the tensor fields
\[
L_{\mu \nu}, \,\,\,\,\,
L_{\mu \nu} n^{\nu}, \,\,\,\,\,
W_{\mu \nu \lambda \rho}, \,\,\,\,\, 
W_{\mu \nu \lambda \rho}n^{\nu}n^{\rho}, \,\,\,\,\, 
W^*_{\mu \nu \lambda \rho}n^{\nu}n^{\rho}, \,\,\,\,\, 
W_{\mu \nu \lambda \rho}n^{\nu},
\]
are projected orthogonally into $\tilde{S}$ and expressed in terms of the
frame $\{e_a\}_{a = 1, 2, 3}$ on $\tilde{S}$, they are given by (the left
hand sides of) 
\[ 
L_{ab},\,\,\,\,\,
L_a \equiv L_{a0},
\]
\[
w_{abcd}  \equiv W_{abcd},\,\,\,\,\,
w_{ab} \equiv W_{a0b0},\,\,\,\,\,
w^*_{ab}  \equiv W^*_{a0b0}, \,\,\,\,\,
w_{abc}  \equiv W_{a0bc},
\]
respectively and satisfy the relations
\begin{equation}
\label{RLrel}
R = 6\,L_{\mu}\,^{\mu} = 6\,(L_{00} + L_a\,^a),
\end{equation}
\[
w_{abcd} = - 2\{ h_{a[c} w_{d]b} + h_{b[d} w_{c]a} \}
,\,\,\,\,\,
w^*_{ad}\,\epsilon^d\,_{bc}   = w_{abc}
,\,\,\,\,\,
w^*_{ad}   = - \frac{1}{2}\,w_{abc}\,\epsilon_d\,^{bc},
\]
\[
w_{ab} = w_{ba},\,\,\,\,\, 
w_a\,^a =  0,\,\,\,\,\, 
w^*_{ab} = w^*_{ba},\,\,\,\,\, 
w^*_a\,^a =  0, 
\]
\[
w_{abc} = - w_{acb},\,\,\,\,\,
w^a\,_{ac} = 0,\,\,\,\,\,
w_{[abc]} = 0,
\]
where indices are moved with $h_{ab}$ and $\epsilon_{abc}$ is totally
antisymmetric with $\epsilon_{123} = 1$. 
The tensors $w_{ab}$ and $w^*_{ab}$ represent the $n$-electric and the 
$n$-magnetic part of $W^i\,_{jkl}$ on $\tilde{{\cal S}}$ respectively.

Equation (\ref{curvedecomp}) implies Gauss' and Codazzi's equation
on $\tilde{{\cal S}}$
\begin{equation}
\label{bzw9}
r_{ab} = - \,\Omega\,\,w_{ab} + L_{ab} + L_c\,^c\,h_{ab}
+ \chi_c\,^c\,\chi_{ab} - \chi_{ac}\,\chi_b\,^c,  
\end{equation}
\begin{equation}
\label{bzw8}
D_b \,\chi_{ca} - D_c \,\chi_{ba} = \Omega\,\,w_{abc} 
+ h_{ab}\,L_c - h_{ac}\,L_b,
\end{equation}
while equations (\ref{boce5}) - (\ref{boce4}) imply the following interior
equations which only involve derivatives in the directions of $e_a$, $a =
1, 2, 3$, tangent to $\tilde{{\cal S}}$
\begin{equation}
\label{bzw10}
2\,\Omega\,S - \Sigma^2 - D_a\Omega \,D^a\Omega = 0, 
\end{equation}
\begin{equation}
\label{bzw1}
D_a\,D_b\,\Omega = - \Sigma\,\chi_{ab} - \Omega\,L_{ab} 
+ S\,h_{ab},
\end{equation}
\begin{equation}
\label{bzw2}
D_a\,\Sigma = \chi_a\,^c D_c\Omega - \Omega\,L_a,
\end{equation}
\begin{equation}
\label{bzw3}
D_a \,S = - D^b\,\Omega \, L_{ba} - \Sigma \, L_a, 
\end{equation}
\begin{equation}
\label{bzw4}
D_a \,L_{bc} - D_b \,L_{ac} = D^e\Omega\,\,w_{ecab} -
\Sigma\,w_{cab} - \chi_{ac}\,L_b + \chi_{bc}\,L_a,
\end{equation}
\begin{equation}
\label{bzw5}
D_a \,L_b - D_b \,L_a = D^e\Omega\,\,w_{eab} + \chi_a\,^c L_{bc} 
- \chi_b\,^c L_{ac}, 
\end{equation}
\begin{equation}
\label{bzw6}
D^c \,w_{cab} = \chi^c\,_a \,w_{bc} - \chi^c\,_b \,w_{ac},
\end{equation}
\begin{equation}
\label{bzw7}
D^a \,w_{ab} = \chi^{ac}\, w_{abc},
\end{equation}
where $r_{ab}$ denotes the Ricci tensor of $h_{ab}$. These equations can
be read as {\it conformal constraints}\index{conformal constraints} for the
fields 
\[
\Omega,\,\,\,\Sigma,\,\,\,S,\,\,\,h_{ab},\,\,\,\chi_{ab},\,\,\, 
L_a,\,\,\,L_{ab},\,\,\,w_{ab},\,\,\,w^*_{ab}.
\]
Alternatively, if a `physical' solution $\tilde{h}_{ab}$,
$\tilde{\chi}_{ab}$ to the vacuum constraints is given and a conformal
factor $\Omega$ and functions $\Sigma$, $R$ have been chosen, which are gauge
dependent functions at our disposal, the equations above
can be used to calculate
\[
S,\,\,\,L_{\mu \nu},\,\,\,W^{\mu}\,_{\nu \lambda \rho},
\]
from the conformal first and second fundamental forms 
$h_{ab}$, $\chi_{ab}$ of $\tilde{{\cal S}}$,
which are related to the physical data by 
\begin{equation}
\label{confreddata}
h_{ab} = \Omega^2\,\tilde{h}_{ab},\,\,\,\,\,
\chi_{ab} = \Omega\,(\tilde{\chi}_{ab} + \Sigma\,\tilde{h}_{ab}).
\end{equation}
The equations above will be discussed in more detail in section 
\ref{afcdat}.

\section{Asymptotic simplicity}
\label{assimple}

To characterize the fall-off behaviour of asymptotically flat solutions
at null infinity in terms of geometric concepts Penrose introduced the
notion of {\it asymptotic simplicity}\index{asymptotic simplicity}
(\cite{penrose:scri:let},
\cite{penrose:scri}, cf. also \cite{penrose:rindler:I} for further
discussions and references). 
\begin{definition}
\label{assidef}
A smooth space-time $(\tilde{{\cal M}}, \tilde{g})$ is called 
{\it asymptotically simple} if there exists a 
smooth, oriented, time-oriented, causal space-time
$({\cal M}, g)$ and a smooth function $\Omega$ on ${\cal M}$ such that:\\ 
(i) ${\cal M}$ is a manifold with boundary ${\cal J}$,\\
(ii) $\Omega > 0$ on ${\cal M} \setminus {\cal J}$ and $\Omega = 0$,
$d\,\Omega \neq 0$ on ${\cal J}$,\\
(iii) there exists an embedding $\Phi$ of $\tilde{{\cal M}}$ onto 
$\Phi(\tilde{{\cal M}}) = {\cal M} \setminus {\cal J}$ which is conformal 
such that $\Omega^2\,\Phi^{-1*}\tilde{g} = g$.\\  
(iv) each null geodesic of $(\tilde{{\cal M}}, \tilde{g})$ acquires two
distinct endpoints on ${\cal J}$. 
\end{definition}

We note that only the conformal class of $(\tilde{{\cal M}}, \tilde{g})$
enters the definition and it is only the conformal structure of 
$({\cal M}, g)$ which is determined here. The set ${\cal J}$ is referred to
as the {\it conformal boundary of $(\tilde{{\cal M}}, \tilde{g})$ at null
infinity}\index{conformal boundary}. This definition is the mathematical
basis for the 

\vspace{.2cm}

\noindent
{\bf Penrose proposal}: {\it Far fields
of isolated gravitating systems behave like that of asymptotically simple
space-times in the sense that they can be smoothly extended to null
infinity, as indicated above, after suitable conformal rescalings}. 

\vspace{.2cm}

Since gravitational fields are governed by Einstein's equations, 
the proposal suggests a sharp characterization 
of the fall-off behaviour implied by the field equations in terms of the
purely geometrical definition (\ref{assidef}). 

We will be interested in the following in solutions to Einstein's vacuum
field equations (\ref{vaceinstequ}) which satisfy the conditions of
definition  (\ref{assidef}) (or suitable generalizations). The two
assumptions have important consequences for the structure of 
$({\cal M}, g)$. We shall only quote those which will be used in the
following discussion   and refer the reader for further information to the
references given above.  

If the vacuum field equations hold near ${\cal  J}$, the latter
defines a smooth null hypersurface of ${\cal M}$ (cf. equation
(\ref{boce5})). It splits into two components, ${\cal J}^+$ and ${\cal
J}^-$, which are generated by the past and future endpoints of null
geodesics in ${\cal M}$ and are thus called {\it future and past null
infinity} (or $scri\,\pm$) respectively.  Each of
${\cal J}^{\pm}$ is ruled by null generators, each set of null generators
has topology $S^2$, and ${\cal J}^{\pm}$ have the topology of 
$\mathbb{R} \times S^2$. For the applications one will have to relax the
conditions of definition (\ref{assidef}). In particular condition $(iv)$,
which is important to obtain the result about the topology of ${\cal
J}^{\pm}$, must be replaced by a different completeness condition if one
wants to discuss space-times with black holes.

One of the main difficulties in developing a well defined concept of {\it outgoing
radiation} in the time-like cut model is related to the fact that there
exists in general no distinguished null direction field along
the time-like boundary ${\cal T}$. In contrast, the null generators of ${\cal
J}^{+}$ define a unique causal direction field on ${\cal J}^{+}$, which is
represented by $\nabla^{AA'}\Omega$. It turns out that the field
$\phi_{ABCD}\,o^A\,o^B\,o^C\,o^D$ on ${\cal J}^{+}$, where $o^A$ denotes a
spinor field satisfying $o^A\,\bar{o}^{A'} = - \nabla^{AA'}\,\Omega$ on
${\cal J}^{+}$ and $\phi_{ABCD}$ the rescaled conformal Weyl spinor field,
has a natural interpretation as the 
{\it outgoing radiation field}\index{radiation field}.  
Further important physical concepts can be
associated with the hypersurface ${\cal J}^{+}$ or subsets of it and
questions of interpretation have been extensively analysed
(cf. \cite{ashtekar}, \cite{chrusciel:jezierski:kijowski}, \cite{geroch} and
the references given there).

Critical however, and in fact a matter of controversy, have been the
smoothness assumptions in the definition, which encode the fall-off
behaviour imposed on the physical fields. It is far from immediate that
they are in harmony with the fall-off behaviour imposed by the field
equations. No problem arises if the proposal can be
justified with $C^{\infty}$ replaced by $C^k$ with sufficiently large integer
$k$. But there is a lower threshold for the differentiability, which is not
easily specified, at which the definition will loose much of its elegance
and simplicity.

In \cite{penrose:scri} it is assumed that 
\begin{equation}
\label{scriksmooth}
{\cal M}\,\,\,\mbox{is of class}\,\,\,C^{k + 1}\,\,\, \mbox{and}\,\,\,g,
\Omega \in C^k({\cal M}),\,\,\,k \ge 3.
\end{equation} 
The conformal Weyl spinor $\Psi_{ABCD} = \Omega\,\phi_{ABCD}$ 
is then in $C^{k - 2}({\cal M})$. Under the further assumption 
\begin{equation}
\label{1addsmooth}
\Omega\,\nabla_{EE'}\,\nabla^A\,_{A'}\,\Psi_{ABCD}
\rightarrow 0 \quad\mbox{at}\quad {\cal J}^+,
\end{equation}
which will certainly be satisfied if $k \ge 4$ in (\ref{scriksmooth}), 
and the natural assumption 
\begin{equation}
\label{2addsmooth}
\mbox{the set of null generators of}\,\, {\cal J}^+ \,\,
\mbox{has topology}\,\,S^2,
\end{equation}
it is then shown that $\Psi_{ABCD}$ vanishes on ${\cal J}^+$. The solution
is thus asymptotically flat in the most immediate sense and the rescaled
conformal Weyl spinor $\phi_{ABCD}$ extends in a
continuous fashion to ${\cal J}^+$. As a consequence, it follows that the
space-time satisfies the {\it Sachs peeling property}\index{Sachs peeling}
(\cite{penrose:scri},
\cite{sachs:waves VI}, \cite{sachs:waves VIII}) which says that in a
suitably chosen spin frame the components of the conformal Weyl spinor
fall-off as $\Psi_{ABCD} = O(\tilde{r}^{A+B+C+D - 5})$ (where $A, B, C,
D$ take values $0, 1$) along an outgoing null geodesic when its (physical) affine
parameter $\tilde{r} \rightarrow \infty$ at ${\cal J}^+$.

Remarkable as it is that such a conclusion can be drawn for the spin-2
nature of the field $\Psi_{ABCD}$ and its governing field equation
$\tilde{\nabla}^F\,_{A'}\,\Psi_{ABCF} = 0$, there remains the question
whether the long time evolution by the field equations is such that
assumptions  (\ref{scriksmooth}), (\ref{1addsmooth}) or the conclusion drawn
from them can be justified. 

As discussed in the introduction, we know by now that these conditions can
be satisfied by non-trivial solutions to the vacuum field equations. What
is not known, however, is how the solutions satisfying these conditions
are to be characterized in terms of their Cauchy data, whether these conditions
exclude solutions modelling important physical phenomena, and if they do, what
exactly goes wrong. Obviously, these questions can only be answered by analysing
the Cauchy problem for Einstein' field equations with asymptotically flat Cauchy
data in the large with the goal to derive sharp results on the behaviour of the
field at null infinity.

The results obtained so far on the existence of solutions which admit
(partial) smooth boundaries at null infinity make it clear that the key
problem here is the behaviour of the fields near space-like infinity. We
shall not consider any further the results which lead to this conclusion
(cf. \cite{friedrich:tueb} for a discussion and the relevant references) but
concentrate on this particular problem.

To begin with we have a look at the asymptotic region of interest here in
the case of Minkowski space. 
If the latter is given in the form 
$(\tilde{{\cal M}} \simeq  \mathbb{R}^4,
\tilde{g}  = \eta_{\mu \nu}\,d\,y^{\mu}\,d\,y^{\nu})$, the coordinate
transformation $\Phi: y^{\mu} \rightarrow 
x^{\mu} = (- y_{\lambda}\,y^{\lambda})^{-1}\,y^{\mu}$ 
renders the metric in
the domain ${\cal D} \equiv \{y_{\lambda}\,y^{\lambda} < 0\} =
\{x_{\lambda}\,x^{\lambda} < 0\}$ in the form
$\tilde{g}  = \Omega^{- 2}\,\eta_{\mu \nu}\,d\,x^{\mu}\,d\,x^{\nu}$
with $\Omega = - x_{\lambda}\,x^{\lambda}$. The metric 
\begin{equation}
\label{sinfmetr}
g = \Omega^2\,\tilde{g} =  \eta_{\mu \nu}\,d\,x^{\mu}\,d\,x^{\nu}, 
\end{equation}
thus extends smoothly to the domain $\bar{{\cal D}}$ of points in 
$\{x_{\lambda}\,x^{\lambda} \le 0\}$ which
are obtained as limits of sequences in ${\cal D}$.
The point $x^{\mu} = 0$ in this set then represents space-like infinity for
Minkowski space. With this understanding it is denoted by $i^0$.
The hypersurfaces ${\cal J}^{'\pm} = \{x_{\lambda}\,x^{\lambda} = 0,\,\,\pm
x^0 > 0\} \subset \bar{{\cal D}}$ represent parts of future and past null
infinity of Minkowski space close to space-like infinity and are generated
by the  future and past directed null geodesics of $g$ through $i^0$.

Consider the Cauchy hypersurface $\tilde{{\cal S}} = \{y^0 = 0\}$ of
Minkowski space. The subset $\tilde{{\cal S}} \cap {\cal D}$ is mapped by
$\Phi$ onto $\{x^0 = 0,\, x^{\mu} \neq 0\}$. Extending the latter to include
the point $x^{\mu} = 0$  amounts to a smooth compactification 
$\tilde{{\cal S}} \rightarrow {\cal S} = \tilde{{\cal S}} \cup \{i\} \sim
S^3$ such that the point
$i$ with coordinates $x^{\mu} = 0$ represents space-like infinity with
respect the metric induced on $\tilde{{\cal S}}$ by $\tilde{g}$. The
distinction of space-like infinity $i$ with respect to a Cauchy data set and
space-like infinity $i^0$ with respect to the solution space-time will
become important and much clearer later on.

Denote by $\tilde{h}_{\alpha \beta}$ and $\tilde{\chi}_{\alpha \beta}$ the
metric and the extrinsic curvature induced by $\tilde{g}$ on 
$\tilde{{\cal S}}$. 
A global representation of the conformal structure induced on ${\cal S}$ by
$\tilde{h}_{\alpha \beta}$ is obtained by using a slightly different
conformal rescaling than before. Set $h' = \Omega^{'2}\,\tilde{h}$
with $\Omega' = 2\,(1 + |y|^2)^{-1}$ where $|y| = \sqrt{(y^1)^2 + (y^2)^2 +
(y^3)^2}$. In terms of standard spherical coordinates $\theta$, $\phi$ 
on $\tilde{S}$ and the coordinate $\chi$ defined by $\cot \frac{\chi}{2}
= |y|$, $\,\,0 \le \chi \le \pi$, the conformal metric takes the form
$h' = - (d\,\chi^2 + \sin^2 \chi\,d\,\sigma^2)$ of the standard metric on
the 3-sphere and extends smoothly to the point $i$, which is given by the
coordinate value $\chi = 0$ and distinguished by the property that 
$\Omega = 0$, $d\,\Omega = 0$, $Hess\,\Omega = c\,h'$, with $c \neq 0$
at $i$. Here
$d\,\sigma^2 \equiv d\,\theta^2 + \sin^2 \theta\,d\,\phi^2$ denotes the
standard line element on the 2-sphere $S^2$. 

Since $\tilde{\chi}_{\alpha \beta} = 0$ and we are free to choose $\Sigma =
0$ in (\ref{confreddata}), we get $\chi'_{\alpha \beta} = 0$. 
By the formulas given in the previous section one can derive from the 
conformal Minkowski data $({\cal S}, h', \chi')$ and a suitable choice of
initial data for the gauge defining fields (\ref{Thetaexpl}), (\ref{bexpl}) a
conformal initial data set for the reduced conformal field equations.
These allow us then to recover the well known conformal embedding of
Minkowski space into the Einstein cosmos (\cite{penrose:scri}) as a smooth
solution to the regular conformal field equations (\cite{friedrich:tueb},
\cite{friedrich:cg on vac}).

We would like to control what happens if the conformal
Minkowski data are subject to finite perturbations. Under which assumptions
will the corresponding solutions preserve asymptotic simplicity ?
This or the apparently simpler question under which
conditions the solutions will preserve near space-like infinity a reasonable
amount of smoothness of the conformal boundary
cannot be answered by immediate applications of the conformal field
equations. The reason is that the conformal data
will not be smooth at the point $i$. The structure of the conformal initial
data as well as the initial value problem for the conformal field equations
near space-like infinity thus require a careful and detailed analysis.
This will be carried out to some extent in the next section.

\section{Asymptotically flat Cauchy data}
\label{afcdat}

As indicated in the case of Minkowski space above we will assume that the data
for the conformal field equations are given on a $3$-dimensional manifold
${\cal S} = \tilde{{\cal S}} \cup \{i\}$ which is obtained from a
`physical' $3$-manifold $\tilde{{\cal S}}$ with an asymptotically flat end
by adjoining a point $i$ which represents space-like infinity. The
data $h_{ab}$, $\chi_{ab}$ on ${\cal S}$ are thought as being obtained
from the physical data $\tilde{h}_{ab}$, $\tilde{\chi}_{ab}$ by equations
(\ref{confreddata}) with suitable choices of $\Omega$ and $\Sigma$ such
that all fields extend with an appropriate behaviour (to be specified more
precisely below) to $i$ and $\Sigma(i) = 0$,
$\Omega > 0$ on $\tilde{{\cal S}}$, $\Omega(i) = 0$, $D_a\,\Omega(i) = 0$,
$D_a\,D_b\,\Omega(i) = -2\,h_{ab}$, where we assume the notation of
subsection \ref{confconstr}.

The constraint equations (\ref{bzw9}) - (\ref{bzw7}) contain analogues of 
the vacuum constraints. The form of these equations suggests a solution
procedure which does not require us to go back to the physical data. By
taking the trace of equation (\ref{bzw9}), using (\ref{bzw10}) and the
trace of (\ref{bzw1}), and writing $\Delta_h \equiv D_a\,D^a$, one gets
\begin{equation}
\label{OconfHamconstr}
\Omega^2\,r = - 4\,\Omega\,\Delta_h\,\Omega
+ 6\,D_a\Omega\,D^a\Omega - 4\,\Omega\,\Sigma\,\chi_c\,^c
+ \Omega^2((\chi_c\,^c)^2 - \chi_{ac}\,\chi^{ac}), 
\end{equation}
where $r$ denotes the Ricci scalar of $h$. With 
$\theta = \Omega^{-\frac{1}{2}}$ this equation takes the form
of Lichnerowicz' equation 
\begin{equation}
\label{confHamconstr}
(\Delta_h - \frac{1}{8}\,r)\theta =
- \frac{1}{8}\,\theta\,\left((\chi_c\,^c)^2 - \chi_{ab}\,\chi^{ab}\right)
+ \frac{1}{2}\,\theta^3\,\Sigma\,\chi_c\,^c.
\end{equation}
By taking the trace of (\ref{bzw8}) and using (\ref{bzw2}) one gets
\begin{equation}
\label{confmomconstr}
D^b(\Omega^{-2}\,\chi_{bc}) = \Omega^{-2}\,D_c\,\chi_b\,^b
- 2\,\Omega^{-3}\,D_c\Sigma.
\end{equation}
Equations (\ref{confHamconstr}) and (\ref{confmomconstr}) correspond to the
Hamiltonian and the momentum constraint respectively.
Assuming now 
\begin{equation}
\label{gaugemax}
\chi_a\,^a = 0\,\,\,\mbox{and (the choice of gauge)}\,\,\,\Sigma = 0\,\,\,
\mbox{on}\,\,\,{\cal S},
\end{equation}
which imply  
$\tilde{\chi}_a\,^a = 0$, equations (\ref{confHamconstr}) and
(\ref{confmomconstr}) suggest to proceed as follows:
(i) prescribe $h$ on ${\cal S}$ and solve the equation $D^a\,\psi_{ab} =
0$ for a symmetric $h$-trace free tensor field $\psi_{ab}$ on ${\cal S}$,
(ii) solve equation (\ref{confHamconstr}) with $\chi_{ab} = \theta^{-4}
\psi_{ab}$ for a positive function $\theta$, i.e. solve
\begin{equation}
\label{redconfHamconstr}
(\Delta_h - \frac{1}{8}\,r)\theta 
= \frac{1}{8}\,\theta^{-7}\,\chi_{ab}\,\chi^{ab},\,\,\,\,\,
\,\,\,\theta > 0.
\end{equation}
The fields $\Omega = \theta^{-2}$, $h_{ab}$, and 
$\chi_{ab} = \Omega^2\psi_{ab}$ then solve (\ref{OconfHamconstr}) and
(\ref{confmomconstr}).
If it is required that 
\begin{equation}
\label{falloffati}
\rho\,\Theta \rightarrow 1,\,\,\,\,\,
\psi_{ab} = O(\frac{1}{\rho^4})\,\,\,\,\,\,\,\mbox{as}\,\,\,
\rho \rightarrow 0,
\end{equation}
where $\rho(p)$ denotes near $i$ the $h$-distance of a point $p$ from
$i$, the fields $\tilde{h}_{ab}$ and $\tilde{\chi}_{ab}$ related by
(\ref{confreddata}) to
$h_{ab}$ and
$\chi_{ab}$ satisfy the vacuum constraints and 
are asymptotically flat (\cite{friedrich:static}).

Using the conformal constraints to determine the remaining conformal fields
one gets
\begin{equation}
\label{sonS}
S = \frac{1}{3}\,\Delta_h\,\Omega + \frac{1}{12}\,\Omega\left(r +
\chi_{ab}\,\chi^{ab}\right) 
= \frac{1}{2\,\Omega}\,D_c\Omega\,D^c\Omega,
\end{equation}
\begin{equation}
\label{LaonS}
L_a = \frac {1}{\Omega}\,D^c\Omega\,\chi_{ca},
\end{equation}
\begin{equation}
\label{XtrfreeLabonS}
L_{ab} - \frac{1}{3}\,L_c\,^c\,h_{ab} =
- \frac {1}{\Omega}\left(D_a\,D_b\,\Omega 
- \frac{1}{3}\,\Delta_h \Omega\,h_{ab}\right),
\end{equation}
\begin{equation}
\label{XL00onS}
L_{00} = \frac{1}{6}\,R - L_c\,^c = \frac{1}{6}\,R 
- \frac{1}{4}\left(r + \chi_{ab}\,\chi^{ab}\right),
\end{equation}
\begin{equation}
\label{XwabonS}
w_{ab} = - \frac{1}{\Omega^2}\left(D_a\,D_b\Omega 
- \frac{1}{3}\,\Delta_h\Omega\,h_{ab}\right)
- \frac{1}{\Omega}\left(\chi_{ac}\,\chi_b\,^c 
- \frac{1}{3}\,\chi_{ce}\,\chi^{ce}\,h_{ab}
+ s_{ab}\right),
\end{equation}
\begin{equation}
\label{wstarabonS}
w^*_{ab} = -
\frac{1}{\Omega}\,D_c\,\chi_{e(a}\,\epsilon_{b)}\,^{ce},
\end{equation}
where we set $s_{ab} = r_{ab} -  \frac{1}{3}\,r\,h_{ab}$. The differential
identities (\ref{bzw3}) - (\ref{bzw7}), which are not needed to get these
expressions, will be then also be satisfied (cf. \cite{friedrich:hypivp}).

In view of conditions (\ref{falloffati}) most of these fields will in general be
singular at $i$. One will have $w_{ab} = O(r^{-3})$ near $i$ whenever the
ADM mass $m$ of the initial data set $\tilde{h}_{ab}$, $\tilde{\chi}_{ab}$
does not vanish. Controlling the time evolution of these data
requires a careful analysis of these singularities. 
As a simplifying hypothesis we assume, as in
\cite{friedrich:i-null}, that the data are time reflection symmetric and
define a smooth conformal structure, i.e.
\begin{equation}
\label{tiresydata}
h_{ab} \in C^{\infty}({\cal S}),\,\,\,\,\,\,\,\,\chi_{ab} = 0.
\end{equation} 
We note that much of the following discussion can be extended to 
more general data such as those considered in \cite{dain:friedrich}
and the more general class of data discussed in \cite{dain:in prepr}, which
includes the stationary data. 

The Ricci scalar $R$ is at our disposal. With
$R = \frac{3}{2}\,r$ one gets on ${\cal S}$ 
\begin{equation}
\label{trfreeLabonS}
L_{ab} =
- \frac {1}{\Omega}\left(D_a\,D_b\,\Omega 
- \frac{1}{3}\,\Delta_h \Omega\,h_{ab}\right)
+ \frac{1}{12}\,r\,h_{ab} ,
\end{equation}
\begin{equation}
\label{L00onS}
L_{00} = 0,\,\,\,\,\,
L_{0a} = 0,\,\,\,\,\,w^*_{ab} = 0, 
\end{equation}
\begin{equation}
\label{wabonS}
w_{ab} = - \frac{1}{\Omega^2}\left(D_a\,D_b\Omega 
- \frac{1}{3}\,\Delta_h\Omega\,h_{ab}
+ \Omega\,s_{ab}\right).
\end{equation}
In spite of this simplification the crucial problem is still present; one
finds that $w_{ab} = O(\rho^{-3})$ near $i$ if $m \neq 0$
(cf. (\ref{rescWeylsing})).

\subsection{Time reflection symmetric asymptotically flat Cauchy data.}
\label{trsafcdat}

To allow for more flexibility in the following analysis, we also want to
admit non-trivial cases with vanishing or negative mass. The positive mass
theorem (\cite{schoen:yau}) then tells us that we must allow for
non-compact ${\cal S}$. This will create no problems because we are
interested only in the behaviour of the fields near space-like
infinity. 

Let $x^a$, $a = 1, 2, 3$, denote $h$-normal coordinates defined on some
convex open normal neighbourhood ${\cal U}$ of $i$ so that with 
$h = h_{ab}(x^c)\,d\,x^a\,d\,x^b$  
\begin{equation}
\label{hnormal}
x^a(i) = 0,\,\,\,\,\,\,\,\,\,\,\,\, 
x^a\,h_{ab}(x^c) = - x^a\,\delta_{ab}\,\,\,\,\,\,\,\,
\mbox{on}\,\,\,\,\,\,\,\,{\cal U}. 
\end{equation}
All equations of this subsection will be written in these coordinates.
We set  $|x| = \sqrt{\delta_{ab}\,x^a\,x^b}$ and
$\Upsilon = |x|^2 = \delta_{ab}\,x^a\,x^b$ so that
\begin{equation}
\label{eik}
h^{ab}\,D_a\,\Upsilon\,D_b\,\Upsilon = - 4\,\Upsilon,
\end{equation}
and 
\begin{equation}
\label{ddGammai}
\Upsilon(i) = 0,\,\,\,\,\,\,\,
D_a\,\Upsilon(i) = 0,\,\,\,\,\,\,\,
D_a\,D_c\,\Upsilon(i) = - 2\,h_{ac}.
\end{equation}
By taking derivatives of (\ref{eik}) and using (\ref{ddGammai}) one obtains
\begin{equation}
\label{Gammaexp}
D_a\,D_b\,D_c\,\Upsilon(i) = 0,\,\,\,\,\,\,\,\,
D_a\,D_b\,D_c\,D_d\,\Upsilon(i) = - \frac{4}{3}\,r_{a(cd)b}[h](i),
\end{equation}
where the curvature tensor of $h$ is given by 
\[
r_{abcd}[h] = 2 \{h_{a[c} l_{d]b} + h_{b[d} l_{c]a}\} 
\]
with $l_{ab}[h] = r_{ab}[h] - \frac{1}{4}\,r[h]\,h_{ab}$ because
dim$({\cal S}) = 3$. Proceeding further in this way on can determine an
expansion of $\Upsilon$ in terms curvature terms at $i$. 
The relations above imply in particular
\begin{equation}
\label{lapups}
(\Delta_h \Upsilon + 6)(i) = 0,\,\,\,\,\,\,\,\,
D_a(\Delta_h \Upsilon + 6)(i) = 0,\,\,\,\,\,\,\,\,
D_a\,D_b(\Delta_h \Upsilon + 6)(i) = \frac{4}{3}\,r_{ab}(i).
\end{equation}

\vspace{.3cm}

Equation (\ref{redconfHamconstr}) and the first of equations
(\ref{falloffati}) can be combined under our assumptions into
\[
(\Delta_h - \frac{1}{8}\,r)\,\theta = 4\,\pi\,\delta_i,
\]
where in the coordinates $x^a$ the symbol $\delta_i$ denotes the
Dirac-measure with weight $1$ at $x^a = 0$. In a neighbourhood of $i$ there
exists then a representation
$\theta = \frac{U}{|x|} + W$ with functions $U$, $W$ which satisfy 
\begin{equation}
\label{colap}
(\Delta_h - \frac{1}{8}\,r)\,\left(\frac{U}{|x|}\right) =
4\,\pi\,\delta_i,\,\,\,\,\,\,\,
(\Delta_h - \frac{1}{8}\,r)\,W = 0\,\,\,\,\,\,
\mbox{near}\,\,\,i,
\end{equation}
and 
\begin{equation}
\label{UWival}
U(i) = 1,\,\,\,\,\,W(i) = \frac{m}{2},
\end{equation}
where $m$ denotes the ADM-mass of the solution.  The functions $U$, $W$ are
analytic on ${\cal U}$ if $h$ is analytic (\cite{garabedian}) and smooth if $h$
is $C^{\infty}$ (\cite{dain:friedrich}).

The function $\sigma \equiv \frac{\Upsilon}{U^2}$ is characterized uniquely by
the conditions that it is smooth, satisfies the equation
$(\Delta_h - \frac{1}{8}\,r)\,\sigma^{-1/2} = 4\,\pi\,\delta_i$, and the
relations
\begin{equation}
\label{sigmaval}
\sigma(i) = 0,\,\,\,\,\,\,D_a\sigma(i) = 0,\,\,\,\,\,\,
D_a\,D_b\,\sigma(i) = - 2\,h_{ab},
\end{equation}
hold, which follow from (\ref{ddGammai}) and (\ref{UWival}). 
If $\sigma'$ is another function satisfying these conditions, then
$\sigma' = \Upsilon\,U^{'-2}$ with $U' = 1 + O(|x|)$
by (\ref{sigmaval}) and $U' \in C^{\infty}({\cal U})$ by the results of
\cite{dain:friedrich}. The function $f = \sigma^{-1/2} - \sigma^{'-1/2}$ then
solves $(\Delta_h - \frac{1}{8}\,r)\,f = 0$ and it follows that
$f \in C^{\infty}({\cal U})$ and
$|x|\,f = U - U' \in C^{\infty}({\cal U})$. Expanding $f$ and $U - U'$ in terms
of spherical harmonics it follows from the last equation that $f$ vanishes at
$i$ at any order. Since $f$ satisfies the conformal Laplace equation it follows
that $f = 0$ on ${\cal U}$ by theorem 17.2.6 of \cite{hoermander:III}.
This implies that $\sigma' = \sigma$ on ${\cal U}$.

The first of equations (\ref{colap}) can be rewritten in the form 
\begin{equation}
\label{Uconstr}
2\,D^a\,\Upsilon\,D_aU + (\Delta_h\,\Upsilon + 6)\,U 
- 2\,\Upsilon\,(\Delta_h - \frac{1}{8}\,r)\,U = 0.
\end{equation}
This allows us to determine from (\ref{sigmaval}) recursively an asymptotic
expansion of $U$, which is convergent if $h$ is real analytic.
The Hadamard parametrix construction is based on an ansatz
\begin{equation}
\label{Uexp}
U = \sum_{p = 0}^{\infty} U_p\,\Upsilon^p,
\end{equation} 
by which the calculation of $U$ is reduced to an ODE problem.
The functions $U_p$ are obtained recursively by 
\[
U_0 = \exp\left\{ \frac{1}{4}\,\int_0^{\Upsilon^{\frac{1}{2}}}
(\Delta_h\,\Upsilon + 6)\, \frac{d\,\rho}{\rho} \right\},
\]
\[
U_{p + 1} = - \frac{U_0}{(4p - 2)\,\Upsilon^{\frac{p + 1}{2}}}\,
\int_0^{\Upsilon^{\frac{1}{2}}}\frac{\Delta_h[U_p]\,\rho^p}{U_0}\,d\,\rho,
\,\,\,\,\,\,p = 0, 1, 2, \dots,
\]  
where the integration is performed in terms of the affine parameter 
$\rho = \Upsilon^{\frac{1}{2}} = |x|$ along the geodesics emanating from
$i$. It follows that
\begin{equation}
\label{Ubehave}
U(i) = 1,\,\,\,\,\,D_a\,U(i) = 0,\,\,\,\,\, 
D_a\,D_b\,U(i) = \frac{1}{6}\,l_{ab}[h],
\end{equation} 
which implies
\begin{equation}
\label{Bsigmaival}
D_a\,D_b\,D_c\sigma(i) = 0,\,\,\,\,
D_a\,D_b\,D_c\,D_d\sigma(i) = 2\,(h_{cd}\,l_{ab} + h_{ab}\,l_{cd}). 
\end{equation}

\vspace{.3cm}

Given $h$ and the solution $W$ of the conformal Laplace equation in (\ref{colap}),
the considerations above show us how to determine an expansion of the function
\begin{equation}
\label{Omegalocal}
\Omega = \theta^{-2} = \left( \frac{1}{\sqrt{\sigma}} + W \right)^{-2}
= \frac{\Upsilon}{(U + \rho\,W)^2},
\end{equation}
in terms of $\rho$ at all orders. Corresponding expansions can be obtained
for the conformal data (\ref{sonS}), (\ref{trfreeLabonS}), (\ref{L00onS}),
(\ref{wabonS}).

\vspace{.3cm}

While $U$ is thus seen to be determined locally by the metric $h$, the
function $W$ carries non-local information. Cases where
$\partial^{\alpha}_{x^a}\,W(i) = 0$ for all multiindices 
$\alpha = (\alpha^1, \alpha^2, \alpha^3) \in \mathbb{N}^3$ with
$|\alpha| \equiv \alpha^1 + \alpha^2 + \alpha^3 \le N$ for some
non-negative integer $N$ or for $N = \infty$ will also be of interest in
the following. In the latter case we have in fact (\cite{hoermander:III})
\begin{equation}
\label{massless}
W = 0\,\,\,\,\,\,\mbox{near}\,\,\,\,i.
\end{equation} 
For convenience this case will be  referred to 
as the {\it massless case}.

\vspace{.3cm}

A rescaling  
\[
h \rightarrow h' = \vartheta^4\,h,\,\,\,\,\,\,
\Omega \rightarrow \Omega' = \vartheta^2\,\Omega,
\] 
with a smooth positive factor $\vartheta$ satisfying $\vartheta(i) = 1$,
leaves $\tilde{h} = \Omega^{-2}\,h$ unchanged but implies changes
\[
\theta \rightarrow \theta' = \vartheta^{-1}\,\theta,\,\,\,\,\,\,
U \rightarrow U' = \frac{|x'|}{|x|}\,\vartheta^{-1}\,U,\,\,\,\,\,\,
W \rightarrow W' = \vartheta^{-1}\,W,
\] 
where $|x'|$ is defined in terms of $h'$-normal coordinates
$x^{a'}$ as described above. Due to the conformal covariance of the
operator on the left hand sides of equations (\ref{colap}), relations
(\ref{colap}), (\ref{UWival}) will then also hold with all fields replaced by
the primed fields.

To reduce this freedom it has been assumed in \cite{friedrich:i-null}
that the metric $h$ is given near $i$ on ${\cal S}$ in the 
{\it cn-gauge}\index{cn-gauge}. 
By definition, this conformal gauge is satisfied by $h$ if there exists
a 1-form $l_*$ at $i$ such that the following holds. If $x(\tau)$, $l(\tau)$
solve the conformal geodesic equations (with respect to $h$) with $x(0) =
i$, $l(0) = l_*$, and  
$h(\dot{x}, \dot{x}) = 1$ at $i$, then a frame $e_a$ which is $h$-orthonormal 
at $i$ and satisfies $\hat{D}_{\dot{x}}\,e_a = 0$ (with
$\hat{D} - D = S(l)$), stays $h$-orthonormal near $i$. This gauge can be
achieved without restrictions on the mass and fixes the scaling uniquely up to
a positive real number and a 1-form given at $i$. It admits an easy discussion
of limits where $m \rightarrow 0$. 

If $m > 0$, it is convenient to set above $\vartheta = \frac{2}{m}\,W$. It
follows then that $W' = \frac{m}{2}$, whence
$0 = (\Delta_{h'} - \frac{1}{8}\,r[h'])\,W' = - \frac{m}{16}\,r[h']$.
Thus, if $m > 0$, we can always assume $h$ to be given such that
\begin{equation}
\label{mgauge} 
r[h] = 0,\,\,\,\,\,\,\Omega = \frac{\sigma}{(1 + \sqrt{\mu\,\sigma})^2}    
\,\,\,\,\,\,\mbox{with}\,\,\,\,\,\,\sqrt{\mu} = \frac{m}{2}.
\end{equation}
In this gauge the function $\sigma$ satisfies near $i$ the equation
$\Delta_h\,(\sigma^{-1/2}) = 4\,\pi\,\delta_i$, which implies by
(\ref{sigmaval})
\begin{equation}
\label{an2fequ}
2\,\sigma\,s =
D_a\sigma\,D^a\sigma
\,\,\,\,\,\,\,\,\mbox{with}\,\,\,\,
s \equiv \frac{1}{3}\,\Delta_h\,\sigma,
\end{equation}
(note that an analogous equation holds with $\sigma$ replaced by
$\Omega$). Equation (\ref{an2fequ}) implies in turn together with
(\ref{sigmaval}) the Poisson equation above.

For later reference we note the form of the conformal
Schwarzschild data in this gauge. In isotropic coordinates the
Schwarzschild line element is given by
\[
d\,\tilde{s}^2 = \left(\frac{1 - m/2\,\tilde{r}}{1 +
m/2\,\tilde{r}}\right)^2d\,t^2 - (1 + m/2\,\tilde{r})^4\,(d\,\tilde{r}^2 +
\tilde{r}^2\,d\,\sigma^2).
\]
Expressing the initial data $\tilde{h}$, $\tilde{\chi}$
induced on $\{t = 0\}$ in terms of the
coordinate $\rho = 1/\tilde{r}$, one finds that $\tilde{\chi} = 0$ 
and $\tilde{h} = \Omega^{-2}\,h$ with
\begin{equation}
\label{confschwarzschilddata}
h = -\,(d\,\rho^2 + \rho^2\,d\,\sigma^2),\,\,\,\,\,\,\,\,
\Omega = \frac{\rho^2}{(1 + \frac{m}{2}\,\rho)^2},
\end{equation}
so that $\sigma = \Upsilon = \rho^2$ resp. $U = 1$. The metric $h$ also
satisfies a cn-gauge.

\subsection{Static asymptotically flat Cauchy data.}
\label{conformallystatic}

Static solutions to the vacuum field equations can be written in
the form
\[
\tilde{g} = v^2\,dt^2 + \tilde{h},
\]
with $t$-independent negative definite metric 
$\tilde{h}$ and
$t$-independent norm $v = \sqrt{\tilde{g}(K, K)} > 0$ of the time-like
Killing field $K = \partial_t$. With the $\tilde{g}$-unit normal of a slice
$\{t = const.\}$ being given by $\tilde{n} = v^{-1}\,K$ and the associated
orthogonal projector by 
$\tilde{h}_{\mu}\,^{\nu} = \tilde{g}_{\mu}\,^{\nu} -
\tilde{n}_{\mu}\,\tilde{n}^{\nu}$, one gets for  the second fundamental
form on this slice  
\[
\tilde{\chi}_{\mu\nu} = 
v^{-1}\,\tilde{h}_{\mu}\,^{\rho}\,\tilde{h}_{\nu}\,^{\delta}\,
\tilde{\nabla}_{\rho}\,K_{\delta} = 0,
\]
because it is symmetric by $\tilde{n}$ being hypersurface
orthogonal while the second term is anti-symmetric by the Killing equation.
The solutions are thus time reflection symmetric.

For these solutions the vacuum field equations are equivalent to the
requirement that the {\it static vacuum field equations}\index{static vacuum
field equations} 
\begin{equation}
\label{vacstaequ}
r_{ab}[\tilde{h}] = \frac{1}{v}\,\tilde{D}_a\,\tilde{D}_b\,v,
\,\,\,\,\,\,\,\,\,\,\,\,\,\,\,\,
\Delta_{\tilde{h}}\,v = 0,
\end{equation}
hold on one and thus on any slice $\{t = const.\}$. In harmonic
coordinates these equations become elliptic and $\tilde{h}$ and $v$ thus
real analytic. 

We consider solutions $\tilde{h}$, $v$ to equations (\ref{vacstaequ})
with non-vanishing ADM-mass which are given on a 3-manifold 
$\tilde{{\cal S}}$
which is mapped by suitable coordinates $\tilde{x}^a$ diffeomorphically
to $\mathbb{R}^3 \setminus \bar{{\cal B}}$, where $\bar{{\cal B}}$ is a
closed ball in
$\mathbb{R}^3$. We assume that $\tilde{h}$ satisfies in these coordinates the
usual condition of asymptotic flatness and $v \rightarrow 1$ as 
$|\tilde{x}| \rightarrow \infty$. The work in \cite{beig:simon} (cf. also
\cite{kennefick:o'murchadha} for a strengthening of this result) then
implies that the conformal structure defined by $\tilde{h}$ 
can be extended analytically to space-like infinity. The physical
3-metric $\tilde{h}$ therefore belongs to the class of data considered
above.

For such solutions it follows from the discussion in
\cite{beig:schmidt:2000} that the gauge (\ref{mgauge}) is achieved if
any of the equivalent equations 
\begin{equation}
\label{vomsig}
v = 1 - m\,\sqrt{\Omega} = 
\frac{1 - \sqrt{\mu\,\sigma}}{1 + \sqrt{\mu\,\sigma}}
,\,\,\,\,\,\,\,\,\,\,\,\,\,\,
\sigma = \left(\frac{2}{m}\,\,\frac{1 - v}{1 + v}\right)^2,
\end{equation}
holds. The set ${\cal S} = \tilde{{\cal S}} \cup \{i\}$ can then be
endowed with a differential structure such that the metric $h =
\Omega^2\,\tilde{h}$ extends as a real analytic metric to $i$. We shall
consider in the following $h$-normal coordinates as in
(\ref{hnormal}) such that the functions $\sigma(x^c)$, $h_{ab}(x^c)$ are
then real analytic on ${\cal U}$.
The first of the static vacuum field equations (\ref{vacstaequ})
then implies
\begin{equation}
\label{n1fequ}
0 = \Sigma_{ab} \equiv D_a\,D_b\,\sigma - s\,h_{ab}
+ \sigma\,(1 - \mu\,\sigma)\,r_{ab}[h], 
\end{equation}
where $s$ is defined as in (\ref{an2fequ}). The second of equations
(\ref{vacstaequ}) implies $r[\tilde{h}] = 0$ and can thus be read as a
conformally covariant Laplace equation for $v$. Using the transformation
rule for this equation and observing (\ref{vomsig}), we find that it
transforms into
\[
0 = (\Delta_h - \frac{1}{8}\,r[h])(\theta\,v)
= (\Delta_h - \frac{1}{8}\,r[h])(\theta - m)
\,\,\,\,\,\,\,\mbox{on}\,\,\,\,
\tilde{{\cal S}},
\]
and is thus satisfied by our assumption $r[h] = 0$.

We shall repeat some of the considerations of
\cite{friedrich:static} in the present conformal gauge. The fact that
solutions to the conformal static field equations are real analytic and can
be extended by analyticity into the complex domain allows us to use some very
concise arguments.  We note that the statements obtained here can also be
obtained by recursive arguments. This will become important if
some of the following considerations are to be transferred to 
$C^{\infty}$ or $C^k$ situations. 

From (\ref{n1fequ}) one gets  
\[
D_c\,\Sigma_{ab} = D_c\,D_a\,D_b\,\sigma - D_c\,s\,h_{ab}
+ \sigma\,(1 - \mu\,\sigma)\,D_c\,r_{ab}
+ (1 - 2\,\mu\,\sigma)\,D_c\,\sigma\,r_{ab}.
\]
With the Bianchi identity, which takes in the present gauge the form
$D^a\,r_{ab} = 0$, follow
the integrability conditions
\begin{equation}
\label{1intco}
0 = \frac{1}{2}\,\,D^c\,\Sigma_{ca} =
D_a\,s + (1 - \mu\,\sigma)\,r_{ab}\,D^b\,\sigma,
\end{equation}
and 
\begin{equation}
\label{cott}
0 = D_{[c}\,\Sigma_{a]b} + \frac{1}{2}\,\,D^d\,\Sigma_{d[c}\,h_{a]b}
\end{equation}
\[
= \sigma \left\{
(1 - \mu\,\sigma)\,D_{[c}r_{a]b} 
- \mu\,(
2\,D_{[c}\sigma\,r_{a]b} + D^d\,\sigma\,r_{d[c}\,h_{a]b})
\right\}.
\]
Equation (\ref{n1fequ}) thus implies an expression for the 
Cotton tensor\index{Cotton tensor}, 
which is given in the present gauge by 
$b_{bca} = D_{[c}r_{a]b}$,
and for its dualized version, which is given by
\begin{equation}
\label{twoindCott}
b_{ab} =
\frac{1}{2}\,b_{acd}\,\epsilon_b\,^{cd}
= \frac{\mu}{1 - \mu\,\sigma}(
D_c\,\sigma\,r_{da}\,\epsilon_{b}\,^{cd} 
- \frac{1}{2}\,r_{de}\,D^e\,\sigma
\,\epsilon_{ba}\,^d).
\end{equation}
It follows that
\[
D_a(2\,\sigma\,s - D_c\,\sigma\,D^c\,\sigma) = 
\sigma\,\,D^c\Sigma_{ca}
- 2\,D^c\sigma\,\Sigma_{ca},
\]
which shows that equation (\ref{an2fequ}) is a consequence of
equations (\ref{sigmaval}) and (\ref{n1fequ}) and that the latter contain
the complete information of the conformal static field equations.

Let $e_{\bf a} = e^c\,_{\bf a}\,\partial_{x^c}$, ${\bf a} = 1, 2, 3$, now
denote the $h$-orthonormal frame field on ${\cal U}$ which is parallely
transported along the $h$-geodesics through $i$ and satisfies 
$e^c\,_{\bf a} = \delta^c\,_{\bf a}$ at
$i$. In the following we assume all tensor fields, except the frame
field $e_{\bf a}$ and the coframe field $\sigma^{\bf c}$ dual
to it, to be expressed in term of this frame field and set 
$D_{\bf a} \equiv D_{e_{\bf a}}$. The coefficients of $h$
are then given by $h_{\bf ab} = - \delta_{\bf ab}$. Any analytic tensor
field $T^{{\bf b}_1 \ldots {\bf b}_q}_{{\bf a}_1 \ldots {\bf a}_p}$ on $V$
has an expansion of the form (cf. \cite{friedrich:i-null})
\[
T^{{\bf b}_1 \ldots {\bf b}_q}_{{\bf a}_1 \ldots {\bf a}_p}(x) = 
\sum_{k \ge 0} \frac{1}{k!}\,x^{c_k} \ldots x^{c_1}\,
(D_{{\bf c}_k} \ldots D_{{\bf c}_1}\,
T^{{\bf b}_1 \ldots {\bf b}_q}_{{\bf a}_1 \ldots {\bf a}_p})(i),
\] 
(where the summation rule ignores whether indices are bold face or not). 

We want to discuss how expansions of this form can be obtained for the
fields
\[
\sigma,\,\,s,\,\, r_{\bf ab},
\]
which are provided by the solutions to the conformal static field
equations. Once these fields are known, the coefficients of the 1-forms
$\sigma^{\bf a} = \sigma^{\bf a}\,_b\,d\,x^b$, which provide the
coordinate expression of the metric by the relation 
$h = - \delta_{\bf ac}\,\sigma^{\bf a}\,_b\,\sigma^{\bf c}\,_d\,
d\,x^b\,d\,x^d$, and the connection coefficients
$\Gamma_{\bf a}\,^{\bf b}\,_{\bf c}$ with respect to $e_{\bf a}$
can be obtained from the structural equations in
polar coordinates (cf. \cite{helgason})
\[
\frac{d}{d\,\rho}\left(\rho\,\sigma^{\bf a}\,_{b}(\rho\,x)\right)
= \delta^{\bf a}\,_{b} 
+ \rho\,\Gamma_{\bf c}\,^{\bf a}\,_{\bf d}(\rho\,x)\,x^d
\,\sigma^{\bf c}\,_{b}(\rho\,x),
\]
\[
\frac{d}{d\,\rho}\left(\rho\,\Gamma_{\bf a}\,^{\bf c}\,_{\bf e}(\rho\,x)
\,\sigma^{\bf a}\,_{b}(\rho\,x)\right)
= \rho\,r^{\bf c}\,_{\bf e d a}(\rho\,x)\,x^d
\,\sigma^{\bf a}\,_{b}(\rho\,x).
\]

For this purpose we consider the data
\begin{equation}
\label{Tstatdata}
c_{{\bf a}_p \ldots {\bf a}_1 {\bf bc}} 
= {\cal R}(D_{{\bf a}_p} \ldots D_{{\bf a}_1}\,r_{\bf bc})(i),
\end{equation} 
where ${\cal R}$ means `trace free symmetric part of'. 
These data have the following interpretation. Since solutions to the
conformal static field equations are real analytic in the given
coordinates $x^a$, all the fields considered above can be extended into
a complex domain ${\cal U}' \subset \mathbb{C}^3$ which comprises 
${\cal U}$ as the subset of real points. The  subset 
${\cal N} = \{\Upsilon = 0\}$ of ${\cal U}'$, where we denote by
$\Upsilon$ again the analytic extension of the real function denoted
before by the same symbol, then defines the cone which is generated by the
complex null geodesics 
$\mathbb{C} \supset {\cal O} \ni \zeta \rightarrow x^a(\zeta) 
= \zeta\,x^a_*\in {\cal U}'$ through $i$, where 
$x^a_* \neq 0$ is constant with
$h_{ab}\,x^a_*\,x^b_* = 0$ at $i$. On
${\cal N}$ the field $D^a\Upsilon\,\partial_{x^a} =
-2\,x^a\,\partial_{x^a}$ is tangent to the null generators of ${\cal
N}$. 
The derivatives of $r_{\bf ab}\,\dot{x}^a\,\dot{x}^b$ with respect to
$\zeta$ at $i$ are given by the complex numbers 
\[
x^{a_p}_* \ldots x^{a_1}_*x^{b}_*x^{c}_*
D_{{\bf a}_p}\ldots D_{{\bf a}_1}r_{\bf bc}(i)
\]
\[ 
= \iota^{A_p}\iota^{B_p} \ldots \iota^{A_1}\iota^{B_1} 
\iota^{C} \ldots \iota^{F}
D_{A_p\,B_p} \ldots D_{A_1B_1}\,r_{CD EF}(i)
\]
where the term on the left hand side is rewritten on the right hand side
in space spinor notation and it is used that 
$x_*^{AB} \equiv \sigma^{AB}\,_a\,x_*^a = \iota^A\,\iota^B$ 
with some spinor $\iota^A$  because $x_*^a$ is a null
vector. Allowing $x_*^a$ to vary over the null cone at $i$, i.e.
allowing $\iota^A$ to vary over $P^1(\mathbb{C})$, we can extract from the
numbers above the real quantities
\begin{equation}
\label{Sstatdata}
c_{A_p\,B_p \ldots A_1B_1 CD EF} = D_{(A_p\,B_p} \ldots
D_{A_1B_1}\,r_{CD EF)}(i),
\end{equation} 
which are equivalent to (\ref{Tstatdata}). 
Giving the data (\ref{Tstatdata})
is thus equivalent to giving $r_{\bf
ab}(\zeta\,x^a_*)\,\dot{x}^a\,\dot{x}^b$ where $x^a_*$ varies over a cut of
the complex null cone at $i$ or to giving, up to a scaling, the
restriction of $r_{ab}\,D^a\Upsilon\,D^b\Upsilon$ to
${\cal N}$. The data (\ref{Tstatdata}) are in one-to-one correspondence to
the multipole moments considered in \cite{beig:simon}.

We consider now the Bianchi identity $D^{\bf a}\,r_{\bf ab} = 0$ and 
equation (\ref{cott}). In space spinor notation they combine into
the concise form
\begin{equation}
\label{spsptwoindCott}
(1 - \mu\,\sigma)\,D_A\,^E r_{BCDE} =
2\,\mu\,r_{E(BCD}\,D_{A)}\,^E\sigma.
\end{equation}
Note that the contraction and symmetrization on the right hand side
project out precisely the information contained in
$r_{\bf ab}\,D^{\bf a}\Gamma\,D^{ \bf b}\Gamma$  while the contraction
which occurs on the left hand side prevents us from using the equation to
calculate any of the information in (\ref{Sstatdata}). 
We use equations (\ref{an2fequ}), (\ref{1intco}) in frame
notation. By taking formal derivatives of these equations one can determine
from (\ref{sigmaval}) and the data (\ref{Tstatdata}) all derivatives of
$\sigma$, $s$, and $r_{\bf ab}$ at $i$. The complete set of data
(\ref{Tstatdata}) resp. (\ref{Sstatdata}) is required for this and these
data determine the expansion uniquely. This procedure has been formalized
in the theory of `exact sets of fields' discussed in
\cite{penrose:rindler:I}, where equations of the type 
(\ref{spsptwoindCott}) are considered.

The formulation given above suggests proving a Cauchy-Kowalevska type
results for equations (\ref{an2fequ}), (\ref{1intco}),
(\ref{spsptwoindCott}) with data prescribed on ${\cal N}$. Although the
existence of the vertex at $i$ may create some difficulties in the present
case, this problem has much in common with the characteristic initial
value problem for Einstein's field equations for which the existence of
analytic solutions has been shown (\cite{friedrich:1982}).
At present, no decay estimates for the
$c_{{\bf a}_p \ldots {\bf a}_1 {\bf bc}}$ as $p \rightarrow \infty$ are
available which would ensure the convergence of these series. 
To simplify the following discussion we shall assume that the series 
considered above do converge. 

We return to the coordinate formalism and show that this
procedure provides a solution to the original equation (\ref{n1fequ}), i.e.
the quantity
$\Sigma_{ab}$ defined from the fields $\sigma$, $s$, and $r_{ab}$ 
by the procedure above does vanish. We show first that 
$\Sigma_{ab} = 0$ on ${\cal N}$. Since $\sigma = 0$ on ${\cal N}$, this
amounts to showing that $m_{ab} \equiv D_a\,D_b\,\sigma - s\,h_{ab}$
vanishes on ${\cal N}$. Differentiating twice the equation 
$D_a\sigma\,D^a\sigma -2\,\sigma\,s = 0$, which has been solved as part
of the procedure above, observing that $D^c\,\Sigma_{cd} = 0$ and
restricting the resulting equation to ${\cal N}$ gives the linear ODE
\[
D^c\sigma\,D_c\,m_{ab} = - D_aD^c\sigma\,\,m_{cb},
\]
along the null generators of ${\cal N}$.
Observing that $D_a\,D_b\,\sigma = - 2\,h_{ab} + O(\Upsilon)$, this
ODE can be written along the null geodesics $x^a(\zeta) = \zeta\,x^a_*$ 
considered above in the form
\[
\frac{d}{d\,\zeta}(\zeta\,m_{ab}) = A_a^c\,\zeta\,m_{cb},
\] with a smooth function $A_a^c = A_a^c(\zeta)$. This implies the desired
result. In view of (\ref{1intco}), (\ref{cott}) it shows that we solved the
problem
\[
D^c\,\Sigma_{ca} = 0,\,\,\,\,\,
D_{[c}\,\Sigma_{a]b} = 0
\,\,\,\,\,\,\,\mbox{near}
\,\,\,\,i,\,\,\,\,\,\,\,
 \Sigma_{ab} = 0
\,\,\,\,\,\,\,\mbox{on}\,\,\,\,{\cal N}.
\]
The first two equations combine in space spinor notation into
$D_A\,^E\,\Sigma_{BCDE} = 0$ with symmetric spinor field $\Sigma_{ABCD}$.
Following again the arguments of \cite{penrose:rindler:I}, we conclude that
$\Sigma_{ab} = 0$.

Equation (\ref{twoindCott}) implies 
\begin{equation}
\label{stcottcond}
D^a\sigma\,D^b\sigma\,b_{ab} = 0
\,\,\,\,\,\mbox{on}\,\,\,\,V.
\end{equation}
A rescaling $h \rightarrow h' = \vartheta^4\,h$ with a positive
(analytic) conformal factor gives
$\sigma \rightarrow \sigma' = \vartheta^2 \sigma$ and 
$b_{ab} \rightarrow b_{ab}' = \vartheta^{-2}b_{ab}$, whence 
\[
D^a\sigma\,D^b\sigma\,b_{ab} \rightarrow
(D^a\sigma\,D^b\sigma\,b_{ab})' =
\]
\[
\vartheta^{-6}D^a\sigma\,D^b\sigma\,b_{ab}
+ 4\,\sigma\,\vartheta^{-7}
D^a\sigma\,D^b\vartheta\,b_{ab}
+ 4\,\sigma^2\,\vartheta^{-8}
D^a\vartheta\,D^b\vartheta\,b_{ab}.
\]
This shows that (\ref{stcottcond}) is not conformally
invariant, but it also shows that the relation
\begin{equation}
\label{Nstcottcond}
 D^a\Upsilon\,D^b\Upsilon\,b_{ab}|_{{\cal N}} = 0,
\end{equation}
implied by (\ref{stcottcond}), is conformally invariant. Using again 
the argument which allowed us to get the quantities (\ref{Sstatdata}),
we can translate this onto the equivalent relations 
\begin{equation}
\label{Tstaregcond}
{\cal R}(D_{a_p} \cdots D_{a_1}\,b_{bc}(i)) = 0,
\,\,\,\,\,\,p = 0, 1, 2, \ldots,
\end{equation}
which take in space spinor notation
the form (cf. (\ref{regcond}))
\begin{equation}
\label{Sstaregcond}
D_{(A_p\,B_p} \cdots
D_{A_1B_1}\,b_{CD EF)}(i) = 0,
\,\,\,\,\,\,p = 0, 1, 2, \ldots.
\end{equation}
We note that for given integer $p_* > 0$ the string of such conditions
with $0 \le p \le p_*$ is conformally invariant.

Since these condition have a particular bearing on the smoothness of
gravitational fields at null infinity (\cite{friedrich:static},
\cite{friedrich:i-null}) and it is not clear whether static equations
are of a greater significance in this context than expected so
far, we take a closer look at (\ref{stcottcond}).
If we apply the operators $D_aD_b$ and $D_aD_bD_c$ to (\ref{stcottcond})
and restrict the resulting equation to $i$, we get the relations
$b_{ab}(i) = 0$ and $D_{(a}b_{bc)}(i) = 0$ respectively, which agree with
(\ref{Tstaregcond}) at the corresponding orders because $D^a b_{ab} = 0$.
However, if we proceed similarly with 
$D_aD_bD_cD_d$, we get
\begin{equation}
\label{statrestr}
D_{(a}D_b\,b_{cd)}(i) = 0.
\end{equation}

Since (\ref{Tstaregcond}) with $p = 2$ can be written in the form
\[
D_{(a}D_b\,b_{cd)}(i) = \frac{1}{7}\,h_{(ab}\,\Delta_h\,b_{cd)}(i),
\]
the relation (\ref{statrestr}) implies in particular that
$\Delta_h\,b_{cd}(i) = 0$. It appears that in general this equation cannot
be deduced in the present gauge from known general identities and
(\ref{Nstcottcond}) alone. There will be similar such conditions at higher
orders. While the particular form of them may depend on the conformal
gauge, the existence of properties which go beyond (\ref{Tstaregcond})
does not. In any case, these observations  show that there is a gap
between $h$ satisfying the regularity conditions (\ref{Tstaregcond}) and
$h$ being conformally static.

This situation is also illustrated by the following observation. If the data
provided by $h$ are conformally flat in a neighbourhood of $i$ they trivially
satisfy conditions (\ref{Tstaregcond}). Without further assumptions the solution
$\theta$ to the Lichnerowicz equation which relates $h$ to the induced vacuum
data $\tilde{h} = \theta^4\,h$ can still be quite general. However, if
$\tilde{h}$ is static the function $\theta$ must be very
special. 

\begin{lemma}
\label{stcflat}
An asymptotically flat, static initial data set for the vacuum field equations
with conformal metric $h$ and positive ADM mass $m$ is locally conformally flat if
and only if it satisfies near $i$ in the gauge (\ref{mgauge}) the equation
$r_{ab}[h] = 0$ and thus in the normal coordinates (\ref{hnormal})
\[
h = - \delta_{ab}\,d\,x^a\,d\,x^b,\,\,\,\,\,\, U = 1,\,\,\,\,\,\, 
\theta = \frac{1}{|x|} + \frac{m}{2}.
\] 
\end{lemma}

\noindent
{\bf Remark}: 
This tells us that {\it the only asymptotically flat, static vacuum data which
are locally conformally flat near space-like infinity are the 
Schwarzschild data} (\ref{confschwarzschilddata}). 
The result of (\cite{valiente kroon:2003}), which suggests that conformal
flatness of the data $h$ near $i$ and the smoothness requirement on the
functions $u^p$ at
$I^{\pm}$ imply that the solution be asymptotically Schwarzschild, can thus be
reformulated as saying that for the given data the smoothness requirement implies
the solution to be asymptotically static at space-like infinity.

\vspace{.3cm}

Proof: By (\ref{cott}) the solution is locally conformally flat if and
only if
$2\,D_{[c}\sigma\,r_{a]b} = h_{b[c}\,r_{a]d}D^d\sigma$. 
Applying $D_e$ to this equation and observing (\ref{n1fequ}) and 
again $D_{[c}\,r_{a]b} = 0$, one gets after a contraction 
\begin{equation}
\label{ricode}
D^c\sigma\,D_c\,r_{ab} = - 3\,s\,r_{ab}  
+ \sigma\,(1 - \mu\,\sigma)\,(h_{ab}\,
r_{cd}\,r^{cd} - 3\,r_{ac}\,r_b\,^c). 
\end{equation}
This equation can be read as an ODE along the integral curves of the vector
field $D^c\,\sigma$. It follows from (\ref{an2fequ}) that  
$u^a = (2\,\sigma\,|s|)^{-\frac{1}{2}}\,D^a\,\sigma$ is a unit
vector field (with direction dependent limits at $i$). Because of
\[
u^a\,D_a\,\Upsilon = 
- \left(\frac{2\,\Upsilon}{|s|}\right)^{1/2}\,
(4\,U^{-1} + 2\,U^{-2}\,D^a\,U\,D_a\,\Upsilon) < 0, 
\]
its integral curves run into $i$ and cover in fact a (possibly
small) neighbourhood ${\cal U}'$ of $i$. 
Equation (\ref{ricode}) can be rewritten in the form
\[
u^a\,D_a(\Upsilon^{3/2}\,r_{bc}) =
A_{bc}^{de}\,(\Upsilon^{3/2}\,r_{de}),
\]
with the matrix valued function
\[
A_{bc}^{de} = 
- \frac{3}{\sqrt{2\,\sigma\,|s|}}\,
(s + 2\,U^{-1} + U^{-2}\,D^aU\,D_a\,\Upsilon)\,
h^d\,_b\,h^e\,_c
\]
\[
+ \frac{\sigma\,(1 - \mu\,\sigma)}{\sqrt{2\,\sigma\,|s|}}\,
(h_{bc}\,r^{d e} -
3\,r^d\,_b\,h^e\,_c),
\]
which is continuous on ${\cal U}'$. This implies that $r_{ab} = 0$ on
${\cal U}'$. 
The remaining statements follow immediately from (\ref{hnormal}) and
(\ref{mgauge}). 


\noindent
{\bf Remark}: 
We note that these data may be obtained in a different form if 
locally conformally flat data are given in the cn-gauge and
one asks under which conditions they are conformally
static. The data are then of the form
\[
h_{ab} = - \delta_{ab},\,\,\,\,\,\,
\Omega^{- \frac{1}{2}} = \theta = \frac{1}{|x|} + W,\,\,\,\,\,\,
\Delta_h\,W = 0,\,\,\,\,\,\,W(i) = \frac{m}{2} > 0.
\]
By a rescaling 
$h \rightarrow \vartheta^4\,h$,
$\theta \rightarrow \vartheta^{-1}\,\theta = \frac{1}{\vartheta\,|x|} +
\frac{m}{2}$ with $\vartheta = \frac{2}{m}\,W$ they are transformed into the
present gauge. Assuming that these data satisfy the conformal static field
equations and expressing the resulting equation again in terms of
$h_{ab} = - \delta_{ab}$ one finds that the solution is static if and
only if $2\,W\,D_a\,D_b\,W - 6\,\,D_a\,W\,D_b\,W
+ 2\,h_{ab}\,D_c\,W\,D^c\,W = 0.$
Since $W > 0$ the equation can be rewritten in terms of $w = W^{-2}$,
which gives 
\begin{equation}
\label{stobstrvan}
2\,w\,D_a\,D_b\,w = h_{ab}\,D_c\,w\,D^c\,w.
\end{equation}
Applying $D_c$, multiplying with $w$, and using twice (\ref{stobstrvan})
again, we conclude that $D_a\,D_b\,D_cw = 0$, whence
$w = k + k_a\,x^a + k_{ab}\,x^a\,x^b$ with some coefficients $k > 0$,
$k_a$, $k_{ab}$. This function satisfies
(\ref{stobstrvan}) if $k_{ab} = h_{ab}\,k_c\,k^c/4\,k$.
With $j_a = k_a/2\,k$ and $m = 2/\sqrt{k}$ this gives
\[
W = \frac{m}{2}\,\frac{1}{\sqrt{1 + 2\,j_a\,x^a
+ j_a\,j^a\,x_b\,x^b}},
\]
with constant $j^a$. 

That these data are equivalent to the ones considered above is seen by
rescaling with $\vartheta = \frac{2}{m}\,W$. By this one achieves 
$W = \frac{m}{2}$. For the metric $\vartheta^4\,h$ to acquire the flat
standard form one needs to perform a coordinate transformation which is
given by a special conformal transformation 
$x \rightarrow (I \circ T_c \circ I)(x)$ where $I$ denotes the inversion
$x^a \rightarrow x^a (\delta_{bc}\,x^b\,x^c)^{-1}$
and $T_c$ a translation $x^a \rightarrow x^a + c^a$ with suitably
chosen constant $c^a$.

\section{A regular finite initial value problem at space-like infinity}
\label{Rfivp}

In the conformal extension of Minkowski space described in section
\ref{assimple} neighbourhoods of space-like infinity, which are swept
out by future complete outgoing and past complete incoming null
geodesics, are squeezed into arbitrarily small neighbourhoods of the point
$i^0$. From the point of view of the causal structure it is 
natural to indicate space-like infinity by a point. The
discussion in section \ref{afcdat} shows, however, that in general 
$i^0$ cannot be a regular point of any smooth conformal extension. 
The condition for an extension to $i^0$ to be $C^{\infty}$  (under
our assumption (\ref{tiresydata})) is that the data are massless in
the sense of (\ref{massless}) and that the free datum $h$ satisfies the
conditions  (\ref{regcond}) with $p_* = \infty$
(\cite{friedrich:static}, \cite{friedrich:i-null}). Thus, smoothness
at $i^0$  excludes the physically interesting cases.

A direct discussion of the initial value problem for the conformal field equations
with initial data on an initial hypersurface ${\cal S} = \tilde{\cal S} \cup
\{i\}$ such that $W^i\,_{jkl} = O(\rho^{-3})$ at $i$ as discussed in section 
\ref{afcdat} faces considerable technical problems. Not only the functional
analytical treatement of a corresponding PDE problem poses enormous difficulties
but already the choice of gauge becomes very subtle. 

The setting described below has been arrived at by attempts to describe
the structure of the singularity as clearly as possible and to deduce from
the conformal field equations a formulation of the PDE problem which still
preserves `some sort of hyperbolicity' at space-like infinity.  It is based
on conformally invariant concepts so that possible singularities should be
identifiable as defects of the conformal structure.

In a conformal Gauss gauge based on a Cauchy hypersurface
$\tilde{{\cal S}}$ it turns out that after blowing up the
point $i$ into a sphere ${\cal I}^0$ and choosing 
the gauge suitably, one arrives at a formulation of the initial value problem
near space-like infinity in which the data can be smoothly extended to and
across ${\cal I}^0$. In that gauge also the evolution equations admit a
smooth extension to space-like infinity. The evolution and extension
process then generates from the set ${\cal I}^0$ a cylindrical piece of
space-time boundary diffeomorphic to $]-1, 1[ \times {\cal I}^0$, which is
denoted by ${\cal I}$. It represents space-like infinity and can be
considered as a blow-up of the point $i^0$. This boundary is neither
postulated nor attached `by hand'.

In this gauge the hypersurfaces ${\cal J}^{\pm}$  representing null infinity
near space-like infinity are given by finite values of the coordinates
which are explicitly known (it has to be shown, of course, that the
evolution extends far enough). These hypersurfaces touch the cylinder
${\cal I}$ at sets ${\cal I}^{\pm}$ diffeomorphic to ${\cal I}^0$, which can be
thought of as boundaries of ${\cal I}$ and of ${\cal J}^{\pm}$ respectively.
The structure of the conformal field equations near the {\it critical sets}
${\cal I}^{\pm}$ appears to be the key to the question of asymptotic
smoothness.

It may appear odd to sqeeze space-time regions of infinite
extend into arbitrarily small neighbourhoods of a point $i^0$ and then  
perform a complicated blow-up to resolve the singularity on the initial
hypersurface which has been created by the first step. The point of the
construction is that the finiteness of the sets ${\cal I}^{\pm}$ allow us to
disclose, to an extent that we can put our hands on it, a subtle feature of
the field equations which otherwise would be hidden at infinity (in the
standard vacuum representation) or in the singularity at
$i^0$ (in the standard conformal rescaling). 

In the following the setting indicated above and its various implications
will be discussed in detail.  While we shall add more recent results
we shall follow to a large extent the original article
\cite{friedrich:i-null}. For derivations and details we refer the reader to
this or the articles quoted below.

\subsection{The gauge on the initial slice and the blow-up at $i$}
\label{initialgaugeandblowup}

The non-smoothness of the conformal data (\ref{trfreeLabonS}),
(\ref{sonS}), (\ref{L00onS}), (\ref{wabonS}), (\ref{Omegalocal}) at $i$
arises from the presence of various factors $\rho$ in the explicit 
expressions. To properly take care of the specific radial and angular
behaviour of the various fields it is natural to choose the frame field in
the general conformal field equations such that the spatial vector fields
$e_a$, $a = 1, 2, 3$, are tangent to the initial hypersurface ${\cal S}$
and one of them, $e_3$ say, is radial. Since there is no preferred
direction at $i$, this only makes sense if the frame is chosen on
$\tilde{{\cal S}}$ such that it has direction dependent limits at $i$.
This singular situation finds a well organized description in terms of a
smooth submanifold of the bundle of frames.  To discuss the field equations
in the spin frame formalism, we will consider in fact a submanifold ${\cal
C}_e$ of the bundle of normalized spin frames over ${\cal S}$ near $i$.
While the use of spinors leads to various simplifications, it should be
mentioned that the construction could be carried out similarly in the
standard frame formalism (cf. \cite{friedrich:tueb}).

\subsubsection{The construction of ${\cal C}_e$}
\label{constrlift}

Consider now ${\cal S}$ as a space-like Cauchy hypersurface of a 4-dimensional
solution space-time $({\cal M}, g)$ with induced metric $h$ on ${\cal S}$.
Denote by
$SL({\cal S})$ the set of spin frames $\delta = \{\delta_A\}_{A = 0,1}$ on
${\cal S}$ which are normalized with respect to the alternating form
$\epsilon$, such that 
\begin{equation}
\label{epsilonnorm}
\epsilon(\delta_A, \delta_B) = \epsilon_{AB},\,\,\,\,\,\epsilon_{01} = 1.
\end{equation}
The group 
\[
SL(2, \mathbb{C}) = 
\{ t^A\,_B \in GL(2, \mathbb{C}) \,\,| \, \, \epsilon_{AC}t^A\,_B\,t^C\,_D 
= \epsilon_{BD} \}, 
\]
acts on $SL({\cal S})$ by $\delta \rightarrow \delta \cdot t =
\{\delta_A\,t^A\,_B\}_{B = 0,1}$.
The vector field $\tau = \sqrt{2}\,e_0$, with $e_0$ the future directed unit
normal of ${\cal S}$, defines a subbundle $SU({\cal S})$ of $SL({\cal S})$
which is given by the spin frames in $SL({\cal S})$ with
\begin{equation}
\label{taunorm}
g(\tau,\delta_A \bar{\delta}_{A'}) =
\epsilon_A\,^0 \epsilon_{A'}\,^{0'} + \epsilon_A\,^1 \epsilon_{A'}\,^{1'}
\equiv \tau_{AA'}.
\end{equation}
It has structure group
\[
SU(2) = 
\{ t^A\,_B \in SL(2, C) \,\,| \, \, \tau_{AA'}t^A\,_B\,\bar{t}^{A'}\,_{B'} 
= \tau_{BB'} \}. 
\]
In any frame in $SU({\cal S})$ the vector $\tau$ is given by
$\tau^{AA'}$. In the following we use the space spinor formalism in the
notation of \cite{friedrich:AdS}. Using the van der Waerden symbols for
space spinors  
\[
\sigma_a\,^{AB} = \sigma_a\,^{(A}\,_{A'}\tau^{B)A'},\,\,\,\,
\sigma^c\,_{AB} = \tau_{(B}\,^{A'} \sigma^c\,_{A)A'},\,\,\,\,c = 1,2,3,
\]
which satisfy
\[
h_{ab} = \sigma_{a\,AB}\sigma_b\,^{AB},\,\,\,\,\,\,\,\,
\epsilon_A \, ^B \, \epsilon_{A'}\,^{B'} 
= \frac{1}{2}\,\tau_{AA'} \tau^{BB'} +
\sigma^a\,_{AF}\tau^F\,_{A'}\tau^{EB'}\sigma_a\,_E\,^B,  
\]
where
\[
h_{ab}\,\sigma^a\,_{AB}\sigma^b\,_{CD}
= - \epsilon_{A(C} \epsilon_{D)B} \equiv h_{ABCD} 
\quad\mbox{with}\quad h_{ab} = - \delta_{ab}, 
\] 
the covering map onto the connected component $SO(3)$ of the rotation group is
given by 
\[
SU(2) \ni t^A \, _B \stackrel{\Psi}{\rightarrow} t^a\,_b = 
\sigma^a \, _{AB}\, t^A \, _C \, t^B \, _D \,\sigma_b \, ^{CD} 
\in SO(3).
\]
The induced isomorphism of Lie algebras will be denoted by $\Psi_{*}$.

The covering morphism of $SU({\cal S})$ onto the bundle $O_{+}({\cal S})$ of
positively oriented orthonormal frames on ${\cal S}$ maps the frame
$\delta \in SU({\cal S})$ onto the frame with vectors
$e_a = e_a(\delta) =
\sigma_a\,^{AB}\,\delta_A\,\tau_B\,^{B'}\,\bar{\delta}_{B'}$ such that 
$h(e_a, e_b) = h_{ab}$.  We use
this map to pull back to $SU({\cal S})$ the $h$-Levi-Civita connection form on 
$O_+({\cal S})$.
Combining this with the map $\Psi_{*}^{-1}$, the connection is
represented by an $su(2)$-valued connection form $\check{\omega}^A\,_B$ on
$SU({\cal S})$. Similarly, pulling back the
$\mathbb{R}^3$-valued solder form on $O_{+}({\cal S})$ and contracting with the
van der Waerden symbols results in a 1-form $\sigma^{AB}$ on $SU({\cal S})$
which is referred to as solder form on $SU({\cal S})$.   

Let $\check{H}$ denote the real horizontal vector field on $SU({\cal S})$
satisfying 
$<\sigma^{AB}, \check{H} >\, = \epsilon_0 \, ^{(A}  \, \, \epsilon_1 \, ^{B)}$ 
or, equivalently,
\begin{equation}
\label{Hcheckdef}
T_{\delta}(\pi) \, \check{H}(\delta) = 
\delta_{(0} \, \tau_{1)}\,^{B'} \bar{\delta}_{B'}
= \frac{1}{2}(\delta_0 \, \bar{\delta}_{0'} - \delta_1 \, \bar{\delta}_{1'})
, \, \, \, \delta \in SU({\cal S}). 
\end{equation}
It follows that 
$T_{\delta \, t}(\pi) \, \check{H}(\delta \, t) 
= T_{\delta}(\pi) \, \check{H}(\delta)$
if and only if
\[
t \in U(1) \equiv \{ t \in SU(2) \, | \,  t = 
 \left( \begin{array}{cc}
e^{i \phi} & 0 \\
0 & e^{- i \phi}
\end{array} \right), \phi \in \mathbb{R} \}.
\]
The field $\check{H}$ will essentially correspond to the `radial' vector field
mentioned above.  

We consider again the normal coordinates satisfying (\ref{hnormal}) near $i$,
set ${\cal B}_e = \{p \in {\cal U}\,|\,\,|x(p)| < e\}$ with $e > 0$ chosen
such that the closure of ${\cal B}_e$ in ${\cal S}$ is contained in ${\cal U}$,
and denote by $(SU({\cal B}_e), \pi)$ the restriction of $(SU({\cal S}), \pi)$
to ${\cal B}_e$. Let $\delta^{*}$ be in the fiber 
$\pi^{-1}(i) \subset SU({\cal B}_e)$ over $i$. The map 
$SU(2) \ni t \rightarrow \delta(t) \equiv \delta^{*} \cdot t \in \pi^{-1}(i)$ 
defines a smooth parametrization of $\pi^{-1}(i)$. 
We denote by  $]-e, e[\, \ni \rho \rightarrow \delta(\rho, t) \in SU({\cal
B}_e)$ the integral curve of the vector field $\sqrt{2} \, \check{H}$
satisfying  $\delta(0, t) = \delta(t)$ and set 
${\cal C}_e 
= \{\delta(\rho, t) \in SU({\cal B}_e)\,|\,\,\,|\rho| < e,\, t \in
SU(2) \}$. 
This set defines a smooth submanifold of $SU({\cal B}_e)$ which is
diffeomorphic to $]-e, e[ \times SU(2)$. The restriction of $\pi$ to this set
will be denoted by $\pi'$. 

The symbol $\rho$, which has been introduced already in section
\ref{trsafcdat}, is used here for the following reason. The integral
curves of
$\sqrt{2}\,\check{H}$ through $\pi^{-1}(i)$ project onto geodesics 
through $i$ with $h$-unit tangent vector. Thus, the projection $\pi'$ maps
${\cal C}_e$ onto ${\cal B}_e$. The action of
$U(1)$ on $SU({\cal B}_e)$ induces an action on ${\cal C}_e$.  While
${\cal I}^0 \equiv \pi^{-1}(i) =
\{\rho = 0 \}$  is diffeomorphic to $SU(2)$, the fiber $\pi^{'-1}(p) \subset
{\cal C}_e$  over a point $p$ in the punctured disk 
$\tilde{{\cal B}}_e \equiv {\cal B}_e \setminus \{i\}$  
coincides with an orbit of $U(1)$ in $SU({\cal B}_e)$ on which 
$\rho = |x(p)|$ and another one on which $\rho = - |x(p)|$.

The map $\pi'$ factorises as ${\cal C}_e \stackrel{\pi_1}{\rightarrow} 
{\cal C}_e' \stackrel{\pi_2}{\rightarrow} {\cal B}_e$ with 
${\cal C}_e' \equiv {\cal C}_e/U(1)$ diffeomorphic
to $]-e, e[ \times S^2$. 
For $\rho_*$ with $0 < |\rho_*| < e$ the subsets $\{\rho = \rho_*\}$ of
${\cal C}_e$ are diffeomorphic to $SU(2)$ and the restrictions of the map
$\pi_1$  to these sets define Hopf fibrations of the form
\begin{equation}
\label{ahopfmap} 
SU(2) \ni t \rightarrow \sqrt{2} \sigma^a\,_{AB}
t^A\,_0 t^B\,_1 \in S^2 \subset \mathbb{R}^3. 
\end{equation}
The set $\pi_2^{-1}(\tilde{{\cal B}}_e)$
(resp. $\pi^{'-1}(\tilde{{\cal B}}_e)$) consists of two components
${\cal C}_e^{'\pm}$ (resp. ${\cal C}_e^{\pm}$) on which $\pm \rho > 0$
respectively. Each of the sets ${\cal C}_e^{'\pm}$ is mapped by $\pi_2$
diffeomorphically onto the punctured disk.  If $\tilde{{\cal B}}_e$ is now
identified via $\pi_2$ with ${\cal C}_e^{'+}$ the manifold $\tilde{{\cal
B}}_e$ is embedded into
${\cal C}_e'$ such that it acquires the set $\pi_1({\cal I}^0) =
\pi_2^{-1}(i)$ as a boundary. The set $\bar{{\cal B}}_e \equiv
\tilde{{\cal B}}_e \cup \pi_2^{-1}(i) \simeq [0, e[ \times S^2$
is a smooth manifold with boundary. 
Viewing $\tilde{{\cal B}}_e$ again as the subset of 
$\tilde{{\cal S}} = {\cal S} \setminus \{i\}$, we get an extension
$\bar{{\cal S}}$ of $\tilde{{\cal S}}$ which can be thought of as being
obtained from ${\cal S}$ by blowing up the point $i$ into a sphere. 
This is our desired extension of the physical initial manifold
and the following discussion could be carried out in terms of the
3-dimensional manifold $\bar{{\cal B}}_e$. 

It turns out more convenient,
however, to use the 4-dimensional $U(1)$ bundle
$\bar{{\cal C}}^+_e = {\cal C}_e^+ \cup {\cal I}^0
= \{ \delta \in {\cal C}_e\,|\,\,\rho(\delta) \ge 0\}
\simeq [0, e[ \times SU(2)$.
It is a manifold with boundary smoothly embedded into $SU({\cal B}_e)$,  
from which it inherits various structures.
The set ${\cal C}_e$ is conveniently parametrized by
$\rho$ and the parallelizable group $SU(2)$. The solder and the connection
form on $SU({\cal B}_e)$ pull back to smooth 1-forms on ${\cal C}_e$. We
denote the latter again by $\sigma^{ab}$ and $\check{\omega}^a\,_b$
respectively. Any smooth spinor field $\xi$ on ${\cal B}_e$ defines on
${\cal C}_e$  a smooth `spinor valued function' which is given at
$\delta \in {\cal C}_e$ by the components of $\xi$ in the frame defined by 
$\delta$ and denoted (in the case of a covariant field) by 
$\xi_{A_1 \ldots A_k, {A'}_1\ldots {A'}_j}$. 
We shall refer to this function as to the `lift' of $\xi$.

The structure equations induce on ${\cal C}_e$ the equations
\begin{equation} 
\label{1stre}
d\sigma^{AB} = - \check{\omega}^A\,_E\,\wedge\,\sigma^{EB}
- \check{\omega}^B\,_E\,\wedge\,\sigma^{AE}, 
\end{equation}
\begin{equation}
\label{2stre}
d\check{\omega}^A\,_B = - \check{\omega}^A\,_E\,\wedge\,\check{\omega}^E\,_B
+ \check{\Omega}^A\,_B,
\end{equation}
where
\[
\check{\Omega}^A\,_B = \frac{1}{2}\,
r^A\,_{BCDF}\,\sigma^{CD}\,\wedge\,\sigma^{EF}
\]
denotes the curvature form determined by the curvature spinor $r_{ABCDEF}$.  
It holds
\begin{equation}
\label{3cuexp}
r_{ABCDEF} = 
\left( \frac{1}{2}\,s_{ABCE} - \frac{r}{12}\,h_{ABCE} \right) \,\epsilon_{DF}
+ \left( \frac{1}{2}\,s_{ABDF} - \frac{r}{12}\,h_{ABDF} \right) \,\epsilon_{CE}
\end{equation}
where $ s_{ABCD} = s_{(ABCD)}$ is the trace free part of the Ricci tensor of $h$
and $r$ its Ricci scalar. The curvature tensor of $h$ is given by 
\[
r_{AGBHCDEF} = - r_{ABCDEF}\epsilon_{GH} - r_{GHCDEF}\epsilon_{AB}.
\]
and the Bianchi identity reads $6\,D^{AB}\,s_{ABCD} = D_{CD}r$.

We use $t \in SU(2, C)$ and $x^1 \equiv \rho$ as `coordinates' on ${\cal C}_e$.
The vector field $\check{H}$ tangent to ${\cal C}_e$ then takes the form 
$\sqrt{2}\,\check{H} = \partial_{\rho}$. Consider now the basis 
\begin{equation}
\label{su2basis} 
u_1 = \frac{1}{2} \left( \begin{array}{cc}
0 & i \\
i & 0
\end{array} \right), \, \, 
u_2 = \frac{1}{2} \left( \begin{array}{cc}
0 & - 1 \\
1 & 0
\end{array} \right),\,\,
u_3 = \frac{1}{2} \left( \begin{array}{cc}
i & 0 \\
0 & - i
\end{array} \right),  
\end{equation}
of the Lie algebra $su(2)$. Here $u_3$ is the generator of the group $U(1)$. 
We denote by  $Z_{u_i}, \, i = 0,1,2$, the Killing vector fields generated on
$SU({\cal B}_e)$ by $u_i$ and the action of $SU(2)$. These fields are
tangent to ${\cal I}^0$. We set there    
\[ 
X_{+} \equiv - (Z_{u_2} + i Z_{u_1}), \, \,
X_{-} \equiv - (Z_{u_2} - i Z_{u_1}), \,\, X \equiv - 2\,i\,Z_{u_3},     
\]
and extend these fields smoothly to ${\cal C}_e$ by requiring 
\begin{equation}
\label{1comm}
[\check{H}, X] = 0, \, \,\, \,\, \,
[\check{H}, X_{\pm}] = 0.
\end{equation}
The vector fields $\check{H},\, X,\, X_{+},\,X_{-}$ constitute a frame
field on ${\cal C}_e$ which satisfies besides (\ref{1comm}) the commutation
relations 
\begin{equation}
\label{2comm}
[X,\, X_{+}] = 2 \, X_{+},\,\,  [X,\, X_{-}] = - 2 X_{-}, \, \,
[X_{+},\,X_{-}] = -X.
\end{equation}
The vector field $iX$ is tangent to the fibers defined by $\pi_1$. 
The complex vector fields $X_{+},\, X_{-}$ are complex conjugates of each other
such that $\overline{X_{-}\,f} = X_{+}\,f$ for any real-valued function $f$.

These vector fields are related to the 1-forms above by 
\begin{equation}
\label{siHXval}
<\sigma^{AB}, \check{H}>\, = \epsilon_0\,^{(A}\,\epsilon_1\,^{B)},\,\,\,\,\,
<\sigma^{AB}, X>\, = 0
\quad\mbox{on}\quad {\cal C}_e,
\end{equation}
\begin{equation}
\label{siXpmval}
<\sigma^{AB}, X_{+}>\, = \rho \, \epsilon_0\,^A \,\epsilon_0\,^B + O(\rho^2),
\,\,\, 
<\sigma^{AB}, X_{-}>\, = - \rho \, \epsilon_1\,^A \,\epsilon_1\,^B 
+ O(\rho^2),  
\end{equation}
\begin{equation}
\label{omHXval}
<\check{\omega}^A\,_B, \check{H}>\, = 0,\,\,\,\,\, 
<\check{\omega}^A\,_B, X>\, = 
\epsilon_0\,^A\,\epsilon_B\,^0 - \epsilon_1\,^A\,\epsilon_B\,^1
\quad\mbox{on}\quad {\cal C}_e,
\end{equation}
\begin{equation}
\label{omXpmval}
<\check{\omega}^A\,_B, X_{+}>\, = \epsilon_0\,^A \,\epsilon_B\,^1 +
O(\rho^2),\,\,\,
<\check{\omega}^A\,_B, X_{-}>\, = - \epsilon_1\,^A \,\epsilon_B\,^0 
+ O(\rho^2),
\end{equation}
as $\rho \rightarrow 0$. 

To transfer the tensor calculus on ${\cal B}_e$ to ${\cal C}_e$
we define vector fields $c_{AB} = c_{(AB)}$ on 
${\cal C}_e \setminus {\cal I}^0$ by requiring
\begin{equation}
\label{scval}
< \sigma^{AB}, c_{CD} >\, = \epsilon_{(C}\,^A \,\epsilon_{D)}\,^B,\,\,\,\,\,
c_{CD} = c^{1}\,_{CD} \partial_{\rho} + 
c^{+}\,_{CD} \, X_{+} + c^{-}\,_{CD}\, X_{-}.
\end{equation}
The first condition implies that 
$T_{\delta}(\pi')\, c_{AB} = \delta_{(A} \tau_{B)}\,^{B'}\bar{\delta}_{B'}$ 
for $\delta \in {\cal C}_e \setminus {\cal I}^0$, while the second removes the
freedom for the vector fields to pick up an arbitrary
component in the direction of $X$. It follows that
\begin{equation}
\label{csing}
c^{1}\,_{AB} = x_{AB},\,\,\,\,
c^{+}\,_{AB} = \frac{1}{\rho}\,z_{AB} + \check{c}^{+}\,_{AB},\,\,\,\,
c^{-}\,_{AB} = \frac{1}{\rho}\,y_{AB} + \check{c}^{-}\,_{AB},
\end{equation}
with smooth functions which satisfy
\begin{equation} 
\check{c}^{\alpha}\,_{AB} = O(\rho),\,\,\,\,\,\, 
\check{c}^{\alpha}\,_{01} = 0,\,\,\,\,\,\alpha = 1, +, -,
\end{equation}
and
\begin{equation}
\label{unve}
x_{AB} \equiv \sqrt{2} \epsilon_{(A}\,^0 \,\epsilon_{B)}\,^1,\,\,\,
y_{AB} \equiv - \frac{1}{\sqrt{2}}\epsilon_A\,^1\epsilon_B\,^1,\,\,\,
z_{AB} \equiv  \frac{1}{\sqrt{2}}\epsilon_A\,^0\epsilon_B\,^0.
\end{equation}
The connection coefficients with respect to $c_{AB}$ satisfy 
\begin{equation}
\label{gsing}
\gamma_{CD}\,^A\,_B \equiv \, <\check{\omega}^A\,_B, \,c_{CD}>\, 
= \frac{1}{\rho} \, \gamma^{*}_{CD}\,^A\,_B 
+  \check{\gamma}_{CD}\,^A\,_B
\end{equation}
with
\[
\gamma^{*}\,_{ABCD} = 
\frac{1}{2} (\epsilon_{AC} x_{BD} + \epsilon_{BD} x_{AC}),
\,\,\,\,\,
\check{\gamma}_{01CD} = 0,\,\,\,\,\,\check{\gamma}_{ABCD} = O(\rho).
\]
The smoothness of the 1-forms and the vector fields
$\check{H},\, X_{+},\,X_{-}$ implies that the vector fields $\rho\,c_{CD}$
and the functions 
\[
c^{1}\,_{CD},\,\,\,\,
\rho \, c^{+}\,_{CD},\,\,\,\,\rho \, c^{-}\,_{CD},\,\,\,\,  
\rho\,\gamma_{CDAB},
\]
extend smoothly to all of ${\cal C}_e$.

A smooth function $F$ on an open subset of ${\cal C}_e$ is said to have
spin weight $s$ if 
\begin{equation}
\label{spinw}
X(F) = 2s\,F 
\end{equation} 
on this set with $2s$ an integer. Any spinor valued function 
induced by a spinor field on ${\cal B}_e$ has a well defined spin weight,
it holds e.g.  
\begin{equation}
\label{rweylspinweight}
X\,\phi_{ABCD} = 2\,(2 - A - B - C - D)\,\phi_{ABCD}.
\end{equation}
It follows from the construction of ${\cal C}_e$ that this is also
true for the functions considered above, it turns out that 
\[
X\,c^1\,_{AB} = 2\,(1 - A - B)\,c^1\,_{AB},\,\,\,\,\,
X\,c^{\pm}\,_{AB} = 2\,(1 - (\pm 1) - A - B)\,c^{\pm}\,_{AB},
\]
\[
X\,\gamma_{ABCD} = 2\,(2 - A - B - C - D)\,\gamma_{ABCD}
\,\,\,\,\mbox{for}\,\,\,\,A, B, C, D = 0, 1.
\]

\vspace{.3cm}

By our construction, equation (\ref{Hcheckdef}), and the
formula for $e_a(\delta)$ given above the vectors
$T_{\delta}(\pi') \,(\sqrt{2}\,\check{H}(\delta)) = e_3(\delta)$
are tangent to and the frame $e_a(\delta(\rho, t))$ is parallely propagated
along the geodesics 
$[-e, e[ \ni \rho \rightarrow \pi'(\delta(\rho, t))$ through $i$. Thus we
have constructed the type of frame field asked for in the beginning.
Working on ${\cal C}_e$ has the advantage that $\rho$ and
$\sqrt{2}\,\check{H}$ define smooth fields and the
smoothness of the various fields considered above can easily be discussed.

The transition from ${\cal B}_e$ to $\bar{{\cal C}}'_e$ respectively 
to $\bar{{\cal C}}^+_e$ amounts to a new choice of differential structure
at space-like infinity. This change is reflected in the drop of rank of the
map $\pi'$ at the set ${\cal I}^0$.  
It follows from (\ref{siHXval}), (\ref{siXpmval}),
that at points over $i$ the vectors 
$X$, $X_{\pm}$ project onto the zero vector while
at points in $\pi^{'-1}(p)$ the real and imaginary parts of
$\check{H},\, X_{+},\, X_{-}$ have non-vanishing projections which
span the tangent space $T_p\,\tilde{{\cal B}}_e$ if 
$p \in \tilde{{\cal B}}_e$. The relations
(\ref{siHXval}), (\ref{siXpmval}), (\ref{omHXval}), (\ref{omXpmval}) 
show that the behaviour of the map $\pi'$ near ${\cal I}^0$ is encoded in
the behaviour of the solder and the connection form.

\vspace{.3cm}

With the structures given above we can perform tensor calculations defined
on
$\tilde{{\cal B}}_e$ now also on ${\cal C}_e \setminus {\cal I}^0$ and they
follow the `usual' rules of the spin frame formalism. If $F$ denotes the lift of
a smooth function $f$ on ${\cal B}_e$, the covariant differential $Df$ is
represented on ${\cal C}_e \setminus {\cal I}^0$ by the invariant function
$D_{AB}f \equiv c_{AB}(F)$. In the following we shall use the same symbol for a
function and its lift. 
If $\mu_{AB}$ is the invariant function induced by a spatial spinor field $\mu$
on $\tilde{{\cal B}}_e$
its covariant differential is given on ${\cal C}_e \setminus {\cal I}^0$ by the
expression
\[ D_{AB} \mu_{AD}
= c_{AB}(\mu_{AD}) - \gamma_{AB}\,^E\,_C\,\mu_{ED}
- \gamma_{AB}\,^E\,_D\,\mu_{CE}.
\]
Analogous formulas hold for covariant differentials of spinor fields of 
higher valence.

\vspace{.3cm}

In terms of $\rho$ and 
$t = (t^A\,_B)$ on $\bar{{\cal C}}_e$ and
the normal coordinates $x^a$ satisfying (\ref{hnormal}) on ${\cal B}_e$,
the projection $\pi'$ has the local expression
\begin{equation}
\label{bpilocexpr} 
\pi': (\rho, t) \rightarrow x^a(\rho, t) =
\rho\,\sqrt{2}\,\sigma^a\,_{CD}\,t^C\,_0\,t^D\,_1.
\end{equation}
This can be used to pull back the functions $\Omega$, $U$, and $W$, which
are related by (\ref{Omegalocal}), to functions of spin weight
zero on $\bar{{\cal C}}_e$. The metric in (\ref{tiresydata}) is built into
our formalism and the second fundamental form lifts to a symmetric spinor
valued function 
$\chi_{ABCD}$ which vanishes everwhere. Using the fields
\begin{equation}
\label{backgroundonC}
\check{c}^{\pm}\,_{AB},
\,\,\,\,\,\,\,\,
\check{\gamma}_{CDAB},\,\,\,\,\,\,\,\,
S_{ABCD},\,\,\,\,\,\,\,\,
r,
\end{equation}
given by (\ref{csing}), (\ref{gsing}), and (\ref{3cuexp}), one can
determine 
\begin{equation}
\label{ddOm}
D_{AB}\,D_{CD}\,\Omega,
\end{equation}
on ${\cal C}_e^+$ and thus also the derived
data (\ref{trfreeLabonS}), (\ref{L00onS}), (\ref{wabonS}). 

\vspace{.3cm}

In particular, a detailed expression for the rescaled conformal Weyl spinor
$\phi_{ABCD}$ is obtained on ${\cal C}^+_e$ by using
(\ref{wabonS}) and (\ref{Omegalocal}). It takes the form
\begin{equation}
\label{rescweyldetexpr}
\phi_{ABCD} = \phi'_{ABCD} + \phi^W_{ABCD},
\end{equation}
where
\begin{equation}
\label{rescweyl'detexpr}
\phi'_{ABCD} =
\sigma^{-2} \left\{ D_{(AB}\,D_{CD)}\,\sigma 
+ \sigma\,s_{ABCD} \right\}
\end{equation}
\[ 
= \frac{1}{\rho^4}\,\left\{ U^2\,D_{(AB}\,D_{CD)}\,(\rho^2) 
- 8\,\rho\,U\,D_{(AB}\rho\,D_{CD)}U\right\}
\]
\[
- \frac{1}{\rho^2}\,\left\{2\,U\,D_{(AB}\,D_{CD)}\,U 
- 6\,D_{(AB}U\,D_{CD)}U
- U^2\,s_{ABCD} \right\}
\]
is derived from $\sigma = \rho^2\,U^{-2}$ and thus from the local
geometry near $i$, while
\begin{equation}
\label{rescweylWdetexpr}
\phi^W_{ABCD} =
\frac{1}{\rho^3}\,\,
\left\{ -
6\,U\,W\,D_{(AB}\,\rho\,D_{CD)}\,\rho  +
U\,W\,D_{(AB}\,D_{CD)}\,(\rho^2) \right\} 
\end{equation} 
\[
+ \frac{4}{\rho^2}\,\,(W\,D_{(AB}\rho\,D_{CD)}U -
3\,U\,D_{(AB}\rho\,D_{CD)}W)
\]
\[
- \frac{2}{\rho}\,(U\,D_{(AB}\,D_{CD)}\,W + W\,D_{(AB}\,D_{CD)}\,U
- 6\,D_{(AB}U\,D_{CD)}\,W - U\,W\,s_{ABCD})
\]
\[
- 2\,W\,D_{(AB}\,D_{CD)}\,W + 6\,D_{(AB}W\,D_{CD)}\,W
+ W^2\,s_{ABCD},
\]
is the part of the rescaled conformal Weyl spinor which depends on the 
non-local information in $W$ and which vanishes in the massless case.
Observing that
\[ 
D_{AB}\,\rho = x_{AB},\,\,\,
D_{AB}\,D_{CD}\,(\rho^2) 
= - 4\,\rho\,\check{\gamma}_{(AB}\,^E\,_{C}\,x_{D)E} = O(\rho^2),\,\,\,
D_{AB}\,U = O(\rho),
\]
one finds that
\begin{equation}
\label{rescWeylsing}
\phi_{ABCD}' = O(\frac{1}{\rho^2}),\,\,\,\,\,
\phi_{ABCD}^W = - \frac{6\,m}{\rho^3}\,\epsilon^2_{ABCD} +
O(\frac{1}{\rho^2}),
\end{equation}
where we set $\epsilon^j_{ABCD} \equiv 
\epsilon_{(A}\,^{(E}\,\epsilon_{B}\,^{F}\,
\epsilon_{C}\,^{G}\,\epsilon_{D)}\,^{H)_j}$ for $j = 0, \ldots, 4$.

\subsubsection{Normal expansions at ${\cal I}^0$ and the functions
$T_m\,^j\,_k$}
\label{normalexp}

To analyse in detail the behaviour of the various fields near space-like
infinity it is convenient to study a particular type of expansion. It
will be discussed here for an unprimed spinor field, similar expansions
hold for other fields. In terms of the normal coordinates $x^a$
on ${\cal B}_e$ define the radial vector field $V = x^a\,\partial_{x^a}$.
Let $\delta^* = \delta^*(x^a)$ be the smooth spin frame field on ${\cal
B}_e$ which satisfies $D_V\,\delta^* = 0$ on ${\cal B}_e$ and coincides
with the spin frame at $i$ chosen as the starting point for our
construction of ${\cal C}_e$. Denote by $e^*_{AB}$ the orthonormal frame
associated with $\delta^*$ and write $V = V^{AB}\,e^*_{AB}$.

Suppose $\xi$ is a smooth spinor field on ${\cal B}_e$ which is given 
in terms of the spin frame field $\delta^*$ by 
$\xi^{*}_{A_1 \ldots A_l} = \xi^{*}_{A_1 \ldots A_l}(x^a)$. Then its Taylor
expansion at $i$ is of the form  
\begin{equation}
\label{basisspexp}
\xi^{*}_{A_1 \ldots A_l}(x^a) =
\sum_{p = 0}^{p = \infty} \frac{1}{p!}\,
V^{B_p C_p}(x^a) \ldots V^{B_1 C_1}(x^a)\,
D_{B_pC_p} \ldots D_{B_1C_1}\xi^{*}_{A_1 \ldots A_l}(i).
\end{equation}

To determine the lift $\xi_{A_1 \ldots A_l}$ of this field to ${\cal
C}^+_e$ one has to observe its transformation behaviour  
$\xi^{*}_{A_1 \ldots A_l} \rightarrow \xi^{*}_{A_1 \ldots A_l}\,
t^{B_1}\,_{A_1} \ldots t^{B_l}\,_{A_l}$ under changes of the frame and the
fact that the pull back of the functions $V^{AB}$ are given in view of 
(\ref{bpilocexpr}) by
\begin{equation}
\label{Vtexpre}
V^{AB}(x^a(\rho, t)) = \sqrt{2}\,\rho\,t^{(A}\,_0\,t^{B)}\,_1.
\end{equation}
If the expansion coefficients $D_{B_pC_p} \ldots D_{B_1C_1}\xi^{*}_{A_1
\ldots A_l}(i)$ are then decomposed into products 
of $\epsilon_{ab}$'s and symmetric spinors at $i$, the essential
components 
$\xi_j = \xi_{(a_1 \ldots a_{l})_j}$,  $0 \le j \le l$,
$0 \le j \le l$, which are of spin weight  $s = \frac{l}{2} - j$,
are obtained as expansion of the form     
\begin{equation}
\label{bCspexp} 
\xi_j = \sum_{p = 0}^{\infty} \xi_{j,p}\,\rho^p 
\end{equation}
where
\begin{equation}
\label{bCcoeffexp}
\xi_{j,p} = \sum_{m = \max \{|l - 2j|, l - 2p\} }^{2p + l} \sum_{k = 0}^{m}  
\xi_{j,p;m,k}\, T_{m}\,^k\,_{\frac{m - l}{2} + j}
\end{equation}
with complex coefficients $\xi_{j,p;m,k}$ and 
functions $T_m\,^j\,_k$ of $t$ as discussed below.

We refer to this type of expansion as to the {\it normal expansion of
$\xi$ at $I^0$}\index{normal expansion}. In the case considered above the
lift of $\xi$ to
${\cal C}^+_e$ has smooth limits at ${\cal I}^0$. Corresponding expansions in
terms of $\rho^k$, $k \in \mathbb{Z}$, can  also be obtained for fields
such as
$\phi_{ABCD}$ on
${\cal C}^+_e$ which are given by algebraic expressions of regular fields
but which become singular at ${\cal I}^0$. 

The functions $T_m\,^j\,_k$, arise (apart from some normalizing
factors) naturally by the procedure indicated above. They are matrix
elements of unitary representations
\[
SU(2) \ni t \rightarrow T_m(t) = (T_m\,^j\,_k(t)) \in SU(m + 1),
\]
which are given by
\[
T_0\,^0\,_0(t) = 1,\,\,\,\,\,\,\,\,
T_m\,^j\,_k(t) 
= {m \choose j}^{\frac{1}{2}}\,{m \choose k}^{\frac{1}{2}}\,
t^{(b_1}\,_{(a_1} \ldots t^{b_m)_j}\,_{a_m)_k},
\]
\[
j,k = 0, \ldots ,m,\,\,\,\,\,\,\,m = 1,2,3, \ldots .
\]
The brackets with lower index now indicate symmetrization and taking
`essential components'. The expansions
obtained above make sense under quite general assumptions; the functions 
$\sqrt{m + 1}\,T_m\,^j\,_k(t)$ form a complete orthonormal set in the
Hilbert space $L^2(\mu, SU(2))$ where $\mu$ denotes the normalized Haar
measure on $SU(2)$. 

Using the identification of ${\cal I}^0$ with $SU(2)$ built into
our construction, we consider the $T_m\,^j\,_k$ as functions on 
${\cal I}^0$ and extend them as $\rho$-independent functions to 
$\bar{\cal C}_e$. The vector fields $X_{\pm}$, $X$ then act  
as left invariant  vector fields and it holds 
\begin{equation}
\label{XTder}
X\,T_m\,^k\,_j = (m - 2j)\,T_m\,^k\,_j,\,\,\,\,
\end{equation}
\begin{equation}
\label{XpmTder}
X_{+}\,T_m\,^k\,_j = \beta_{m,j}\,T_m\,^k\,_{j - 1},\,\,\,\,
X_{-}\,T_m\,^k\,_j = - \beta_{m,j + 1}\,T_m\,^k\,_{j + 1}
\end{equation}
for $0 \le k, j \le m$, $m = 0, 1, 2, \ldots$, with
$\beta_{m,j} = \sqrt{ j\,(m - j + 1)}$. It follows that functions
$f$ with spin weight $s$ have expansions of the form
\begin{equation}
\label{swexp}
f = \sum_{m \ge |2s|} \sum_{k = 0}^{m} 
f_{m,k}\,T_{m}\,^k\,_{\frac{m}{2} - s},
\end{equation}
where the $m$'s are even if $s$ is an integer and odd if $s$ is a
half-integer. All functions considered in the following have integer spin
weight.

\subsection{The regularizing gauge for the evolution equations}
\label{evolutiongauge}

To obtain definite expressions for the expansions of the data
at $i$ and because the terms of lower order are then simplified, it has
been assumed in \cite{friedrich:i-null} that the metric $h$ is given in a
cn-gauge near $i$. This will be assumed also here, though the discussion of
the static case given below will show that this is not necessary for our
construction. The coordinates
$\rho$, $t$ and the frame field constructed above depend on the choice
of scaling of the metric $h$ on ${\cal S}$. Most important is the fact
that $\Omega = O(\rho^2)$  near ${\cal I}^0$, it affects the
definition of $\rho$ in an essential way.

In analysing the evolution of our data in time it turns out
convenient to use a different conformal factor $\Theta$
which is related to the conformal factor $\Omega$ by  
\begin{equation}
\label{confacSrel} 
\Theta = \kappa^{-1}\,\Omega \quad\mbox{on}\quad \bar{{\cal C}}_e,
\end{equation} 
with a function
\begin{equation}
\label{kappaspec} 
\kappa = \rho\,\kappa' 
\quad\mbox{with}\quad 
\kappa' \in C^{\infty}(\bar{{\cal C}}_e),
\,\,\,\,\,\,\kappa' > 0,
\,\,\,\,\,\,X\,\kappa' = 0,  
\,\,\,\,\,\,\kappa'|_{{\cal I}^0} = 1.
\end{equation} 
The value of $\kappa'$ on ${\cal I}^0$ is chosen for convenience here,
nothing is gained in the following by requiring a different (positive)
boundary value for it. 

The change of the conformal factor implies a map 
$\Xi:  \delta \rightarrow \kappa^{\frac{1}{2}}\,\delta$ which maps the set 
${\cal C}^+_e$ bijectively onto a smooth submanifold 
${\cal C}^*$ of the bundle of conformal spin frames over
$\tilde{\cal B}_e$. 
We use the diffeomorphism $\Xi$ to carry the coordinates $\rho$ and $t$
and  the vector fields $\partial_{\rho}, X, X_{+}, X_{-}$ to 
${\cal C}^*$. The projection of ${\cal C}^*$ onto
$\tilde{\cal B}_e$ will be denoted again by $\pi'$.     

Assuming a conformal Gauss system for the evolution in time as described
in section \ref{genconffieldequ}, the evolution of the spin frames
constituting ${\cal C}^*$ defines in the the bundle of conformal frames over
the space-time manifold $\tilde{{\cal M}}$ a smoothly embedded 5-dimensional
manifold $\tilde{{\cal N}}$ which is again a $U(1)$ bundle over the
space-time and whose projection onto $\tilde{{\cal M}}$ we denote again
by $\pi'$. The manifold $C^*$ represents a smooth hypersurface of 
$\tilde{{\cal N}}$. 

By pushing forward the coordinates $\rho$, $t$ and the vector
fields $\partial_{\rho}$, $X$, $X_{\pm}$ with the flow of the conformal
geodesics ruling $\tilde{{\cal N}}$, these structures can be extended to
$\tilde{{\cal N}}$ such that $i\,X$ generates the kernel of $\pi'$. The
parameter $x^0 \equiv \tau$ of the conformal geodesics defines a further
independent coordinate with $x^0 = \tau = 0$ on ${\cal C}^*$, so that the
tangent vector field of this congruence can be denoted by
$\partial_{\tau}$.

The reduced field equations (\ref{reds1e}), (\ref{redd2e}),
(\ref{redd1e}), (\ref{sysyhy}) (the latter specialization of
(\ref{redsymhypbianchi}) is chosen here for only definiteness) are now
interpreted as equations on
$\tilde{{\cal N}}$ by assuming that the
$e_{AA'}$ are vector fields on $\tilde{{\cal N}}$ which are defined at a
spin frame 
$\delta \in \tilde{{\cal N}}$ by the requirement that they project onto
the frame defined by $\delta$ on $\tilde{{\cal M}}$, i.e.
$T_{\delta}\,\pi'(e_{AA'}) = \delta_A\,\bar{\delta_{A'}}$, and whose
$X$-component is fixed by requiring an expansion of the form
\begin{equation}
\label{ivpconfframe} 
e_{AA'} = \frac{1}{\sqrt{2}}\,\tau_{AA'}\,\partial_{\tau} -
\tau^B\,_{A'}\,e_{AB},
\end{equation} 
with `spatial vectors' 
\begin{equation}
\label{ivpspaceconfframe} 
e_{AB} = e^0\,_{AB}\,\partial_{\tau} + 
e^1\,_{AB}\,\partial_{\rho} + e^+\,_{AB}\,X_+ + e^-\,_{AB}\,X_-. 
\end{equation}
The unknowns in the reduced field equations are then interpreted as
spinor valued functions on $\tilde{{\cal N}}$. 
It can be shown that spin weights are preserved under the evolution by the
reduced system.

We have to express the initial data for the conformal field equations
in terms of the new scaling. With $\kappa$, the fields
(\ref{backgroundonC}), (\ref{ddOm}), and the associated covariant
derivatives (carried over to ${\cal C}^*$, observing that the
local expression of $\Xi$  in the given coordinates is the identity) 
one gets for the curvature fields
\begin{equation}
\label{kpdat}
\phi_{ABCD} = \frac{\kappa^3}{\Omega^2}\,\left(D_{(AB}\,D_{CD)}\Omega +
\Omega\,s_{ABCD}\right), 
\end{equation}
\begin{equation}
\label{kPdat}
\Theta_{AA'CC'} = - 
\kappa^2\,\left(\frac{1}{\Omega}\,D_{(AB}\,D_{CD)}\Omega
+ \frac{1}{12}\,r\,h_{ABCD}\right)\,\tau^B\,_{A'}\,\tau^B\,_{C'}. 
\end{equation}

For the frame (\ref{ivpconfframe}), one gets by (\ref{csing})
\begin{equation}
\label{ke01dat}
e^0\,_{AB} = 0,\,\,\,\,\,  
e^1\,_{AB} = \rho\,\kappa'\,x_{AB},\,\,\,\,\,
\end{equation}
\begin{equation}
\label{kepmdat}
e^{+}\,_{AB} = \kappa'\,z_{AB} + \kappa\,\check{c}^{+}\,_{AB}
,\,\,\,\,\,
e^{-}\,_{AB} = \kappa'\,y_{AB} + \kappa\,\check{c}^{-}\,_{AB}.
\end{equation}

For the conformal factor $\Theta$ we get
\begin{equation}
\label{ivpconfac} 
\Theta = \Theta_* \equiv \kappa^{-1}\,\Omega
= \frac{\rho}{\kappa'\,(U + \rho\,W)^2} 
\quad\mbox{on}\quad 
{\cal C}^*.
\end{equation} 
We assume that initial data for the 1-form $f$, which will
be related in the end to $\tilde{f}$ by the relation
$f = \tilde{f} - \Theta^{-1}\,d\,\Theta$, 
satisfy 
\begin{equation}
\label{ivpctildef} 
<f, \partial_{\tau}>\, = 0,\,\,\,\,\,\,\,\,\,\,\,\,
\mbox{pull back of}
\,\,\,f\,\,\,\mbox{to}\,\,\, {\cal C}^* 
= \kappa^{-1}\,d\,\kappa.
\end{equation} 
It follows then that from (\ref{Thetaexpl}) that $\Theta$ takes 
the form
\begin{equation}
\label{ivpThetaexpl}
\Theta
= \Theta_*\,\left(1 
- \tau^2\,\frac{\kappa^2_*}{\omega^2_*}\right)
\,\,\,\mbox{on}\,\,\,\tilde{{\cal N}},
\end{equation}
with a function $\omega$ which is given by
\begin{equation}
\label{Aomegadef}
\omega = \frac{2\,\Omega}{\sqrt{|D_{a}\Omega D^{a}\Omega|}}
=
\rho\,(U + \rho\,W)\, \left\{ U^2 + 2\,\rho\,U\,x^{AB}D_{AB}U 
- \rho^2\,D^{AB}U\,D_{AB}U 
\right.
\end{equation}
\[
\left.
+ 2\,\rho^2\,U\,x^{AB}D_{AB}W 
- 2\,\rho^3\,D^{AB}U\,D_{AB}W 
- \rho^4\,D^{AB}W\,D_{AB}W \right\}^{- \frac{1}{2}}
\,\,\mbox{on}\,\,\, {\cal C}^*.
\]
Here the second member is given in the notation of section
\ref{trsafcdat} while the term on the right hand side is given in the
notation of section \ref{constrlift}. In (\ref{ivpThetaexpl}) and in the
following formulas the subscripts $*$ are saying that the corresponding
functions are constant along the conformal geodesics.

For $d_{AA'}$ we get by (\ref{bexpl}) the explicit expression
\begin{equation}
\label{dexpl}
d_{AA'} = \frac{1}{\sqrt{2}}\,\tau_{AA'}\,\dot{\Theta}
- \tau^B\,_{A'}\,d_{AB}\,\,\,\,\,\mbox{on}\,\,\,\,\tilde{{\cal N}},
\end{equation}
where the dot denotes the derivative with respect to $\tau$ and 
\begin{equation}
\label{dABexpl}
d_{AB} = 2\,\rho\,\left(\frac{U\,x_{AB} - \rho\,D_{AB}U 
- \rho^2\,D_{AB}W}
{(U + \rho\,W)^3}\right)_{*},
\end{equation}
where the notation of section \ref{constrlift} is used on the right
hand side.

If one uses (\ref{ke01dat}) and (\ref{kepmdat}) to write 
for a given smooth function $\mu$ on ${\cal C}^*$   
\[
\mu_{AB} \equiv \kappa^{-1}\,(e^1\,_{AB}\,\partial_{\rho}
+ e^{+}\,_{AB}\,X_{+} + e^{-}\,_{AB}\,X_{-})\,\mu,
\]
one gets with the 1-form (\ref{ivpctildef}) and the spatial connection
coefficients (\ref{gsing}) the space-time connection coefficients in the form
\begin{equation}
\label{kGammadat}
\Gamma_{AA'CD} = 
\left(\frac{1}{2}\,\rho\,(\epsilon_{AC}\,\kappa'_{BD} 
+ \epsilon_{'BD}\,\kappa'_{AC})
- \rho\,\kappa'\,\check{\gamma}_{ABCD}
+ \frac{1}{2}\,\epsilon_{AB}\,\kappa_{CD}
\right)\,\tau^B\,_{A'}.
\end{equation}
Note that the $\hat{\Gamma}_{AA'BC}$ in the reduced equations can be
expressed by (\ref{hatGambyGam}) in terms of the $\Gamma_{AA'BC}$. 

Most important for us is the observation that {\it the functions given by}
(\ref{kpdat}), (\ref{kPdat}), (\ref{ke01dat}), (\ref{kepmdat}),
(\ref{ivpThetaexpl}), (\ref{dexpl}), (\ref{kGammadat}) 
{\it have smooth limits as $\rho
\rightarrow 0$ and can in fact be smoothly extended into the 
coordinate range} $\rho \le 0$. For the unknowns in the new scaling we
thus obtain normal expansion in terms of non-negative powers of $\rho$. 
In particular, one has
\begin{equation}
\label{newrescweyldetexpr}
\phi_{ABCD} = \kappa^3(\phi'_{ABCD} + \phi^W_{ABCD}),
\end{equation}
with (\ref{rescweyl'detexpr}), (\ref{rescweylWdetexpr})
on the right hand side. Pushing the expansion (\ref{rescWeylsing}) a bit
further and using  (\ref{confacSrel}) one gets in the cn-gauge (in which 
$h_{ab} = - \delta_{ab} + O(\rho^3)$)

\begin{equation}
\label{newrescWeylsingnexp}
\phi_{ABCD} = - \kappa^{'3}\,6\,m\,\epsilon^2_{ABCD} 
\end{equation}
\[
- \rho\,\kappa^{'3}\,12\,\left(
X_{+}\,W_1\,\epsilon^1\,_{ABCD}
+ 3\,W_1\,\epsilon^2\,_{ABCD}
- X_{-}\,W_1\,\epsilon^3\,_{ABCD}\right)
\]
\[
- \frac{\rho^2\,\kappa^{'3}}{2}
\sum_{k, j = 0}^4 {4 \choose j} \left( 
4\,\sqrt{6\,{4 \choose j}}\,W_{2;4,k}
- \frac{2 - j}{3}\,
\sqrt{2\,{4 \choose k}}\,\,b^{*}_k(i)
\right)T_4\,^k\,_j\,\epsilon^j_{ABCD}, 
\]
\[
+ O(\rho^3).
\]
It is assumed here that $W$ is an arbitrary solution to 
$(\Delta_h - \frac{1}{8}\,r)\,W = 0$ on ${\cal B}_e$. Its normal expansion
takes in the cn-gauge the form 
\[
W = \sum_{p = 0}^2 \rho^p\,W_p + O(\rho^3) =
\sum_{p = 0}^2 \rho^p\,(\sum_{k = 0}^{2p} W_{p;2p,k}\,T_{2p}\,^k\,_p) + O(\rho^3)
\]
with 
\[
W_{0;0,0} = W(i) = \frac{m}{2},\,\,\,\,\,\,\,
W_{1;2,k} = { 2 \choose k}^{\frac{1}{2}}\,D_{(ab)_k}W^{*}(i),
\]
\[
W_{2;4,k} = {4 \choose 2}^{-\frac{1}{2}}\,{4 \choose k}^{\frac{1}{2}}
\,D_{(ab}D_{cd)_k}W^{*}(i).
\]
In the case where $\kappa'$ is constant the right hand side of 
(\ref{newrescWeylsingnexp}) provides the terms of a normal expansion up to
the quadrupole term. If $\kappa$ depends on $\rho$ and $t$ the terms given
above need to be expanded further to obtain the normal expansion.

The transition (\ref{confacSrel}) to the conformal factor $\Theta$
corresponds to a transition $h \rightarrow h' = \kappa^{-2}\,h$ of the
metric on ${\cal B}_e$ (assuming that $\kappa'$ arise as a lift of a
smooth positive function on ${\cal B}_e$ with $\kappa'(i) = 1$)
in the sense that then $\Omega^{-2}\,h = \tilde{h} = \Theta^{-2}\,h'$. The
coordinate $\rho$ is then not
adapted to the geometry defined by the metric $h'$. To illustrate the
situation assume that $h$ is flat. Then 
\begin{equation}
\label{flatinconf}
h' = - \kappa^{'-2}\,\rho^{-2}\,(d\,\rho^2 +
\rho^2\,d\,\sigma^2) = - \kappa^{'-2}\,(d\,r^2 + d\,\sigma^2),
\end{equation}
with $r = - \log \rho$ near $i$. With respect to the new coordinate $r$,
which is adapted to the geometry of $h'$, the point $i$ is shifted to
infinity but the surface measure of any sphere around $i$ remains finite
and positively bounded from below. This behaviour is reflected by the fact
that the frame coefficient $e^1\,_{AB}$ in (\ref{ke01dat}) vanishes while
the frame coefficients $e^{\pm}\,_{AB}$ in (\ref{kepmdat}) have finite and
non-vanishing limits at ${\cal I}^0$. We shall keep the coordinate $\rho$
because it ensures the finite coordinate representation of the boundary
${\cal I}^0$ as well as the smoothness of the data near ${\cal I}^0$.

With the gauge defined above the functions $\Theta$ and
$d_{AA'}$ in equations (\ref{redd2e}), (\ref{redd1e}) are given 
by (\ref{ivpThetaexpl}) and (\ref{dexpl})
and the finite regular initial value problem near space-like
infinity for the reduced field equations  (\ref{reds1e}), (\ref{redd2e}),
(\ref{redd1e}), (\ref{sysyhy}) is completely determined. 
We write this system schematically as system of equations
for the unknown $u = (w, \phi)$ 
with $\phi = (\phi_{ABCD})$ and
$w = (e_{AA'}, \hat{\Gamma}_{AA'BC}, \Theta_{AA'BB'})$ or,
alternatively,
$w = (e_{AA'}, \Gamma_{AA'BC}, \Theta_{AA'BB'})$. 
It takes the form 
\begin{equation}
\label{schemredequ}
\partial_{\tau}\,w = F(x, w, \phi),\,\,\,\,\,
A^{\mu}\,\partial_{x^{\mu}}\,\phi = H(w)\,\phi,
\end{equation}
where the $x$-dependence in
the first equation comes in here via the functions $\Theta$ and
$d_{AA'}$.

Important for the following is that {\it with any choice of $\kappa$
satisfying (\ref{kappaspec}) the functions $\Theta$ and $d_{AA'}$ take
smooth limits as $\rho \rightarrow 0$ and can be extended smoothly into a
range where $\rho \le 0$}. With the smooth extensibility of the initial
data observed before, we find that {\it the initial value problem for the
reduced field equations with the data prescribed above can be extended
smoothly into a range where $\rho \le 0$ so that the reduced equations
form still a symmetric hyperbolic system}. It may be noted finally that the
congruence of conformal geodesics (considered as point sets) underlying
our gauge does not depend on the choice of $\kappa$, whereas the parameter
$\tau$ depends on it in an essential way.

\subsection{Specific properties of the regular finite initial value
problem at space-like infinity}
\label{specificsofrfivpsli}

The nature of the initial value problem formulated above is conveniently
discussed by considering certain subsets of 
$\mathbb{R} \times \mathbb{R} \times SU(2)$ which are defined by the
range admitted for the coordinates $(\tau, \rho, t)$. We define
5-dimensional subsets 
\[
\tilde{{\cal N}} \equiv \{|\tau| < \frac{\omega}{\kappa},\,\, 
0 < \rho < e,\,\,t \in SU(2)\},
\]
\[
\bar{{\cal N}} \equiv \{|\tau| \le \frac{\omega}{\kappa},\,\,
0 \le \rho < e,\,\,t \in SU(2)\},
\]
where $\frac{\omega}{\kappa}$ is a function of
$\rho$ and $t$. It holds then 
\[
\bar{{\cal N}} = \tilde{{\cal N}} \cup {\cal J}^-
\cup {\cal J}^+ \cup {\cal I} \cup {\cal I}^- \cup {\cal I}^+,
\]
with 4-dimensional submanifolds 
\[
{\cal J}^{\pm} \equiv \{\tau = \pm \frac{\omega}{\kappa},\,\,
0 < \rho < e,\,\,t \in SU(2)\},\,\,\,\,\,\,
{\cal I} \equiv \{|\tau| < 1,\,\,
\rho = 0,\,\,t \in SU(2)\},
\]
and 3-dimensional submanifolds
\[
{\cal I}^{\pm} \equiv \{|\tau| \pm 1,\,\,
\rho = 0,\,\,t \in SU(2)\},\,\,\,\,\,\,\,
{\cal I}^0 \equiv \{\tau = 0,\,\,\rho = 0,\,\,t \in SU(2)\},
\]
where it has been observed that $\frac{\omega}{\kappa} \rightarrow 1$ 
as $\rho \rightarrow 0$. We note that 
\[
\Theta > 0\,\,\,\mbox{on}\,\,\,\tilde{{\cal N}},\,\,\,\,\,\,\, 
\Theta = 0, \,\,d\Theta \neq 0\,\,\,\mbox{on}\,\,\, 
{\cal J}^- \cup {\cal J}^+ \cup {\cal I},\,\,\,\,\,\,\,
\Theta = 0, \,\,d\Theta = 0 \,\,\,\mbox{on}\,\,\, {\cal I}^{\pm}.
\]
The set $C^* = \{\tau = 0,\,\,0 < \rho < e,\,\,t \in SU(2)\}$
defines a hypersurface of $\tilde{{\cal N}}$. Its closure in 
$\bar{{\cal N}}$ is given by
\[
\bar{C} \equiv \{\tau = 0,\,\,0 \le \rho < e,\,\,t \in SU(2)\} =
C^* \cup {\cal I}^0.
\]

Factoring out the group $U(1)$ implies projections (denoted again by
$\pi'$) onto subsets
$\mathbb{R} \times \mathbb{R} \times S^2$ which are of one dimension lower
than the sets above. In particular, $\tilde{{\cal N}}$ projects onto a set
$\tilde{{\cal M}}$ which represents the `physical space-time'. For
convenience we will usually work with the manifolds above and use for them
the same words as for the projections, so that $\tilde{{\cal N}}$ will be
referred to as the `physical space-time' etc.

\vspace{.3cm}

For suitable $\epsilon > 0$ consider a smooth extension of the data given on
$C^*$ to the set
$\bar{C}_{ext} = \{\tau = 0,\,\,- \epsilon < \rho < e,\,\,t \in SU(2)\}$
and an extension of the functions $\Theta$, $d_{AA'}$ to the domain
$\bar{{\cal N}}_{ext} = \{|\tau| < \frac{\omega}{\kappa},\,\, 
- \epsilon < \rho < e,\,\,t \in SU(2)\}$, so that the reduced
conformal field equations (\ref{reds1e}), (\ref{redd2e}), (\ref{redd1e}),
(\ref{sysyhy}) still represent a symmetric hyperbolic system of
the form (\ref{schemredequ}). Then there
exists a neighbourhood ${\cal V}$ of $\bar{C}_{ext}$ in  
$\bar{{\cal N}}_{ext}$ on which there exists a unique smooth
solution $e_{AA'}$, $\hat{\Gamma}_{AA'BC}$ (resp. $\Gamma_{AA'BC}$),
$\Theta_{AA'BB'}$, $\phi_{ABCD}$ to our extended initial value problem
which satisfies the gauge conditions (\ref{spinwgaugecond}). 

It turns out, that {\it the restriction of this solution to the set 
${\cal V} \cap \bar{{\cal N}}$ is uniquely determined by the data on
$C^*$}. 
The data on $C^*$ have a unique smooth extension to $\bar{C}$ and it
follows from (\ref{ivpconfframe}), (\ref{ivpspaceconfframe}), 
(\ref{ke01dat}), and (\ref{kepmdat}) 
that $e^1\,_{CC'} \rightarrow 0$ as $\rho \rightarrow 0$. 
Equations (\ref{reds1e}) imply in particular 
\begin{equation}
\label{1reds1e}
\sqrt{2}\,\partial_{\tau}\,e^1\,_{CC'} = 
- \Gamma_{CC'}\,^{AA'}\,_{BB'}\,\tau^{BB'}\,e^1\,_{AA'}.
\end{equation}
It follows that $e^1\,_{CC'} = 0$ on ${\cal V} \cap {\cal I}$ and 
as a consequence that the matrices $A^{\mu}$ in (\ref{schemredequ}) are
such that
\begin{equation}
\label{Arvanishes}
A^1 = 0 \,\,\,\mbox{on}\,\,\,{\cal I},
\end{equation}
if the solution extends far enough. 
One can apply to the system (\ref{schemredequ}) on subsets of 
${\cal V} \cap \bar{{\cal N}}$ the standard method of deriving energy
estimates. Without further information on the system the partial
integration would yield contributions from boundary integrals over
parts of ${\cal V} \cap {\cal I}$. Because of (\ref{Arvanishes})
these boundary integrals vanish and one obtains energy estimates which
allow one to show the asserted uniqueness property.
The extension above has been considered to simplify the argument.
Alternatively, the space-time
$\tilde{{\cal N}}$ can be thought of as a solution of a very specific
`maximally dissipative' initial boundary value problem where initial data
are prescribed on $\bar{{\cal C}}$ and no data are prescribed on
${\cal I}$ because of (\ref{Arvanishes}) (cf. \cite{friedrich:nagy} and
the existence theory in \cite{gues},
\cite{secchi:II}). 

In the present gauge the set ${\cal I}$, which is generated from 
${\cal I}^0$ by the extension and evolution process, can be considered as
being obtained by performing limits of conformal geodesics.  
It represents a boundary of the space-time
$\tilde{{\cal N}}$ which may be understood as a blow-up of the point $i^0$.
We refer to it as the {\it cylinder at space-like infinity}\index{cylinder
at space-like infinity}. 

Suppose that there exists on $\tilde{{\cal N}}$ a smooth solution 
$e_{AA'}$, $\hat{\Gamma}_{AA'BC}$ (resp. $\Gamma_{AA'BC}$),
$\Theta_{AA'BB'}$, $\phi_{ABCD}$ of the reduced conformal field equations
(\ref{reds1e}), (\ref{redd2e}), (\ref{redd1e}), (\ref{sysyhy})
which satisfies the gauge conditions (\ref{spinwgaugecond}) on
$\tilde{{\cal N}}$ and coincides on the initial hypersurface 
$C^* = \{\tau = 0\} \subset \tilde{{\cal N}}$ with the
data given above. The projections $T\pi'(e_{AA'})$ then
define a frame field on $\tilde{{\cal M}}$ for which exists a unique
smooth metric $g$ on $\tilde{{\cal M}}$ such that 
$g(T\pi'(e_{AA'}), T\pi'(e_{AA'})) = \epsilon_{AB}\,\epsilon_{A'B'}$.   
Denote by $D'$ the domain of dependence  in $\tilde{{\cal M}}$ with respect
to $g$ of the set $\pi'({\cal C}^*)$ and set $D = \pi^{'-1}(D')$. By
the discussion above we can assume that the closure of $D$ in 
$\bar{{\cal N}}$ contains the set ${\cal I}$ and the solution extends
smoothly to ${\cal I}$. It follows from the
structure of the characteristics of the reduced equations, that the
solution is determined on $D$ uniquely by the data on 
${\cal C}^*$ and it follows from the discussion in \cite{friedrich:AdS}
and the fact that the data satisfy the constraints that the complete set
(\ref{s1e}), (\ref{d2e}), (\ref{d1e}), (\ref{bieq}) of conformal field
equations is satisfied on $D$. Since $\Theta$ has spin weight zero it
descends to a function on $\tilde{{\cal M}}$ and 
$\tilde{g} = \Theta^{-2}\,g$ satisfies the vacuum field equations.

The restriction to $D$ arises here because we only considered the data
on ${\cal C}^*$. Observing that the latter were obtained by restricting 
the data given on the initial hypersurface ${\cal S}$ to ${\cal B}_e$ it
is reasonable to assume that the conformal field equations hold everywhere
on $\tilde{{\cal N}} \cup {\cal I}$ and $\tilde{g}$ defines a solution to
the vacuum field equations on $\tilde{{\cal M}}$.

Assume $u$ is a solution of a (possibly non-linear) hyperbolic system of
partial differential equations of first order on some manifold. A
hypersurface of this manifold is then called a {\it
characteristic}\index{characteristic} of that system (with respect to $u$),
if the system implies for some components of
$u$ non-trivial interior differential equations on the hypersurface. These
interior equations are called  {\it transport equations}\index{transport
equations} (cf. \cite{courant:hilbert:II}).

Because of (\ref{Arvanishes}) the set ${\cal I}$ is then a characteristic of
the extended field equations. It is in fact of a very special type
(i.e. a {\it total characteristic}\index{total characteristic}), 
because the system (\ref{schemredequ}) reduces on ${\cal I}$ to an
interior symmetric hyperbolic system of transport equations for the
{\it complete} system of unknowns. Together with the data on ${\cal I}^0$ it
allows us to determine $u = (v, \phi)$ on ${\cal I}$. 

Suppose that the solution extends in a $C^1$ fashion to the sets
${\cal J}^{\pm}$. Since $\Theta = 0$, $d\Theta \neq 0$ on ${\cal J}^{\pm}$
the sets $\pi'({\cal J}^{\pm})$ form (part of) the conformal boundary at
null infinity for the vacuum solution $\tilde{g}$. Since $\phi_{ABCD}$ is
$C^1$ one finds Sachs peeling. Of course, {\it it will be one of our
main tasks to control under which assumptions the solutions will extend
with a certain smoothness to the sets ${\cal J}^{\pm}$}.

As remarked before, we can expect the decision about the smoothness of the
solution at null infinity to be made in the area where the latter `touches'
space-like infinity. This location has a precise meaning in the present
setting. It is given by the 
{\it critical sets ${\cal I}^{\pm}$}\index{critical sets}, which
can be considered either as boundaries of ${\cal J}^{\pm}$ or as
the boundary components of ${\cal I}$. The nature of these
sets is elucidated by studying {\it conformal Minkowski space} in the
present setting.  

We start with the line element given by (\ref{flatinconf}) and 
choose $\kappa' = 1$. Since $\omega = \rho$ by (\ref{omegadef}) it
follows that ${\cal J}^{\pm} =  \{|\tau| = \pm 1,\,\,
0 < \rho < e,\,\,t \in SU(2)\}$ and $\bar{{\cal M}} = \{|\tau| \le 1,\,\,
0 \le \rho < e\} \times S^2$. 
It will be useful to express the frames
considered in the following in terms of the specific frame
\begin{equation}
\label{01+-frame}
v_0 = \partial_{\bar{\tau}},\,\,\,\,
v_1 = \rho\,\partial_{\rho},\,\,\,\,
v_{\pm} = X_{\pm}.
\end{equation}
The complete solution to the conformal field equations then is given by
\begin{equation}
\label{lnmmbarvects}
e^{\star}_{AA'} = \frac{1}{\sqrt{2}}\left\{
\left((1 - \tau)\,\epsilon_A\,^0\,\epsilon_{A'}\,^{0'}
+ (1 + \tau)\,\epsilon_A\,^1\,\epsilon_{A'}\,^{1'}\right)\,v_0
\right.
\end{equation}
\[
\left.
+ (\epsilon_A\,^0\,\epsilon_{A'}\,^{0'}
- \epsilon_A\,^1\,\epsilon_{A'}\,^{1'})\,v_1
- \epsilon_A\,^0\,\epsilon_{A'}\,^{1'}\,v_+
- \epsilon_A\,^1\,\epsilon_{A'}\,^{0'}\,v_-\right\}
\]
\begin{equation}
\label{zumkomp}
\Gamma^{\star}_{AA'BC} = - \frac{1}{2}\,\tau_{AA'}\,x_{BC},
\end{equation}
\begin{equation}
\label{Riccikruekomp}
\Theta^{\star}_{AA'BB'} = 0,
\end{equation}
\begin{equation}
\label{Weylkruekomp}
\phi^{\star}_{ABCD} = 0.
\end{equation}
The conformal factor and the metric $g$ implied by
$e^{\star}_{AA'}$ are given by 
\begin{equation}
\label{finconfmink}
\Theta^{\star} = \rho\,(1 - \tau^2),
\,\,\,\,\,\,\,\,\,\,
g^{\star} = d\tau^2 + 2\,\frac{\tau}{\rho}\,d\tau\,d\rho 
- \frac{1 - \tau^2}{\rho^2}\,d\rho^2 - d\,\sigma^2.
\end{equation}
With the coordinate transformation 
\begin{equation}
\label{firstcotrafo}
r = \frac{1}{\rho\,(1 - \tau^2)},\,\,\,\,\,\,\,\,\,\,
t = \frac{\tau}{\rho\,(1 - \tau^2)},
\end{equation}
one gets in fact the standard Minkowski metric
$\tilde{g} = \Theta^{-2}\,g^{\star} = dt^2 - dr^2 - r^2\,d\sigma^2$
in spherical coordinates . The flat metric corresponding
to (\ref{sinfmetr}) is given by
$\Omega^{\star 2}\tilde{g} = \rho^2\,g^{\star} = d(\tau\,\rho)^2 - d\rho^2
-\rho^2\,d\sigma^2$ with $\Omega^{\star} = \rho\,\Theta^{\star} = \rho^2 -
(\tau\,\rho)^2$. For this metric the curves with constant coordinates
$\rho$, $\theta$, and $\phi$ are obviously conformal geodesics and because
of their conformal invariance it follows that the corresponding curves for
$g^{\star}$ are conformal geodesics with parameter $\tau$. Equations  
(\ref{firstcotrafo}) can be read as their parametrized version in
Minkowski space.

The metric $g^{\star}$ given by (\ref{finconfmink}) extends smoothly
across null infinity but it has no reasonable limit at ${\cal I}$.
Its contravariant version
\[
g^{\star \sharp} = (1 - \tau^2)\,\partial^2_{\tau} 
+ 2\,\tau\,\partial_{\tau}\,(\rho\,\partial_{\rho})
- (\rho\,\partial_{\rho})^2 - (d\,\sigma^2)^{\sharp},
\]
does extend smoothly to ${\cal I}$. While it drops rank in the limit, it
does imply a smooth contravariant metric on ${\cal I}$ whose covariant
version $l^{\star} = (1 - \tau^2)^{-1}\,d\tau^2 - d\,\sigma^2$
defines a smooth conformally flat Lorentz metric on ${\cal I}$. The
coordinate transformation $\tau = \sin \xi$ shows that this metric is not
complete. The Killing fields of Minkowski space, which are conformal
Killing fields for $g^{\star}$, extend smoothly to ${\cal I}$ such that
they become tangent to ${\cal I}$, vanish there in the case of the
translational Killing fields, and act as non-trival conformal Killing
fields for the metric $l^{\star}$ in the case of infinitesimal Lorentz
transformations.

The fields (\ref{lnmmbarvects}), (\ref{zumkomp}),
(\ref{Riccikruekomp}) extend smoothly to all of $\bar{{\cal M}}$. The
property  (\ref{Arvanishes}) results from the fact that the fields
$e^{\star}_{00'}$, $e^{\star}_{11'}$ become linear dependent on ${\cal I}$.
Since they do not vanish there, this degeneracy does not cause any
difficulties in the field equations. On ${\cal I}^{+}$ and ${\cal I}^{-}$
however, the field $e^{\star}_{00'}$ and $e^{\star}_{11'}$ respectively
vanishes. This strong degeneracy has important consequences for the
(extended) conformal field equations. To see this, we solve the transport
equations on ${\cal I}$ to determine the matrices $A^{\mu}$ on ${\cal I}$
in the general case. 

Extending the data (\ref{kpdat}), (\ref{kPdat}), (\ref{ke01dat}),
(\ref{kepmdat}), (\ref{kGammadat}), one finds that they
agree on ${\cal I}^0$, irrespective of the choice of $\kappa'$ satisfying
conditions of (\ref{kappaspec}), with the implied Minkowski data.
Since the extensions of the functions $\Theta$ and $d_{AA'}$ vanish on
${\cal I}$, the transport equations for the frame, connection, and Ricci
tensor coefficients are independent of the choice of initial data.
It follows that the restrictions of these coefficients to ${\cal I}$ agree
with those of the Minkowski data given above. It follows in particular
that $e^1\,_{AA'} = 0$ on ${\cal I}$. Applying formally the operator
$\partial_{\rho}$ to equation (\ref{1reds1e}) (which is part of the
reduced field equations), restricting to ${\cal I}$, and observing the
data $\partial_{\rho}e^1\,_{AA'}|_{{\cal I}^0}$, one finds that
$\partial_{\rho}e^1\,_{AA'} = \epsilon_A\,^0\,\epsilon_{A'}\,^{0'}
- \epsilon_A\,^1\,\epsilon_{A'}\,^{1'}$ on ${\cal I}$. Writing 
\[
e_{AA'} = e^i\,_{AA'}\,v_i\,\,\,\,\mbox{with}\,\,\,\,i = 0, 1, +, -,
\] 
and assuming the summation rule,
we find that {\it irrespective of the free datum $h$ given on 
${\cal S}$ and the choice of $\kappa'$ the fields $e_{AA'}$,
$\Gamma_{AA'BC}$, $\Theta_{AA'BB'}$ coincide at lowest order  
with the Minkowski fields above in the sense that
$\Theta_{AA'BB'} = O(\rho)$ and}
\begin{equation}
\label{Aminksplitoff}
e^i\,_{AA'} = e^{\star i}\,_{AA'} + \check{e}^i\,_{AA'},\,\,\,\,\,\,
\Gamma_{AA'BC} = \Gamma^{\star}_{AA'BC} + \check{\Gamma}_{AA'BC},
\end{equation}
{\it with}
\begin{equation}
\label{Bminksplitoff}
\check{e}^i\,_{AA'} = O(\rho),\,\,\,\,\,\,
\check{\Gamma}_{AA'BC} = O(\rho)\,\,\,\,
\mbox{as}\,\,\,\,\rho \rightarrow 0.
\end{equation}
Assuming $\kappa = \omega$ in the general case, which by
(\ref{Aomegadef}) is consistent with (\ref{kappaspec}) if $e$ is chosen
small enough, the similarity with the Minkowski case becomes even closer.
Then $\Theta = f\,\Theta^{\star}$ with proportionality factor 
$f \equiv \frac{\Omega}{\rho\,\omega}$ which
extends smoothly to $\bar{{\cal N}}$ such that
$f \rightarrow 1$ on ${\cal I}$. The set ${\cal J}^{\pm}$ is given as in
the Minkowski case above. The discussion below shows, however, that this
particular choice of $\kappa$ may not always be the most useful one.

In the general case the first deviation from the Minkowski case is found
in the value of the rescaled conformal Weyl spinor on ${\cal I}$. On
${\cal I}^0$ it is given by (\ref{newrescWeylsingnexp}). Restricting the
Bianchi equation to ${\cal I}$ and using the coefficients
determined above one finds that
\begin{equation}
\label{phio}
\phi_{ABCD} = - 6\,m\,\epsilon^2_{ABCD}\,\,\,\,\mbox{on}\,\,\,
{\cal I}.
\end{equation}

The discusssion above shows that the matrices $A^{\mu}$ are determined on
${\cal I}$ by the Minkowski data and the structure of the characteristics
of the evolution equations for the rescaled conformal Weyl tensor agrees
on ${\cal I}$ with that of the equations which are obtained by linearizing
the Bianchi equation on Minkowski space. These (overdetermined) spin-2
equations take in the gauge above the form
\begin{equation}
\label{kconfminkb}
(1 + \tau)\,\partial_{\tau}\psi_k
- \rho\,\partial_{\rho}\psi_k + X_+\,\psi_{k+1}
+ (2 - k)\,\psi_k = 0,
\end{equation}
\begin{equation}
\label{k+1confminkb}
(1 - \tau)\,\partial_{\tau}\psi_{k+1}
+ \rho\,\partial_{\rho}\psi_{k+1} + X_-\,\psi_k
+ (1 - k)\,\psi_{k+1} = 0,
\end{equation}
where $k = 0, 1, 2, 3$ and the $\psi_j$ denote the essential components
of the linearized conformal Weyl spinor. 

The most conspicuous feature of
these equations is the factor $(1 + \tau)$ in (\ref{kconfminkb}), which
vanishes on ${\cal J}^- \cup {\cal I}^-$, and the factor $(1 - \tau)$ in
(\ref{k+1confminkb}), which vanishes on ${\cal J}^+ \cup {\cal I}^+$.
On ${\cal J}^{\pm}$ these factors arise because the coordinate 
$\tau$ is constant on ${\cal J}^{\pm}$ and these sets are
characteristics for the equations. By choosing $\kappa'$ differently,
this degeneracy can be removed on ${\cal J}^{\pm}$ 
(cf. \cite{friedrich:spin-2}). At 
${\cal I}^{\pm}$, however, this degeneracy cannot be removed in the
present setting. Any symmetric hyperbolic system extracted from these
equations, like e.g. 
\[
(1 + \tau)\,\partial_{\tau}\,\psi_0
- \rho\,\partial_{\rho}\,\psi_0 + X_+\psi_1 = - 2\,\psi_0, 
\]
\[ 
(4 + 2\,\tau)\,\partial_{\tau}\,\psi_1
- 2\,\rho\,\partial_{\rho}\,\psi_1 
+ X_-\psi_0 + 3\,X_+\psi_2 = - 4\,\psi_1, 
\]
\[
6\,\partial_{\tau}\,\psi_2 + 3\,X_-\psi_1 
+ 3\,X_+\psi_3 = 0,
\]
\[
(4 - 2\,\tau)\,\partial_{\tau}\,\psi_3
+ 2\,\rho\,\partial_{\rho}\,\psi_3 + 
3\,X_-\psi_2 + \,X_+\psi_4 = 4\,\psi_3,  
\]
\[
(1 - \tau)\,\partial_{\tau}\,\psi_4
+ \rho\,\partial_{\rho}\,\psi_0 + X_+\psi_1 = 2\,\psi_4. 
\]
must contain such factors at least in the equations for $\psi_0$ and 
$\psi_4$.  Writing this in the form $A^{\mu}\,\partial_{\mu}\,\psi =
H\,\psi$, and writing $\xi_{\tau} = \,<\partial_{\tau}, \xi>$,
$\xi_{\rho} = \,<\partial_{\rho}, \xi>$, $\xi_{\pm} = \,<\xi, X_{\pm}>$
we find 
\[
\det(A^{\mu}\,\xi_{\mu}) =
24\,\xi_{\tau}\,(g^{\mu\nu}\,\xi_{\mu}\,\xi_{\nu})\,
(3\,\xi_{\tau}^2 + g^{\mu\nu}\,\xi_{\mu}\,\xi_{\nu})
\]
with
\[
g^{\mu\nu}\,\xi_{\mu}\,\xi_{\nu} =
(1 - \tau^2)\,\xi_{\tau}^2 + 2\,\tau\,\rho\,\xi_{\tau}\,\xi_{\rho}
- \rho^2\,\xi_{\rho}^2 - \frac{1}{2}\,(\xi_+\,\xi_- + \xi_-\,\xi_+).
\]
It follows that characteristics pertaining to the quadratic terms 
which start on ${\cal I}$, stay on ${\cal I}$ and that those starting in
the physical space-time never end on ${\cal I} \cup {\cal I}^- \cup {\cal
I}^+$ but always run out to ${\cal J}^{\pm}$. 
Most importantly however, and this also holds true for
the general system (\ref{schemredequ}), the quadratic form
$g^{\mu\nu}\,\xi_{\mu}\,\xi_{\nu}$ degenerates at ${\cal I}^{\pm}$ and
there is a loss of real characteristics. This follows also
directly from
\begin{equation}
\label{hypbreakdown}
\det(A^{\tau}) = 0\,\,\,\mbox{on}\,\,{\cal I}^{\pm}. 
\end{equation}
It appears that this {\it loss of hyperbolicity at
the critical sets $I^{\pm}$}, is the key to the
smoothness problem for the conformal structure at null infinity.

\subsection{The s-jet at space-like infinity}
\label{sjetsatI}

The relations (\ref{Arvanishes}) and (\ref{hypbreakdown}) are the dominant
features of the regular finite initial value problem at space-like
infinity. The consequences of (\ref{hypbreakdown}) are not deduced by the 
standard textbook analysis, we have to rely on the specific
properties of our problem. It turns out that a considerable amount of
information on the behaviour of the solution near the critical sets can be
obtained by exploiting (\ref{Arvanishes}). We know already that the
solution is smooth in some neighbourhood of $\bar{{\cal C}}$ in 
$\bar{{\cal N}}$ and that $u$ can be calculated on
${\cal I}$ by solving intrinsic equations on ${\cal I}$. It will be shown
in the following that a full formal expansion of $u$ in terms of $\rho$ can
be calculated on ${\cal I}$ by solving certain transport equations. 

The following notation will be convenient in the following.
For $p = 0, 1, 2, \ldots$ and any sufficiently smooth 
(possibly vector-valued) function $f$
defined on ${\cal N} \cup {\cal I}$ we write $f^p$ for the restriction 
to ${\cal I}$ of the $p$-th radial derivative $\partial^p_{\rho}\,f$.
The set of functions $f^0, f^1, \ldots, f^p$ on ${\cal I}$ will be
denoted by $J^p_{\cal I}(f)$ and referred to as the {\it jet of order $p$
of $f$ on ${\cal I}$} (and similarly with ${\cal I}$ replaced by ${\cal
I}^0$.)
If $u = (w, \phi)$ is a {\it solution} of equations (\ref{schemredequ})
we refer to $J^p_{\cal I}(u)$ (respectively
$J^p_{\cal I}(w)$, $J^p_{\cal I}(\phi)$) as to {\it the 
s-jet\index{s-jet} of $u$
(resp. $w$, $\phi$) of order $p$} and to the {\it data}
$J^p_{{\cal I}^0}(u)$ (respectively $J^p_{{\cal I}^0}(w)$, $J^p_{{\cal
I}^0}(\phi)$) on ${\cal I}^0$ as to {\it the d-jet of
$u$ (resp. $w$, $\phi$) of order $p$}\index{d-jet}.
A s-jet $J^p_{\cal I}(u)$ (respectively
$J^p_{\cal I}(w)$, $J^p_{\cal I}(\phi)$) of order $p$ will be called {\it
regular on}
\[
\bar{{\cal I}} \equiv {\cal I} \cup {\cal I}^- \cup {\cal I}^+,
\]
(or simply {\it regular}) if the corresponding functions on
${\cal I}$ extend smoothly to the critical sets ${\cal I}^{\pm}$. 

An initial data set on ${\cal S}$ will be called {\it asymptotically static
of order $p$}\index{asymptotically static}, where $p \in \mathbb{N} \cup
\{\infty\}$, if its d-jet
$J^{p}_{{\cal I}^0}(u)$ coincides with the d-jet of order $p$ of some static
asymptotically flat data set defined on some neighbourhood of $i$ in 
${\cal S}$. It will be seen later that {\it asymptotic staticity} (of
order $p$) is an important feature of initial data sets.

\vspace{.3cm}

Applying the operator $\partial^p_{\rho}$ formally to the first of
equations (\ref{schemredequ}) and restricting to ${\cal I}$, one obtains
for $w^{p}$ an equation of the form
\begin{equation}
\label{wpschemredequ}
\partial_{\tau}\,w^{p} = G(\tau, t, w^0, \ldots w^{p-1}, w^{p}, 
\phi^0, \ldots \phi^{p-1}),\,\,\,p = 1, 2, \ldots,
\end{equation}
where the right hand side is an affine function of $w^{p}$.
The functions $\phi^{p}$ do not appear here, because the rescaled
conformal Weyl spinor occurs in the equations for the frame, connection,
and Ricci coefficients with the factors $\Theta$ and
$d_{AB}$, which vanish on ${\cal I}$. It follows that the s-jet 
$J^{p}_{\cal I}(w)$ can be determined by the integration of an (easily
solvable) linear system of ODE's, if the s-jet $J^{p-1}_{\cal I}(u)$ and
the d-jet $J^{p}_{{\cal I}^0}(w)$ are known. 

\vspace{.3cm}

With the notation (\ref{Aminksplitoff}) the Bianchi equation can be written
\begin{equation}
\label{AcurvflatBian}
\nabla^{\star F}\,_{A'}\,\phi_{BCDF} = - \phi_{A'BCD},
\end{equation}
where 
\begin{equation}
\label{0atIAcurvflatBianpart}
\phi_{A'BCD} = \phi_{A'(BCD)} \equiv
\check{e}^{iF}\,_{A'}\,v_i(\phi_{BCDF})
- 4\,\check{\Gamma}^F\,_{A'}\,^E\,_{(B}\,\phi_{CDF)E}.
\end{equation}
Then
\begin{equation}
\label{pderIrestrbian}
(\sqrt{2}\,\nabla^{\star F}\,_{A'}\,\phi_{BCDF})^p 
= - \sqrt{2}\,\phi^p_{A'BCD},
\end{equation}
provides equations with left hand sides given by
\begin{equation}
\label{jbianonI}
 (1 + \tau)\,\partial_{\tau}\,\phi^p_j
+ X_+\,\phi^p_{j + 1}
+ (2 - j - p)\,\phi^p_j = \ldots,
\end{equation}
\begin{equation}
\label{j+1bianonI}
(1 - \tau)\,\partial_{\tau}\,\phi^p_{j+1}
 + X_-\,\phi^p_j 
+ (1 - j + p)\,\phi^p_{j+1} = \ldots,
\end{equation}
where $j = 0,\ldots, 3$, and right hand sides given by (cf.
(\ref{Bminksplitoff}))
\begin{equation}
\label{rescweylrec}
\phi^p_{A'BCD} = 
\sum_{i = 0, +, -} \,\sum_{j = 1}^p {p \choose j}
\,(\check{c}^{\,i F}\,_{A'})^j\,v_i\,(\phi^{p - j}_{BCDF})
\end{equation}
\[
+ \sum_{j = 1}^p (p - j)\,{p \choose j}
\,(\check{c}^{\,1 F}\,_{A'})^j\,\phi^{p - j}_{BCDF}
- 4\,\sum_{j = 1}^p {p \choose j}
\,(\check{\Gamma}^F\,_{A'}\,^E\,_{(B})^j\,\phi^{p - j}_{CDF)E}. 
\]
We note that these expressions depend on $J^p_{\cal I}(w)$ but only on
$J^{p-1}_{\cal I}(\phi)$. Thus, given these s-jets, the s-jet $J^{p}_{\cal
I}(\phi)$ can be obtained by solving a linear system of ODE's, if
$J^{p}_{{\cal I}^0}(\phi)$ is given. {\it Because the system is singular at
the critical sets it is not clear a priori that $J^{p}_{\cal I}(\phi)$ is
regular, even if
$J^p_{\cal I}(w)$ and $J^{p-1}_{\cal I}(\phi)$ are regular}.

\vspace{.3cm}

To obtain more detailed information on the solutions, it is useful to
consider a system system of second order. From (\ref{AcurvflatBian})
follows
\[
\nabla^{\star}_{EE'}\,\nabla^{\star EE'}\,\phi_{ABCD}
= 2\,\nabla^M_A\,^{E'}\,\nabla^{\star E}\,_{E'}\,\phi_{BCDE}
= 2\,\nabla^M_A\,^{E'}\,\phi_{E'BCD},
\]
which is equivalent to
\begin{equation}
\label{Aflatwaveop}
\nabla^{\star}_{EE'}\,\nabla^{\star EE'}\,\phi_{ABCD}
= f_{ABCD}
\equiv - 2\,\nabla^{\star}_{E'(A}\,\phi^{E'}\,_{BCD)},
\end{equation}
\begin{equation}
\label{Bflatwaveop}
0 = g_{BC} \equiv \nabla^{\star A'A}\,\phi_{A'ABC}.
\end{equation}
While the right hand side of 
\begin{equation}
\label{pderrestIwavebian}
(\nabla^{\star}_{EE'}\,\nabla^{\star EE'}\,\phi_{ABCD})^p
= f^p_{ABCD},
\end{equation}
depends again, similar to (\ref{rescweylrec}), on $J^p_{\cal I}(w)$ and
$J^{p-1}_{\cal I}(\phi)$, the left hand side takes the decoupled form
\begin{equation}
\label{pIphiIpart}
(1 - \tau^2)\,\partial^2_{\tau}\,\phi^p_j
+ 2\,\{(p - 1)\,\tau - j + 2\}\,\partial_{\tau}\,\phi^p_j
+ C\,\phi^p_j - p\,(p - 1)\,\phi^p_j = \ldots
\end{equation}
where the spin weight relations $X\,\phi_j = 2\,(2 - j)\,\phi_j$
and the Casimir operator
$C = - \frac{1}{2}\,(X_+\,X_- + X_-\,X_+) + \frac{1}{4}\,X^2$ on $SU(2)$
have been used to arrive at this expression.

The fields $\phi^p_j$ have expansions
\[
\phi^p_j = \sum_{q = |2-j|}^p \,\phi^p_{j,q}
\,\,\,\,\,\,\mbox{where}\,\,\,\,\,\, 
\phi^p_{j,q} = \sum_{k = 0}^{2\,q} \,
\phi^p_{j, q, k}\,\,T_{2\,q}\,^k\,_{q-2+j},
\]
with coefficients $\phi^p_{j, q, k} = \phi^p_{j, q, k}(\tau)$. Since
the Casimir operator satisfies 
\[
C\,(T_{2\,q}\,^k\,_{q-2+j}) = q\,(q + 1)\,T_{2\,q}\,^k\,_{q-2+j},
\]
equation (\ref{pIphiIpart}) implies for $\phi^p_{j,q}$ ODE's of the
form 
\begin{equation}
\label{phipIequ}
D_{(n, \alpha, \beta)}\,\phi^p_{j,q}
\equiv
(1 - \tau^2)\,\partial^2_{\tau}\,\phi^p_{j,q}
+ \{\beta - \alpha - (\alpha + \beta +
2)\,\tau\}\,\partial_{\tau}\,\phi^p_{j,q} 
\end{equation}
\[
+ n\,(n + \alpha + \beta + 1)\,\phi^p_{j,q} = \ldots
\]
with 
\[
\alpha = j - p - 2,\,\,\,\beta = - j - p + 2,\,\,\,\,
n = n_1 \equiv p + q \,\,\,\mbox{or}\,\,\,
n = n_2 \equiv p - q - 1.
\]

The equations above allow us to calculate recursively a formal expansion
of the solution $u = (w, \phi)$ to (\ref{schemredequ}) in a series of the
form 
\begin{equation}
\label{ITaylorexp}
u = \sum_{n = 0}^{\infty}\frac{1}{n!}\,u^p\,\rho^p,
\end{equation}
on ${\cal I}$ (note the different meanings of the supersripts $p$) with
coefficients
$u^p = u^p(\tau, t) \in C^{\infty}({\cal I})$. In some neighbourhood of
${\cal I}^0$ in $\bar{{\cal N}}$ this series represents in fact the Taylor
series of smooth functions and it converges near ${\cal I}^0$ if the datum
$h$ is real analytic. We shall try to deduce from it information on the
behaviour of $u$ near the critical sets.

\subsection{Behaviour of the s-jets near the critical sets.}
\label{sjetsnearcriticalsets}

Because $J^0_{\cal I}(u)$ is regular, the integration gives a regular s-jet
$J^1_{\cal I}(w)$. The calculation of $J^1_{\cal I}(\phi)$ gives (in the
cn-gauge and with $\kappa' = 1$) the regular solution

\begin{equation}
\label{6Iu1data}
\phi^1_{ABCD} = 
- \{ W_1\,36\,(1- \tau^2) 
+ m^2\,(18\,\tau^2 - 3\,\tau^4\}\,\epsilon^2\,_{ABCD}
\end{equation}
\[
- 12\,(1 - \tau)^2\,X_{+}\,W_1\,\epsilon^1\,_{ABCD}
+ 12\,(1 + \tau)^2\,X_{-}\,W_1\,\epsilon^3\,_{ABCD}.
\]
Thus $J^2_{\cal I}(w)$ will again be regular. It tuns out that
$J^2_{\cal I}(\phi)$ will not necessarily be regular. The integration
(cn-gauge, $\kappa' = 1$) gives
\[
\phi^2_{ABCD} = \phi^{ih\,2}_{ABCD}  
+ \breve{\phi}^{W\,2}_{ABCD}
+ \breve{\phi}^{'\,2}_{ABCD},
\]
with 
\[
\phi^{2\,ih}_{(ABCD)_0} = 0,\,\,\,\,\,
\phi^{2\,ih}_{(ABCD)_2} = c_2(\tau)\,m\,W_1 + c_3(\tau)\,m^3,\,\,\,\,  
\phi^{2\,ih}_{(ABCD)_4} = 0,
\]
\[
\phi^{2\,ih}_{(ABCD)_1} = c_1(\tau)\,m\,X_{+}W_1,\,\,\,\,  
\phi^{2\,ih}_{(ABCD)_3} = - c_1(-\tau)\,m\,X_{-}W_1,  
\]
where the $c_i(\tau)$ are polynomials in $\tau$ of order $\le 8$,
\begin{equation}
\label{1bIu3data}
\breve{\phi}^{W\,2}_{(ABCD)_j}
= - 4\,\sqrt{6\,{4 \choose j}}\,(1 + \tau)^j\,(1 - \tau)^{4 - j}
\sum_{k = 0}^4 W_{2;4,k}\,T_4\,^k\,_j,
\end{equation}
and
\begin{equation}
\label{1cIu3data}
\breve{\phi}^{'2}_{(ABCD)_j} = a_j(\tau)\,\frac{1}{3}\,\sum_{k = 0}^4 
\sqrt{2\,{4 \choose k}}\,b^{*}_{(EFGH)_k}\,T_4\,^k\,_j 
\end{equation}
with
\[
a_0(\tau) = 2\,(1 - \tau)^4\,K(- \tau) = - a_4(- \tau),
\]
\[
a_1(\tau) = 4\,(1 - \tau)^3\,(1 + \tau)\,K(- \tau) 
- \frac{3}{1 - \tau} = - a_3(- \tau),
\]
\[
a_2(\tau) = \sqrt{6}\,\{\frac{2 - \tau}{(1 + \tau)^2} 
- 2\,(1 - \tau)^2\,(1 + \tau)^2\,K(\tau) \} = - a_2(- \tau),
\]
\[
K(\tau) = 1 - 3\,\int_0^{\tau}\frac{ds}{(1 - s)\,(1 + s)^5}.
\]
While the first two terms extend smoothly to ${\cal I}^{\pm}$, the third
term has logarithmic singularities at the critical sets unless the {\it
regularity condition} $b_{ABCD}(i) = 0$ is satisfied
(the quadrupole term $W_2$, which looks so innocent
here, reappears in obstructions to smoothness at higher order
\cite{valiente kroon:2003}, \cite{valiente kroon:2003B}).

It is thus clearly important to control the behaviour of the s-jets at 
${\cal I}^{\pm}$ at all orders. Equations $D_{(n, \alpha, \beta)} u = 0$
are well known from the theory of Jacobi polynomials and they have been
used in \cite{friedrich:i-null} to derive a certain representation of the
solutions in terms of polynomials built from the generalized Jacobi
polynomials $P_n^{(\alpha, \beta)}(\tau)$ (\cite{szego}). By the
overdeterminedness of the system (\ref{jbianonI}), (\ref{j+1bianonI})
the problem can be reduced to the integration of the functions
$\phi^p_{0,q}$, $\phi^p_{4,q}$. The
functions $\phi^p_{1,q}$, $\phi^p_{2,q}$, $\phi^p_{3,q}$ can be
calculated from them algebraically. 

One finds for 
$p \ge 3$ and $q = p$ the representation
\begin{equation}
\label{phi0reprlogterm}
\phi^p_{0,p} = 
(1 - \tau)^{p + 2}\,(1 + \tau)^{p - 2}\,
\left(\phi^p_{0,p\,*} + 
\right.
\end{equation}
\[
\left.
\frac{(p+1)\,(p+2)}{4\,p}\,
(\phi^p_{0,p\,*} - \phi^p_{4,p\,*})    
\,\int_0^{\tau} 
\frac{d\tau'}{(1 + \tau')^{p - 1} \, (1 - \tau')^{p + 3}} \right),
\]
\begin{equation}
\label{phi4reprlogterm}
\phi^p_{4,p} = (1 + \tau)^{p + 2}\,(1 - \tau)^{p - 2}\,
\left(\phi^p_{4,p\,*} - 
\right.
\end{equation}
\[
\left.
\frac{(p+1)\,(p+2)}{4\,p}\,
(\phi^p_{0,p\,*} - \phi^p_{4,p\,*})    
\,\int_0^{- \tau} 
\frac{d\tau'}{(1 + \tau')^{p - 1} \, (1 - \tau')^{p + 3}} \right),
\]
where the substcript $*$ indicates initial data on ${\cal I}^0$.

Denoting by $y_{p, q}$ the column vector formed from $\phi^p_{0,q}$,
$\phi^p_{4,q}$, one obtains for $p \ge 3$ and $0 \le q \le p - 1$ 
\begin{equation}
\label{qlep-1repr}
y_{p, q}(\tau) = X_{p, q}(\tau)\,
\left( X_{p, q\, *}^{-1}\,y_{p, q\, *} + \int_{0}^{\tau}
X_{p, q}(\tau')^{-1}\,B_{p, q}(\tau')\,d\tau'
\right). 
\end{equation}
The functions $B_{p, q}$ are derived from the right hand sides of 
(\ref{pderIrestrbian}) and (\ref{pderrestIwavebian})
and can thus be calculated from  
$J^p_{\cal I}(w)$ and $J^{p-1}_{\cal I}(\phi)$.
The  
matrix-valued valued functions $X_{p, q}$ are given by
\[
X_{p, 0} = 
\left( \begin{array}{cc}
(1 + \tau)^{p - 2}\,(p + \tau) & 0 \\
0 & (1 - \tau)^{p - 2}\,(p - \tau) \\
\end{array} \right),
\]
\[
X_{p, 1} = 
\left( \begin{array}{cc}
(1 + \tau)^{p - 2} & 0 \\
0 & (1 - \tau)^{p - 2}\\
\end{array} \right),
\]
\[
X_{p, q} = 
\left( \begin{array}{cc}
Q_{1; p, q}(\tau) & (-1)^q\,Q_{3; p, q}(\tau) \\
(-1)^q\,Q_{3; p, q}(- \tau) & Q_{1; p, q}(- \tau)\\
\end{array} \right),
\,\,\,\,\,\,\,2 \le q \le p - 1,
\]
with polynomials 
\[
Q_{1; p, q}(\tau) =
(\frac{1 - \tau}{2})^{p + 2}\,P_{q - 2}^{(p + 2, - p + 2)}(\tau),
\]
\[
Q_{3; p, q}(\tau) =
(\frac{1 + \tau}{2})^{p - 2}\,P_{q + 2}^{(- p - 2, p - 2)}(\tau).
\] 
of degree $n_1 = p + q$.

\vspace{.3cm}

The solutions to the transport equations can be 
calculated, order by order, explicitly. The only
difficulty is the calculation of the functions 
$B_{p, q} = B_{p, q}[J^p_{\cal I}(w), J^{p-1}_{\cal I}(\phi)]$ which
become more and more complicated at each step.

The most conspicuous feature of these expressions is the occurrence of
logarithmic singularities at ${\cal I}^{\pm}$. The latter can arise, as a
consequence of the evolution process and the structure of the data, even
under the strongest smoothness assumptions on the conformal datum $h$.
We will have to discuss to what extent the occurrence of such
singularities can be related to the structure of the initial data and
whether it can be avoided by a judicious choice of the latter.

\subsection{Regularity conditions}
\label{regularitycondition}

Expanding the integrals in (\ref{phi0reprlogterm}),
(\ref{phi4reprlogterm}) one finds  
\[
\phi^p_{0,p} \approx (1 - \tau)^{p + 2}\,
(1 + \tau)^{p - 2}\,\log (1 - \tau) +
\,\,analytic\,\,in\,\,\tau\,\,\,\,
\mbox{as} \,\,\,\tau \rightarrow 1
\]
and a similar behaviour for $\phi^p_{4,p}$ as $\tau \rightarrow -1$,
unless the initial data on 
${\cal I}^0$ satisfy the condition
\[
\phi^p_{0,p\,*} = \phi^p_{4,p\,*}.
\]
(Note that the singularities get less severe with increasing $p$.) This
raises the question whether data can be given which satisfy these
conditions. By a lenghty recursion argument it can be shown
(\cite{friedrich:i-null}) that {\it for given integer $p_* \ge 0$ the
fields $\phi^p_{j, p}$ resulting from (\ref{phi0reprlogterm}),
(\ref{phi4reprlogterm})
extend smoothly to ${\cal I}^{\pm}$ for $2 \le p \le p_* +
2$ if and only if the free datum $h$ satisfies the regularity 
condition}\index{regularity condition} 
\begin{equation}
\label{regcond}
D_{(A_q B_q} \ldots D_{A_1 B_1}\,b_{ABCD)}(i) = 0,
\,\,\,\,\,q = 0,1,2, \ldots ,p_*. 
\end{equation}

By (\ref{Sstaregcond}) these conditions
are satisfied for static data with $p_* = \infty$. This allows
one to construct a large class of data satisfying (\ref{regcond})
by gluing with a partition of unity an asymptotically flat static end
to a given time reflection symmetric data set and solving the Lichnerowicz
equation. 

Condition (\ref{regcond}) has been observed as a regularity
condition before. In \cite{friedrich:static} has been derived under the
strong assumption that the solution be massless (cf. (\ref{massless}))  a
{\it necessary and sufficient} condition on $h$ that space-like infinity
can be represented by a regular point $i^0$ in a smooth conformal
space-time extension (so that ${\cal J}^{\pm}$ will be smooth
near space-like infinity). This condition, referred to as {\it
radiativity condition}, implies (\ref{regcond}). It has been shown in
\cite{friedrich:i-null} that these two conditions are in fact equivalent. 

The first term in (\ref{qlep-1repr}) is polynomial and
thus regular. The second term is not so easy to handle. 
If (\ref{regcond}) is not assumed the corresponding $\log$-terms will enter
the integral in a non-linear way and the solution will have at ${\cal
I}^{\pm}$ polyhomogeneous expansions in terms of expressions
$(1 \mp \tau)^k\,\log^j(1 \mp \tau)$ with $k, j \in \mathbb{N}_0$. 
We shall assume therefore that (\ref{regcond}) holds with $p_* = \infty$.

From the expressions above it follows that the Wronskian  $\det(X_{p, q})$
has a factor $(1 - \tau^2)^{p - 2}$. The regularity of the integrals in
(\ref{qlep-1repr}) thus depends on the precise structure of the functions
$B_{p, q}(\tau)$, which get quite complicate with increasing $p$.
It has been shown in \cite{friedrich:i-null} and \cite{friedrich:kannar1} 
that $J^p_{\cal I}(u)$ is regular for $p \le 3$ if (\ref{regcond}) is
satisfied with $p_* \le 1$.

Because the functions $B_{p, q}$ are getting increasingly complicated with
$p$, J. A. Valiente-Kroon studied the case where $h$ is conformallly flat on
${\cal B}_e$  with the help of an algebraic computer program (\cite{valiente
kroon:2003}). In that case condition (\ref{regcond}) is
trivially satisfied but there still exists a large class of non-trivial data
for which $h$ is not conformally flat outside ${\cal B}_e$. In the conformal
factor (\ref{ivpconfac}) one has $U = 1$ on ${\cal B}_e$ but $W$ will
be a non-trivial solution to the conformally covariant Laplace equation
with $m = 2\,W(i) \neq 0$. It turns out that $J^4_{\cal I}(u)$ is again
regular. For  $J^5_{\cal I}(u)$ however, logarithmic terms are observed.
They come with certain coefficients which depend on the data. Choosing the
data such that these coefficients vanish, still new logarithmic terms are
observed for $J^6_{\cal I}(u)$. Restricting to the axially symmetric case
to keep the expressions manageable, new logarithmic terms crop up for $p =
7$ and $p = 8$.

The form of the conditions obtained at these orders suggests
a general formula which needs to be satisfied to excluded logarithmic terms
at any given order $p$ (\cite{valiente kroon:2003}). If this formula is
correct, all derivatives of $W$ must vanish at $i$ if the
logarithmic terms are required to vanish at all orders. As a consequence
the solution must become asymptotically Schwarzschild at $i$ (cf. Lemma
\ref{stcflat}). Since $W$ is governed on ${\cal B}_e$ by an elliptic
equation with analytic coefficients it would follow that the solution is
precisely Schwarzschild near $i$.

How seriously do we need to take the singularities at ${\cal I}^{\pm}$ ?
To answer this question one needs to control the evolution of the field in
a full neighbourhood of $\bar{{\cal I}}$ in $\bar{{\cal M}}$. This has not
been achieved yet. However, the analysis of the linearized
setting, which is given by the spin-2 equations (\ref{kconfminkb}),
(\ref{k+1confminkb}) on Minkowski space in the gauge (\ref{lnmmbarvects}),
gives some insight (\cite{friedrich:spin-2}).

While the functions $B_{p, q}$ vanish in that case, the singularities
arising from (\ref{phi0reprlogterm}), (\ref{phi4reprlogterm}) do in
general survive the linearization process. The analysis then shows that
for prescribed integer $j$ the function
\[
\psi_k - \sum_{p' = 0}^{p - 1} \frac{1}{p'\,!}\,\psi^{p'}_k\,\rho^{p'}
\,\,\,\,\mbox{on}\,\,\,\, \tilde{{\cal M}}
\]
extends to a function of class $C^j$ on $\bar{{\cal M}}$, if one chooses 
$p \ge j + 6$ in the expansion above. Here 
$\psi^{p'}_k$, $p' = 0, 1, \ldots p - 1$, are understood as
$\rho$-independent functions on $\tilde{{\cal M}} \cup {\cal I}$, 
which agree on ${\cal I}$ with the s-jet $J^{p - 1}_{\cal I}(\psi)$ (defined
by equations (\ref{kconfminkb}), (\ref{k+1confminkb})). Note that the sum
above provides the first terms of an asymptotic expansion of the solution
at ${\cal J}^{\pm}$.

It follows that the solution will extend smoothly to all of $\bar{{\cal
M}}$ if the linearized version of (\ref{regcond}) is satisfied with $p_* =
\infty$. If the condition is satisfied only with some finite 
$p_* \ge 2$ but violated at $p = p_* +1$, the solution will develop a
logarithmic singularity at ${\cal I}^{\pm}$ which will be transported along
the null generators of ${\cal J}^{\pm}$ so that the solution will be only in
$C^{p_* - 2}(\bar{{\cal M}})$. While it remains to be seen whether the
solutions to the non-linear equations admit similar asymptotic expansions
at ${\cal J}^{\pm}$, the discussion shows clearly that the regularity of the
s-jets
$J^p_{\cal I}(u)$ is a prerequisite for the smooth extensibility of the
solutions to ${\cal J}^{\pm}$.

If the solutions to the non-linear equations show a singular behaviour
on ${\cal J}^{\pm}$ as indicated above, does it refer to something `real'
or to a failure of the gauge ? If the underlying conformal
structure where smooth at null infinity, the conformal geodesics should
pass through ${\cal J}^{\pm}$ where $\Theta \rightarrow 0$ and the
1-form, the $\hat{\nabla}$-parallely transported frame, and therefore also
the rescaled conformal Weyl spinor in that frame should be represented by
smooth functions of $\tau$ along the conformal geodesics because these
as well as their natural parameter $\tau$ depend only on the conformal
structure. Singularities as indicated above therefore refer to
intrinsic features of the underlying conformal structure. 

The results of (\cite{valiente kroon:2003}) show first of all that the
regularity condition (\ref{regcond}) with $p_* = \infty$ are {\it not
sufficient} for the regularity of $J^p_{\cal I}(u)$, $p = 0, 1, 2, \ldots$.
It appears that the Lichnerowicz equation, which breaks the conformal
invariance by fixing the scaling of the physcial metric $\tilde{h} =
\Omega^{-2}\,h$, does play a role in the smoothness of the conformal
structure at null infinity. This is remarkable because it shows that
besides the local condition (\ref{regcond}) there are other conditions to
be observed which are `not so local'. However, the Lichnerowicz equation
is introduced only as a device to reduce the problem of solving the  {\it
underdetermined elliptic system} of constraints to an {\it elliptic}
problem.  The results of 
\cite{chrusciel:delay:2003}, \cite{corvino}, \cite{corvino:schoen}
exploit the underdeterminedness of the contraints in quite a different way.
They teach us to be careful with the words `local' and `global' in the
present context.

The main purpose of calculating $J^p_{\cal I}(u)$ for the first few $p$
is to get an insight into
(\ref{qlep-1repr}) which would allow us to control
the behaviour of $J^p_{\cal I}(u)$ near ${\cal I}^{\pm}$ in dependence of the
data given on ${\cal S}$. One may speculate that the results above are
telling us that asymptotic staticity, or more generally asymptotic
stationarity, at space-like infinity is of more importance in the present
context than expected so far. Recent generalizations of the calculations in
(\cite{valiente kroon:2003}) to non-conformally flat data seem to support
this view (\cite{valiente kroon:2003B}).

This raises the question whether the setting proposed in
\cite{friedrich:i-null} is for static solutions as smooth as one would
expect. This is far from obvious because of the loss of hyperbolicity at
the critical sets. Giving an answer to this question for general static
solutions will be the purpose of the following chapters.

\section{Conformal extensions of static vacuum space-times}
\label{confextstaticvacuum}

For static asymptotically flat vacuum solutions with positive ADM
mass we shall construct in the following a conformal extension 
which will include null infinity and will also allow us to discuss the
cylinder at space-like infinity. The extension will be defined in terms of 
explicitly given coordinates and conformal rescaling. In section
\ref{stavacatI} will it be shown that it coincides with the extension
(not the coordinates etc.) as defined in section
\ref{specificsofrfivpsli}. 

Because one expects usually `not much to happen at space-like infinity'
for static asymptotically flat solutions, one may wonder
why the detailed discussion of the fields near space-like infinity
should be so complicated. An obvious reason is that a gauge which is chosen
to discuss space-like and null infinity must introduce a `time
dependence', it cannot be adapted to a Killing field whose flow
lines run out to time-like infinity. However, the main reason is 
that the static field equations play an important role in
discussing the regularity of the field near the critical sets; we will
have to make extensive use of them.

\vspace{.3cm}

The static vacuum solution is assumed in the form
\[
\tilde{g} = v^2\,dt^2 + \Omega^{-2}\,h,
\]
with $v = v(x^c)$, $h = h_{ab}(x^c)\,dx^a\,dx^b$ and a conformal factor
$\Omega = \Omega(x^c)$, where we assume $h$-normal coordinates $x^a$ 
which satisfy (\ref{hnormal}) and the conformal gauge which achieves
(\ref{mgauge}) on the set $\mathbb{R} \times {\cal U}$, where 
${\cal U} = \{|x| < \bar{\rho}_*\}$ with a
sufficiently small $\bar{\rho}_* > 0$.
We set 
\[
\Upsilon = |x|^2,\,\,\,\,\,\,
e^a = \frac{x^a}{|x|} = - \frac{1}{2}\,\Upsilon^{- 1/2}\,D^a\,\Upsilon   
\,\,\,\,\,\,
\mbox{for}\,\,\,\,\,\,|x| > 0,\,\,\,\,\,\,
\bar{\rho} = \sqrt{\sum_{a = 0}^3 (x^a)^2}.
\]
Coordinates $\psi^A$, $A = 2, 3$, on the sphere $S^2 = \{|x| = 1\}$
can be used to parametrize $e^a$ and we write then
$e^a = e^a(\psi^A)$ and $d\,e^a = e^a_{\,\,,\psi^A}\,d\,\psi^A$. For
convenience the coordinates $\psi^A$ will be assumed in the following to
be real analytic. If $x^a = \bar{\rho}\,e^a(\psi^A)$, the metric
$h$ takes the form 
\[
h = - d\,\bar{\rho}^2 + \bar{\rho}^2\,k,
\]
with ($\bar{\rho}$-dependent) $2$-metrics 
\[
k = k_{AC}\,d\psi^A\,d\,\psi^C \equiv
h_{ac}(\bar{\rho}\,e^c)\,d\,e^a\,d\,e^c,
\] 
on the spheres $\bar{\rho} = const. > 0$. For 
$\bar{\rho} \rightarrow 0$ the metric $k$ approaches the standard line
element $d\,\sigma^2 = - k(0,\psi^A)$ on the 2-dimensional unit sphere
in the coordinates $\psi^A$.

\vspace{.5cm}

We write now $x^0 = t$ and $x^{0'} = \bar{\tau}$, $x^{1'} = \bar{\rho}$,
$x^{A'} = \psi^A$ and consider the map
$\Phi:  x^{\mu'} \rightarrow x^{\mu}(x^{\mu'})$ 
defined by
\begin{equation}
\label{Bbasemap}
t(x^{\mu'})  = \int\limits_{\bar{\rho}\,(1 -
\bar{\tau})}^{\bar{\rho}}\frac{d\,s}{(v\,\Omega)(s\,e^a(\psi^A))},
\,\,\,\,\,\,\,\,\,\,\,\, 
x^a(x^{\mu'}) = \bar{\rho}\,(1 - \bar{\tau})\,e^a(\psi^A).
\end{equation}
It follows that the four differentials
\begin{equation}
\label{dx2}
d\,x^a = 
((1 - \bar{\tau})\,d\,\bar{\rho} - \bar{\rho}\,d\,\bar{\tau})\,e^a
+ \bar{\rho}\,(1 - \bar{\tau})\,d\,e^a,
\end{equation}
\begin{equation}
\label{dt2a}
dt = \left(\frac{1}{(v\,\Omega)(\bar{\rho}\,e^a)}
- \frac{1 - \bar{\tau}}{(v\,\Omega)(\bar{\rho}\,(1 - \bar{\tau})\,e^a)}\right)
d\,\bar{\rho}
+ \frac{\bar{\rho}}{(v\,\Omega)(\bar{\rho}\,(1 - \bar{\tau})\,e^a)}\,d\,\bar{\tau}
+ l,
\end{equation}
with
\begin{equation}
\label{dt2b}
l =  l_A\,d\,\psi^A,\,\,\,\,\,\,\,\,\,\,\,\,
l_A = \int\limits_{\bar{\rho}\,(1 - \bar{\tau})}^{\bar{\rho}}
\left(\frac{1}{(v\,\Omega)(s\,e^a)}
\right)_{,\psi^A} d\,s,
\end{equation}
are independent for $0 \le \bar{\tau} < 1$ and $0 < \bar{\rho} < \bar{\rho}_*$
and we can consider the $x^{\mu'}$ as
smooth coordinates on an open neighbourhood of space-like
infinity in $\{t \ge 0\}$. 
For $s > 0$ we set 
\begin{equation}
\label{hdef}
h(s\,e^a) \equiv \frac{(v\,\Omega)(s\,e^a)}{s^2} = 
\frac{U(s\,e^a) - s\,\frac{m}{2}}
{(U(s\,e^a) + s\,\frac{m}{2})^3}.
\end{equation}

To indicate the different arguments replacing $s$ in this and other
functions of $s\,e^a$ or of $s$ and $\psi^A$, we write out the argument
replacing $s$ explicitly but suppress the dependence on $e^a$ or $\psi^A$.
Thus $h(s)$ will be written for $h(s\,e^a)$ and $k(\bar{\rho})$ for
$k(\bar{\rho},
\psi^A)$, etc. 

\vspace{.3cm}

With this notation and the conformal factor
\[
\Lambda = \Omega\,\Upsilon^{-1/2},
\] 
a conformal representation of $\tilde{g}$ is defined by 
\begin{equation}
\label{Xconfext}
\bar{g} \equiv \Phi^*\left(\Lambda^2\,\tilde{g}\right) = 
2\,\left( 
\frac{h(\bar{\rho}\,(1 - \bar{\tau}))}{h(\bar{\rho})}
\,\frac{d\,\bar{\rho}}{\bar{\rho}} +
\bar{\rho}\,
h(\bar{\rho}\,(1 - \bar{\tau}))\,l\right)d\,\bar{\tau}
\end{equation}
\[
- 2\,(1 - \bar{\tau})\,\left( 
\frac{h(\bar{\rho}\,(1 - \bar{\tau}))}{h(\bar{\rho})}\,
\frac{d\,\bar{\rho}^2}{\bar{\rho}^2} +
\bar{\rho}\,\,h(\bar{\rho}\,(1 -
\bar{\tau}))\,l\,\frac{d\,\bar{\rho}}{\bar{\rho}}   
\right)
\]
\[
+ (1 - \bar{\tau})^2\,
\left( 
\frac{h(\bar{\rho}\,(1 - \bar{\tau}))}{h(\bar{\rho})}\,
\frac{d\,\bar{\rho}}{\bar{\rho}} +
\bar{\rho}\,\,h(\bar{\rho}\,(1 - \bar{\tau}))\,l\right)^2
+ k(\bar{\rho}\,(1 - \bar{\tau})).
\]

The new coordinates do not reflect the symmetries of the
underlying space-time, but they are sufficient to discuss the part of
the space-time in the future of the initial hypersurface $\{t = 0\}$. 
We replace ${\cal S}$ by the manifold with boundary $\bar{{\cal S}}$ 
introduced in section \ref{constrlift}. The points of 
$\partial \bar{{\cal S}}$ are thought of as ideal end
points attached to the curves 
$\bar{\rho} \rightarrow x^a(\bar{\rho}) = \bar{\rho}\,e^a(\psi^A)$ in
$\tilde{{\cal S}}$ as $\bar{\rho} \rightarrow 0$ for fixed value of
$\psi^A$. The coordinates $\bar{\rho}$ and $\psi^A$ extend (by definition)
to analytic coordinates on $\bar{{\cal S}}$ with $\bar{\rho} = 0$ on 
$\partial \bar{{\cal S}}$.
We set  
\[
\tilde{{\cal M}}' = \{0 \le \bar{\tau} < 1,\,\,0 < \bar{\rho}\},\,\,\,\,
\bar{{\cal M}}' 
= \tilde{{\cal M}}' \cup {\cal J}^{+'} \cup {\cal I}' \cup {\cal I}^{+'},
\]
where it is understood that the unspecified coordinate systems $\psi^A$
`cover' the sphere $S^2$, and
\[
{\cal J}^{+'} = \{\hat{\tau} = 1, \bar{\rho} > 0\},\,\,\,\,
{\cal I}^{0'} = 
\partial \bar{{\cal S}} = \{\bar{\tau} = 0, \bar{\rho} = 0\},
\]
\[
{\cal I}' = \{0 \le \bar{\tau} < 1, \bar{\rho} = 0\},\,\,\,\,
{\cal I}^{+'} = \{\bar{\tau} = 1, \bar{\rho} = 0\},\,\,\,\,
\bar{{\cal I}}' = {\cal I}' \cup {\cal I}^{+'}.
\]
While the notation alludes to related sets introduced in
section \ref{specificsofrfivpsli}, the prime should warn the reader that
the sets defined above differ in various aspects from those considered in
\ref{specificsofrfivpsli}. The range of $\bar{\rho}$ should be also 
bounded from above in these definitions. We leave this bound unspecified
because its specific value is unimportant here, we will be
concerned only with the behaviour of the metric in a neighbourhood of
$\bar{{\cal I}}'$ in 
$\bar{{\cal M}}'$. 

Important for the following are the observations:\\
(i) the function $h(s\,e^a(\psi^A))$ as given by the right hand side of
(\ref{hdef}) and considered as function of $s$ and $\psi^A$
extends as a real analytic function into a domain where $s < 0$.
This follows immediately from the values taken by $U$ and its
analyticity.\\
 (ii) similarly, the 1-form $l$ given by (\ref{dt2b}) extends as a real
analytic function into a domain where $\bar{\rho} \le 0$ and
$\bar{\tau} \ge 1$. This follows from 
\[
\left(\frac{1}{v\,\Omega}\right)_{,\psi^A}(s) = \frac{1}{s^2}\,
\frac{2\,(U(s) - s\,m)(U(s) + s\,\frac{m}{2})^2}
{(U(s) - s\,\frac{m}{2})^2}\,U_{,\psi^A},
\]
and (\ref{Ubehave}) with $s^2 = \Upsilon$.

For the following it is convenient to slightly modify the frame
(\ref{01+-frame}) and set
\begin{equation}
\label{frames}
v_0 = \partial_{\bar{\tau}},\,\,
v_1 = \bar{\rho}\,\partial_{\bar{\rho}},\,\,
v_A = \partial_{\psi^A},\,\,\,\,\,
\end{equation}
\[
\alpha^0 = d\,\bar{\tau},\,\,
\alpha^1 = \frac{1}{\bar{\rho}}\,d\,\bar{\rho},\,\,
\alpha^B = d\,\psi^B,\,\,\,\,\,A, B = 2, 3.
\]
One then gets $\bar{g} =
\bar{g}_{ik}\,\alpha^i\,\alpha^k$ with metric coefficients
\[
\bar{g}_{00} = 0,\,\,\,\,\,
\bar{g}_{01} = \frac{h(\bar{\rho}\,(1 - \bar{\tau}))}{h(\bar{\rho})},\,\,\,\,\,
\bar{g}_{0A} = \bar{\rho}\,h(\bar{\rho}\,(1 - \bar{\tau}))\,l_A,
\]

\[
\bar{g}_{11} = - (1 - \bar{\tau})\,\frac{h(\bar{\rho}\,(1 -
\bar{\tau}))}{h(\bar{\rho})}
\left(2 - (1 - \bar{\tau})\,
\frac{h(\bar{\rho}\,(1 - \bar{\tau}))}{h(\bar{\rho})}\right),
\]

\[
\bar{g}_{1A} = - (1 - \bar{\tau})\,\bar{\rho}\,\,h(\bar{\rho}\,(1 - \bar{\tau}))
\left(1 - (1 - \bar{\tau})\,\frac{h(\bar{\rho}\,(1 -
\bar{\tau}))}{h(\bar{\rho})}\right) l_A,  
\]

\[
\bar{g}_{AB} = 
\left\{\bar{\rho}\,(1 - \bar{\tau})\,\,h(\bar{\rho}\,(1 - \bar{\tau}))\right\}^2
l_A\,l_B + k_{AB}(\bar{\rho}\,(1 - \bar{\tau})).
\]

In terms of the new coordinates the metric given by (\ref{Xconfext}) extends
analytically through the set ${\cal J}^{+'}$. The latter is a null
hypersurface for the extended metric and represents future null infinity
for the space-time defined by $\tilde{g}$. 
By contrast, the right hand side of (\ref{Xconfext}) does not extend smoothly
to $\bar{{\cal I}}'$. However, the frame coefficients $\bar{g}_{ik}$
and their contravariant versions $\bar{g}^{ik}$ do extend analytically to
all of $\bar{{\cal M}}'$. It will be shown later how ${\cal I}'$ 
relates to (part of) the cylinder at space-like infinity denoted in
\ref{specificsofrfivpsli} by ${\cal I}$.

One has $\bar{g}_{ik} = g^*_{ik} + O(\bar{\rho}^2)$ with
\begin{equation}
\label{gapprox}
g^*_{ik} = \left[ \begin{array}{cccc}
0 & 1 + 2\,m\,\bar{\rho}\,\bar{\tau} & 0 & 0\\
1 + 2\,m\,\bar{\rho}\,\bar{\tau} & - (1 - \bar{\tau})\,(1 + \bar{\tau} +
4\,m\,\bar{\rho}\,\bar{\tau}^2) & 0 & 0  \\ 0 & 0 & k_{22}(0) & k_{23}(0) \\
0 & 0 & k_{32}(0) & k_{33}(0)
\end{array} \right],
\end{equation}
so that 
$\det(g^*_{ik}) < 0$ for $\bar{\rho} \ge 0$, $0 \le \bar{\tau} \le 1$ 
and $g^{ik} = g^{*ik} + O(\bar{\rho}^2)$ with
\begin{equation}
\label{congapprox}
g^{*ik} = \left[ \begin{array}{cccc}
\frac{(1 - \bar{\tau})\,(1 + \bar{\tau}
+ 4\,m\,\bar{\rho}\,\bar{\tau}^2)}{(1 + 2\,m\,\bar{\rho}\,\bar{\tau})^2} & 
\frac{1}{1 + 2\,m\,\bar{\rho}\,\bar{\tau}} & 0 & 0\\
\frac{1}{1 + 2\,m\,\bar{\rho}\,\bar{\tau}} & 0 & 0 & 0  \\ 
0 & 0 & k^{22}(0) & k^{23}(0) \\
0 & 0 & k^{32}(0) & k^{33}(0)
\end{array} \right].
\end{equation}

Since the conformal factor $\Lambda$ does not depend on $t$, the 
static Killing vector field represents a Killing field also for the metric
$\bar{g}$. In the new coordinates it takes the form
\begin{equation}
\label{2killingexpr}
K = \frac{(v\,\Omega)(\bar{\rho})}{\bar{\rho}}\,
\{
(1 - \bar{\tau})\,\partial_{\bar{\tau}}  
+ \bar{\rho}\,\partial_{\bar{\rho}}  \} =
\bar{\rho}\,h(\bar{\rho})\,\{(1 - \bar{\tau})\,v_0 + v_1\},
\end{equation}
and extends smoothly to all of $\bar{{\cal M}}'$.

Denote by $\bar{\nabla}$ the Levi-Civita connection of $\bar{g}$.
Since the commutators of the frame fields $v_k$ vanish, the connection
coefficients defined by 
$\bar{\nabla}_i v_j \equiv \bar{\nabla}_{v_i} v_j = \gamma_i\,^k\,_j\,v_k$
are given by the formula
\[
\gamma_i\,^k\,_j = \frac{1}{2}\,\bar{g}^{kl}\left(v_j(\bar{g}_{il})
+ v_i(\bar{g}_{lj}) - v_l(\bar{g}_{ij})\right).
\]
Again, {\it the connection coefficients $\gamma_i\,^k\,_j$ in the
frame $v_k$ extend analytically through $\{\bar{\rho} = 0\}$ and $\{\bar{\tau} =
1\}$}.  One finds 
\[
\gamma_i\,^k\,_j = \frac{1}{2}\,g^{*kl}\left(v_j(g^*_{il})
+ v_i(g^*_{lj}) - v_l(g^*_{ij})\right) + O(\bar{\rho}^2),
\]
which implies 
\begin{equation}
\label{Igammarel}
\gamma_i\,^k\,_j = \bar{\tau}\,\delta^k\,_0\,\left\{  
2\,\delta^0\,_{(i}\,\delta^1\,_{j)}
- (1 - \bar{\tau}^2)\,\delta^1\,_i\,\delta^1\,_j
\right\} 
- \bar{\tau}\,\delta^k\,_1\,\delta^1\,_i\,\delta^1\,_j
\,\,\,\,\,\mbox{on}\,\,\,\,\,\{\bar{\rho} = 0\}.
\end{equation} 

As a consequence of the behaviour of $\bar{g}_{ij}$ and
$\gamma_i\,^k\,_j$ the components of all tensor fields in the frame $v_k$
which are derived by standard formulas from the metric and the
connection coefficients, such as those of the Ricci tensor and the
conformal Weyl tensor of $\bar{g}$, extend analytically through
${\cal J}^{+'}$ and $\bar{{\cal I}}'$ i.e.
{\it the metric $\bar{g}$ and its connection $\bar{\nabla}$ 
imply in the frame $v_i$ a smooth frame formalism on $\bar{{\cal M}}'$.}

It follows that the coordinate expressions of these tensor fields,
such as $R_{\mu' \nu'}[\bar{g}] = R_{ik}\,\alpha^i\,_{\mu'}\,\alpha^k\,_{\nu'}$, 
and, by the argument given in \cite{penrose:scri} (cf. also
\cite{friedrich:tueb}), the rescaled conformal
Weyl tensor $W^{\mu'}\,_{\nu' \lambda' \rho'}[\bar{g}]
= \Lambda^{-1}\,C^{\mu'}\,_{\nu' \lambda' \rho'}[\bar{g}]$ 
extend smoothly to ${\cal J}^{+'}$. Unfortunately, this does not give us
the needed details about the components $R_{jk}$ and it does not tell us
anything about the behaviour of the frame components
$W^{i}\,_{jkl}[\bar{g}]$ of the rescaled conformal Weyl tensor on 
${\cal I}'$ and the critical set ${\cal I}^{+'}$. This requires detailed
calculations. Only the analyticity of $h$ near $i$ is required to
control the smoothness of the fields near ${\cal J}^{+'}$. 
This follows from the {\it ellipticity} of the conformal static field
equations near $i$. To deduce the desired behaviour near $\bar{{\cal I}}'$,
however, one will have to invoke at least, as discussed in section
\ref{regularitycondition}, the regularity condition (\ref{regcond}) with
$p_* = \infty$. The detailed form of the conformal static field equations
will thus become much more important.

\subsection{The Ricci tensor of $\bar{g}$ near $\bar{{\cal I}}'$}
\label{ricciatIbar}

The tensor
\[
L[\bar{g}]_{\rho' \nu'} = \frac{1}{2}\left(R[\bar{g}]_{\rho' \nu'}
- \frac{1}{6}\,R[\bar{g}]\,\,\bar{g}_{\rho' \nu'}\right),
\] 
is needed to integrate the conformal geodesic equations which define the
setting introduced in  section \ref{Rfivp}. The purpose of this section is
to demonstrate 
\begin{lemma}
\label{LIval}  
The frame components 
$L_{jk} = L[\bar{g}]_{\rho' \nu'}\,v^{\rho'}\,_j\,v^{\nu'}\,_k$
extend as real analytic functions to $\bar{{\cal I}}'$ with
\[
L_{0k} \rightarrow \frac{1}{2}\,\delta^1\,_k,\,\,\,\,
L_{11} \rightarrow - \frac{1 - \bar{\tau}^2}{2},\,\,\,\,
L_{1A} \rightarrow 0,\,\,\,\,
L_{AB} \rightarrow - \frac{1}{2}\,k_{AB}(0)
\,\,\,\,\mbox{as}\,\,\,\,\bar{\rho} \rightarrow 0.
\]
\end{lemma}

Proof: 
Under the rescaling
$\tilde{g} \rightarrow \bar{g} = \Lambda^2\,\tilde{g}$ the tensor
\[
L[\tilde{g}]_{\rho \nu} = \frac{1}{2}\left(R[\tilde{g}]_{\rho \nu}
- \frac{1}{6}\,R[\tilde{g}]\,\tilde{g}_{\rho \nu}\right),
\] 
transforms into
\[
L[\bar{g}]_{\rho \nu} = L[\tilde{g}]_{\rho \nu} 
- \frac{1}{\Lambda}\,\bar{\nabla}_{\rho}\,\bar{\nabla}_{\nu}\,\Lambda
+ \frac{1}{2\,\Lambda^2}\,\bar{\nabla}_{\mu}\,\Lambda
\,\bar{\nabla}^{\mu}\,\Lambda\,\,\bar{g}_{\rho\nu}.
\]
Suppose $\tilde{g} = v^2\,dt^2 + \tilde{h}$ is a static vacuum solution and
$\bar{g} = \Lambda^2\,\tilde{g} =  \Lambda^2\,(v^2\,dt^2 + \tilde{h})
= N^2\,dt^2 + h^*$ with
\[
N = \Lambda\,v,\,\,\,\,\,\,
h^* \equiv \Lambda^2\,\tilde{h} = \Lambda^2\,\Omega^{-2}\,h
= h^*_{ab}(x^c)\,d\,x^a\,d\,x^b,
\]
\[
\mu = \mu(x^a),\,\,\,\,\,\,\,
v = v(x^a),\,\,\,\,\,\,\,
\Omega = \Omega(x^a),\,\,\,\,\,\,\,
\Lambda = \Lambda(x^a).
\]
In the following the gauge (\ref{mgauge}) and coordinates satisfying
(\ref{hnormal}) will be assumed.  The connection coefficients of the
metric $\bar{g}$ in the coordinates $t$, $x^a$ are given by  
\[
\Gamma_a\,^b\,_c[\bar{g}] = \Gamma_a\,^b\,_c[h^*] \,\,\,\,\,\, \mbox{(the
Levi-Civita connection of}\,\,h^*),
\]
\[
\Gamma_t\,^a\,_t[\bar{g}] = 0,\,\,\,\,\,\,\,
\Gamma_t\,^a\,_t[\bar{g}] = - N\,h^{*ab}D_b\,N,\,\,\,\,\,\,\,
\Gamma_b\,^a\,_t[\bar{g}] = \Gamma_t\,^a\,_b[\bar{g}] =0,
\]
\[
\Gamma_b\,^t\,_c[\bar{g}] = 0,\,\,\,\,\,\,\,
\Gamma_b\,^t\,_t[\bar{g}] = \Gamma_t\,^t\,_b[\bar{g}] = \frac{1}{N}\,D_b\,N.
\]  
and $L[\bar{g}]_{\rho \nu}$ is given by 

\begin{equation}
\label{Ltt}
L[\bar{g}]_{tt} 
= - \frac{v\,\Omega^2}{\Lambda^2}\,D_a\,N\,D^a\,\Lambda
+ \frac{v^2\,\Omega^2}{2\,\Lambda^2}\,D_a\,\Lambda
\,D^a\Lambda,
\end{equation}
\[
L[\bar{g}]_{ta} = L[\bar{g}]_{at} = 0,
\]
\begin{equation}
\label{Lab}
L[\bar{g}]_{ab} 
= - \frac{1}{\Lambda}\,D^*_a\,D^*_b\,\Lambda 
+ \frac{1}{2\,\Lambda^2}\,D_c\,\Lambda\,D^c\,\Lambda\,h_{ab},
\end{equation}
where $D$ and $D^*$ denote the $h$- and $h^*$-Levi-Civita connections
respectively.  With $\Lambda = \Omega\,\Upsilon^{- 1/2}$ and the map $\Phi$
defined by (\ref{Bbasemap}) one can determine from these formulas the frame
coefficients 
\[
L_{ik} = 
\,<\Phi^* (L[\Lambda^2\,\tilde{g}]);\,v_i,\,v_k>\,=
\]
\[
<(L[\bar{g}]_{tt} \circ \Phi)\,d\,t\,d\,t +
(L[\bar{g}]_{ab} \circ \Phi)\,d\,x^a\,d\,x^b;\,v_i,\,v_k>.
\]
With equations (\ref{n1fequ}) and the relations
\begin{equation}
\label{dOsigma}
D_a\,\Omega = \Omega^{\frac{3}{2}}\,\sigma^{-\frac{3}{2}}\,D_a\,\sigma
= (1 + \sqrt{\mu\,\sigma})^{-3}\,D_a\,\sigma,
\end{equation}
\begin{equation}
\label{ddOsigma}
D_a\,D_b\,\Omega = 
\frac{1}{(1 + \sqrt{\mu\,\sigma})^3}\,\,D_a\,D_b\,\sigma
- \frac{3}{2}\,\sqrt{\frac{\mu}{\sigma}}\,\,
\frac{1}{(1 + \sqrt{\mu\,\sigma})^4}\,
D_a\,\sigma\,D_b\,\sigma, 
\end{equation}
which are implied by (\ref{vomsig}), one gets 
\[
\Upsilon\,D^a\,N\,D_a\,\Lambda =
\]
\[
v\,\Omega\left\{
\frac{2\,s\,(1 - 2\,\sqrt{\mu\,\sigma})}
{(1 - \sqrt{\mu\,\sigma})\,(1 + \sqrt{\mu\,\sigma})^4}
- \frac{(2 -
3\,\sqrt{\mu\,\sigma})\,\Upsilon^{-1}\,D^c\,\Upsilon\,D_c\,\sigma} {2\,(1 -
\sqrt{\mu\,\sigma})\,(1 + \sqrt{\mu\,\sigma})^3} - \frac{1}{U^2\,(1 +
\sqrt{\mu\,\sigma})^2}
\right\},
\]
\[
\Upsilon\,D^a\,\Lambda\,D_a\,\Lambda =
\Omega\left\{
\frac{2\,s}{(1 + \sqrt{\mu\,\sigma})^4}
- \frac{\Upsilon^{-1}\,D^c\,\Upsilon\,D_c\,\sigma}
{(1 + \sqrt{\mu\,\sigma})^3}
- \frac{1}{U^2\,(1 + \sqrt{\mu\,\sigma})^2}
\right\},
\]
\[
\Upsilon^{1/2}\,D^*_a\,D^*_b\,\Lambda 
= \frac{s\,h_{ab} - \sigma(1 - \mu\,\sigma)\,R_{ab}}
{(1 + \sqrt{\mu\,\sigma})^3}
- \frac{3}{2}\,\frac{\sqrt{\mu\,\sigma}\,\,\sigma^{-1}\,
D_a\sigma\,D_b\sigma}
{(1 + \sqrt{\mu\,\sigma})^4}
\]
\[
- \frac{D_a\,D_b\,\Upsilon}
{2\,U^2\,(1 + \sqrt{\mu\,\sigma})^2}
+ \frac{e_a\,e_b}
{U^2\,(1 + \sqrt{\mu\,\sigma})^2}
\]
\[
- h_{ab}\left(
\frac{1}{U^2\,(1 + \sqrt{\mu\,\sigma})^2}
+ \frac{\Upsilon^{-1}\,D^c\,\Upsilon\,D_c\,\sigma}
{2\,(1 + \sqrt{\mu\,\sigma})^3}
\right),
\]
which allow us to obtain the following expressions for the $L_{jk}$.

In the case of $L_{00}$ there occurs a cancellation of the
second terms in (\ref{Ltt}), (\ref{Lab}) respectively, so that 
(with the understanding that $e^a \circ \Phi = e^a(\psi^A)$)
\begin{equation}
\label{L00}
L_{00} = 
\bar{\rho}^2\,
\left(\frac{L[\bar{g}]_{tt}}{(v\,\Omega)^2} +
L[\bar{g}]_{ab}\,e^a\,e^b\right) \circ \Phi
\end{equation}
\[
= - \bar{\rho}^2\,\left(
\frac{1}{v\,\Omega^2}\left\{
\Upsilon\,D_a\,N\,D^a\,\Lambda
+ v\,\Omega\,\Upsilon^{1/2}\,D^*_a\,D^*_b\,\Lambda\,e^a\,e^b
\right\}\right) \circ \Phi,
\]
with
\[
\Upsilon\,D_a\,N\,D^a\,\Lambda
+ v\,\Omega\,\Upsilon^{1/2}\,D^*_a\,D^*_b\,\Lambda\,e^a\,e^b =
\]
\[
v\,\Omega\,\left\{
\frac{(1 - 4\,\sqrt{\mu\,\sigma} + \mu\,\sigma)\,(s + 2)}
{(1 - \sqrt{\mu\,\sigma})\,(1 + \sqrt{\mu\,\sigma})^4}
- \frac{\sigma(1 - \mu\,\sigma)\,R_{ab}\,e^a\,e^b}
{(1 + \sqrt{\mu\,\sigma})^3}
\right.
\]
\[
\left.
- \frac{6\,\sqrt{\mu\,\sigma}}
{(1 + \sqrt{\mu\,\sigma})^4}\left[(\frac{1}{U} 
+ \frac{D^a\,\Upsilon\,D_a\,U}{2\,U^2})^2 - 1\right]
\right.
\]
\[
\left.
+
\frac{(1 - 2\,\sqrt{\mu\,\sigma})}
{(1 - \sqrt{\mu\,\sigma})\,(1 + \sqrt{\mu\,\sigma})^3}
\left[
\frac{2 - 2\,U^2}{U^2} + \frac{D^a\,\Upsilon\,D_a\,U}{U^3}
\right]
\right\}.
\]
Since the term in curly brackets is of the order $O(\Upsilon)$, the
function $L_{00}$ extends smoothly to $\{\bar{\rho} = 0\}$ with
$L_{00} \rightarrow 0$ as $\bar{\rho} \rightarrow 0$.

It holds
\[
L_{01} = 
\frac{\bar{\rho}^2}{(v\,\Omega)(\bar{\rho})}\,
\left(\frac{L[\bar{g}]_{tt}}{v\,\Omega}\right) \circ \Phi - (1 -
\bar{\tau})\,L_{00},
\]
with
\[
\frac{L[\bar{g}]_{tt}}{v\,\Omega} =
\Omega^{-1}\,\Upsilon\,\,(\frac{v}{2}\,D_a\,\Lambda\,D^a\Lambda
- D_a\,N\,D^a\,\Lambda) =
\]
\[
v\,\left\{
\frac{1}{2\,U^2\,(1 + \sqrt{\mu\,\sigma})^2}
- \frac{(1 - 3\,\sqrt{\mu\,\sigma})\,s}
{(1 - \sqrt{\mu\,\sigma})\,(1 + \sqrt{\mu\,\sigma})^4}
\right.
\]
\[
\left.
- \frac{1 - 2\,\sqrt{\mu\,\sigma}}
{(1 - \sqrt{\mu\,\sigma})\,(1 + \sqrt{\mu\,\sigma})^3}
\left[\frac{2}{U^2} + \frac{D^a\,\Upsilon\,D_a\,U}{U^3}\right]
\right\},
\]
so that $L_{01}$ extends smoothly to $\{\bar{\rho} = 0\}$ with
$L_{01} \rightarrow \frac{1}{2}$ as $\bar{\rho} \rightarrow 0$.

\[
L_{0A} = \bar{\rho}\,
\left(\frac{L[\bar{g}]_{tt}}{v\,\Omega}\right) \circ \Phi\,\,l_A - 
\bar{\rho}^2\,(1 - \bar{\tau})\,(L[\bar{g}]_{ab} \circ \Phi)\,e^a\,e^b_{,\psi^A},
\]
with
\[
L[\bar{g}]_{ab}\,e^a\,e^b_{,\psi^A} =
\]
\[ 
(1 - \sqrt{\mu\,\sigma})\,R_{ab}\,e^a\,e^b_{,\psi^A}
+ \frac{\mu\,U_{'\psi^A}}{\sqrt{\mu\,\Upsilon\,\sigma}}
\frac{1}{(1 + \sqrt{\mu\,\sigma})^2}\left[
\frac{1}{U^4} + 2\,\frac{\Upsilon\,D^a\Upsilon\,D_a\,U}{U^6}\right]
\]
so that $L_{0A}$ extends smoothly to $\{\bar{\rho} = 0\}$ with
$L_{0A} \rightarrow 0$ as $\bar{\rho} \rightarrow 0$.
\[
L_{11} = \left(
\frac{\bar{\rho}^2\,(v\,\Omega)(\bar{\rho}
(1 - \bar{\tau}))}{(v\,\Omega)^2(\bar{\rho})}
- 2\,\frac{\bar{\rho}^2\,(1 - \bar{\tau})}{(v\,\Omega)(\bar{\rho})}\,
\right)\left(\frac{L[\bar{g}]_{tt}}{v\,\Omega}\right) \circ \Phi
+ (1 - \bar{\tau})^2\,L_{00},
\]
extends smoothly to $\{\bar{\rho} = 0\}$ with
$L_{11} \rightarrow - \frac{1 - \bar{\tau}^2}{2}$ as 
$\bar{\rho} \rightarrow 0$.
\[
L_{1A} = 
\]
\[
\left(\frac{\bar{\rho}}{(v\,\Omega)(\bar{\rho})}
- \frac{\bar{\rho}\,(1 - \bar{\tau})}{(v\,\Omega)
(\bar{\rho}(1 - \bar{\tau}))}\right)
L[\bar{g}]_{tt} \circ \Phi\,\,l_A - 
\bar{\rho}^2\,(1 - \bar{\tau})^2\,
(L[\bar{g}]_{ab} \circ \Phi)\,e^a\,e^b_{,\psi^A},
\]
extends smoothly to $\{\bar{\rho} = 0\}$ with
$L_{1A} \rightarrow 0$ as $\bar{\rho} \rightarrow 0$.
\[
L_{AB} = 
L[\bar{g}]_{tt} \circ \Phi\,\,l_A\,l_B - 
(\Gamma\,L[\bar{g}]_{ab}) \circ \Phi\,\,
e^a_{,\psi^A}\,e^b_{,\psi^B}.
\]
extends smoothly to $\{\bar{\rho} = 0\}$ with
$L_{AB} \rightarrow - \frac{1}{2}\,k_{AB}(0)$ 
as $\bar{\rho} \rightarrow 0$.


\subsection{The rescaled conformal Weyl tensor of $\bar{g}$ near 
$\bar{{\cal I}}'$}
\label{rescwaylatibar}

In this section we shall make a few general observations concerning the
rescaled conformal Weyl tensor and then specialize to the conformal static
case. After a remark about the radiation field on ${\cal J}^{+'}$ we will
analyse the smoothness of the rescaled conformal Weyl tensor near the set
$\bar{{\cal I}}'$.

Let $\tilde{g}$ be a Lorentz metric and $\tilde{{\cal S}}$ a space-like
hypersurface with unit normal $\tilde{n}$ and induced metric
$\tilde{h}_{\mu \nu} = \tilde{g}_{\mu \nu} -
\tilde{n}_{\mu}\,\tilde{n}_{\nu}$. We set 
$\tilde{p}_{\mu \nu} = \tilde{h}_{\mu \nu} -
\tilde{n}_{\mu}\,\tilde{n}_{\nu}$,
$\tilde{\epsilon}_{\nu  \lambda \rho} =
\tilde{n}^{\mu}\,\tilde{\epsilon}_{\mu \nu  \lambda \rho}$, 
and denote by $\tilde{c}_{\nu \rho} =  
C_{\mu \nu \lambda \rho}[\tilde{g}]\,\tilde{n}^{\mu} \tilde{n}^{\lambda}$
and $\tilde{c}^*_{\nu \rho} =  
C^*_{\mu \nu \lambda \rho}[\tilde{g}]\tilde{n}^{\mu}\,\tilde{n}^{\lambda}$
(the star on  the right hand side indicating the dual) the
$\tilde{n}$-electric and the $\tilde{n}$-magnetic part of the conformal Weyl
tensor respectively. The latter are symmetric, trace-free, and spatial, i.e.
$\tilde{n}^{\nu}\,\tilde{c}_{\nu \rho} = 0$, 
$\tilde{n}^{\nu}\,\tilde{c}^*_{\nu \rho} = 0$.
The conformal Weyl tensor of $\tilde{g}$ is then given in terms of its
electric and the magnetic part by (cf. \cite{friedrich:nagy})
\begin{equation}
\label{weyldecomp}
C_{\mu \nu \lambda \rho}[\tilde{g}] = 2 \left(
\tilde{p}_{\nu[\lambda}\,\tilde{c}_{\rho]\mu} 
- \tilde{p}_{\mu[\lambda}\,\tilde{c}_{\rho]\nu} 
- \tilde{n}_{[\lambda}\,\tilde{c}^*_{\rho] \delta}\,\epsilon^{\delta}\,_{\mu
\nu}  
- \tilde{n}_{[\mu}\,\tilde{c}^*_{\nu] \delta}\,\epsilon^{\delta}\,_{\lambda
\rho} 
\right).
\end{equation} 

Suppose that $\tilde{g}$ is a solution to the vacuum field equations. Then
the first and second fundamental form $\tilde{h}_{ab}$ and
$\tilde{\chi}_{ab}$ induced by $\tilde{g}$ on $\tilde{{\cal S}}$ satisfy
the Gauss and the Codazzi equation (expressing the pull-back of spatial
tensors to $\tilde{{\cal S}}$ in terms of spatial coordinates $x^a$)
\begin{equation}
\label{Bbzw9}
r_{ab}[\tilde{h}] = - \tilde{c}_{ab} 
+ \tilde{\chi}_c\,^c\,\tilde{\chi}_{ab} 
- \tilde{\chi}_{ca}\,\tilde{\chi}_b\,^c, 
\end{equation}
\begin{equation}
\label{Bbzw8}
\tilde{D}_b\,\tilde{\chi}_{d(a}\,\tilde{\epsilon}_{c)}\,^{bd} 
= - \tilde{c}^*_{ac}.
\end{equation}
This allows us to express the conformal Weyl tensor in terms of
$\tilde{h}_{ab}$ and $\tilde{\chi}_{ab}$.
If $\tilde{{\cal S}}$ is a hypersurface of time reflection symmetry, so
that $\tilde{\chi}_{ab} = 0$, these equations imply 
\begin{equation}
\label{trybzw9}
r_{ab}[\tilde{h}] = - \tilde{c}_{ab},\,\,\,\,\,\,\,\,\, 
\tilde{c}^*_{ac} = 0,
\end{equation}
and the Weyl tensor assumes the form
\begin{equation}
\label{Bweyldecomp}
C_{\mu \nu \lambda \rho}[\tilde{g}] = 2 \left(
\tilde{p}_{\nu[\lambda}\,\tilde{c}_{\rho]\mu} 
- \tilde{p}_{\mu[\lambda}\,\tilde{c}_{\rho]\nu}\right)
\equiv - (\tilde{p} \oslash \tilde{c})_{\mu \nu \lambda \rho},
\end{equation} 
where $\oslash$ denotes the bi-linear Kulkarni-Nomizu product of 
two symmetric 2-tensors (cf. \cite{besse}).

If $\Lambda$ is an arbitrary conformal factor, the rescaled conformal Weyl
tensor of $\bar{g} = \Lambda^2\,\tilde{g}$ is given by
$W^{\mu}\,_{\nu \lambda\rho}[\bar{g}] =
\Lambda^{-1}\,C^{\mu}\,_{\nu \lambda \rho}[\bar{g}]$.
In view of the behaviour of the conformal Weyl
tensor under conformal rescalings, one gets (observe the index positions)
\begin{equation}
\label{dc}
W_{\mu \nu \lambda \rho}[\bar{g}] =
\Lambda\,C_{\mu \nu \lambda \rho}[\tilde{g}].
\end{equation}
Its electric part with respect to the $\bar{g}$-unit 
vector $\Lambda^{-1}\,\tilde{n}$ is then given by
\begin{equation}
\label{dtildec}
w_{\mu \nu}[\bar{g}] =
\Lambda^{-1}\tilde{c}_{\mu \nu}[\tilde{g}]
\end{equation}

With $h = \Omega^2\,\tilde{h}$, the gauge (\ref{mgauge}), the general
transformation law
\[
r_{ab}[\tilde{h}] = r_{ab}[h] + \Omega^{-1}\,D_aD_b\Omega 
+ h_{ab}\,(\Omega^{-1}\,D_cD^c\Omega 
- 2\,\Omega^{-2}\,D_c\Omega\,D^c\Omega),
\]
and the equation $2\,\Omega\,\Delta_h\,\Omega = 3\,D_a\Omega\,D^a\Omega$,
one gets from (\ref{trybzw9}) and (\ref{dtildec}) in the general time
reflection symmetric case
\begin{equation}
\label{elrcW}
w_{ab}[\bar{g}] =
- (\Lambda\,\Omega)^{-1}(D_a\,D_b\,\Omega 
- \frac{1}{3}\,h_{ab}\,D_c\,D^c\,\Omega
+ \Omega\,r_{ab}[h])
\,\,\,\,\,\,\mbox{on}\,\,\,\,
{\cal S} = \tilde{{\cal S}} \cup \{i\}. 
\end{equation}

A conformal scaling which represents space-like infinity (with respect to
the initial hypersurface $\tilde{{\cal S}}$ and with respect to the solution
space-time) by a point is achieved by choosing $\Lambda = \Omega$ on 
${\cal S}$. With this particular choice one has
\begin{equation}
\label{winpointgauge}
w_{ab}[\bar{g}] = - \Omega^{-2}(D_a\,D_b\,\Omega 
- \frac{1}{3}\,h_{ab}\,D_c\,D^c\,\Omega + \Omega\,\,r_{ab}[h])
\end{equation}
\[
= O(\Upsilon^{- 3/2}) \,\,\,\,\,\mbox{as}\,\,\,\,\, 
\Upsilon \rightarrow 0\,\,\,\,\,\mbox{unless}\,\,\,\,\,m = 0.
\]
We note that in the massless case the precise behaviour depends on the
freely prescribed metric $h$ on ${\cal S}$ near $i$. In the massless case
one has $\Omega = \sigma$ and the comparison of the expression for
$w_{ab}[\bar{g}]$ with (\ref{n1fequ}) shows that in the case where $h$ 
represents conformally static vacuum data one has
\begin{equation}
\label{statraddat}
w_{ab}[\bar{g}] = - \mu\,r_{ab}[h],
\end{equation}
i.e. the rescaled conformal Weyl tensor is smooth.

We return to the case where $\Lambda = \Omega\,\Upsilon^{-
1/2}$.  With (\ref{dOsigma}), (\ref{ddOsigma}) we get then
\begin{equation}
\label{BelrcW}
w_{ab}[\bar{g}] =
- \frac{\sqrt{\Gamma}}{\sigma^2}\,\left\{
(1 + \sqrt{\mu \,\sigma})\,(\,D_a\,D_b\,\sigma 
- \frac{1}{3}\,\Delta_h\,\sigma\,h_{ab})
\right.
\end{equation}
\[
\left.
- \frac{1}{2}\,\sqrt{\frac{\mu}{\sigma}}\,
(3\,D_a\,\sigma\,D_b\,\sigma - D_c\,\sigma\,D^c\,\sigma\,h_{ab})
+ \sigma\,(1 + \sqrt{\mu \,\sigma})^2\,r_{ab}\right\}
\]
\[
=
\frac{m}{4}\,\frac{U}{\sigma^2}\,
(3\,D_a\,\sigma\,D_b\,\sigma - D_c\,\sigma\,D^c\,\sigma\,h_{ab})
- \frac{m}{2}\,U\,(1 + \sqrt{\mu \,\sigma})^2
\,r_{ab}[h]
\]
\[
- \frac{\sqrt{\Upsilon}}{\sigma^2}\,(1 + \sqrt{\mu \,\sigma})\,\Sigma_{ab},
\]
where we use $\Sigma_{ab}$ as defined by the right hand side of (\ref{n1fequ})
without assuming $h$ to be conformally static. If $h$ is conformally static
the electric part of the rescaled conformal Weyl tensor on
${\cal S}$ is given by
the right hand side of (\ref{BelrcW}) with $\Sigma_{ab} = 0$. In the
present conformal gauge, defined by (\ref{mgauge}), one has
\[
\Sigma_{ab} = O(\Upsilon^{3/2})\,\,\,\,\mbox{near}\,\,\,\,i,
\]
for any time reflection symmetric initial data $h$.

If the solution is static and written again in the form 
$\tilde{g} = v^2\,d\,t^2 + \tilde{h}$, then equations (\ref{weyldecomp}),
(\ref{trybzw9}) hold with $\tilde{n} = \frac{1}{v}\,\partial_t$ and
$t$-independent fields for each slice 
$\tilde{{\cal S}} = \{t = t_*\}$ with 
$t_* = const$. The relations above then imply for all $(t, x^a)$
\begin{equation}
\label{statD}
W_{\mu \nu \lambda \rho}[\bar{g}] = 
- \Upsilon^{-1}(p \oslash w)_{\mu \nu \lambda \rho}
\end{equation}
with
\[
p_{\mu \nu} = h_{\mu \nu} - n_{\mu}\,n_{\nu},\,\,\,\,\,\,\,
n_{\mu} = \Omega\,\tilde{n}_{\mu}.
\]

With $\otimes$ denoting the tensor product, we write for arbitrary 1-forms
$a$,
$c$
\[
a \otimes_s c = a \otimes c + c \otimes a,\,\,\,\,\,\,
a^2 = a \otimes a,
\]
and note that the Kulkarni-Nomizu product
is symmetric, i.e.
\begin{equation}
\label{NKsymm}
m \oslash n = n \oslash m,
\end{equation}
for symmetric 2-tensors $m$, $n$,
and satisfies for  arbitrary 1-forms $a$, $c$, $e$ 
\begin{equation}
\label{NKnil}
(a \otimes a) \oslash (a \otimes_s c) = 0,\,\,\,\,\,\,\,
(a \otimes_s e) \oslash (a \otimes_s c) 
= - (a \otimes a) \oslash (c \otimes_s e).
\end{equation}

\vspace{.3cm}

We show how it follows in the present setting that the {\it radiation
field} vanishes on ${\cal J}^{+'}$. Since the extended Killing vector field
$K$ is tangent to the null generators of ${\cal J}^{+'}$ without vanishing
there, the complete information on the radiation field is contained in the
field
\[
K^{\nu}\,K^{\rho}\,W_{\mu \nu \lambda \rho}[\bar{g}]\,
d\,x^{\mu}\,d\,x^{\lambda}
= - \Upsilon^{-1}\,K^{\nu}\,K^{\rho}\,(p \oslash w)_{\mu \nu \lambda \rho}
\,d\,x^{\mu}\,d\,x^{\lambda}
\]
\[
= - \Upsilon^{-1}\,K^{\nu}\,K^{\rho}\,p_{\nu\rho}\,w_{ab}\,d\,x^a\,d\,x^b
= - \Upsilon^{-1}\,(v\,\Omega)^2\,w_{ab}\,d\,x^a\,d\,x^b
\]
\[
=
\frac{m}{4}\,\frac{(1 - \sqrt{\mu\,\sigma})^2}{(1 +
\sqrt{\mu\,\sigma})^6}\,\left\{
\frac{1}{U}\,
(2\,s\,h_{ab}
- 3\,\sigma^{-1}\,D_a\,\sigma\,D_b\,\sigma)\,d\,x^a\,d\,x^b
\right.
\]
\[
\left.
+ \frac{2\,\sigma}{U}
(1 + \sqrt{\mu \,\sigma})^2
\,r_{ab}[h]\,d\,x^a\,d\,x^b\right\}
\]
with $s$ as given in (\ref{an2fequ}). Because of the relation 
\begin{equation}
\label{ddsigma}
\sigma^{-1}\,D_a\,\sigma\,D_b\,\sigma = 4\,U^{-4}\,(U^2\,e_a\,e_b
- U\,\Upsilon^{1/2}\,(e_a\,D_b\,U + e_b\,D_a\,U)
+ \Upsilon\,D_a\,U\,D_b\,U),
\end{equation}
and the factor $\sigma$ in the second term it
follows that
\begin{equation}
\label{radfieldcomp}
K^{\nu}\,K^{\rho}\,W_{\mu \nu \lambda \rho}[\bar{g}]\,d\,x^{\mu}\,d\,x^{\lambda}
\rightarrow - 2\,m\,\bar{\rho}^2\,d\,\bar{\tau}^2
\,\,\,\,\,\,\,\mbox{as}\,\,\,\,
\bar{\tau} \rightarrow 1,\,\,\,\bar{\rho} > 0.
\end{equation}
Thus, the pull-back of 
$K^{\nu}\,K^{\rho}\,W_{\mu \nu \lambda
\rho}[\bar{g}]\,d\,x^{\mu}\,d\,x^{\lambda}$ to ${\cal J}^+$, which provides the
radiation field up to a scaling, vanishes everywhere on ${\cal J}^+$.

\vspace{.3cm}

\begin{lemma}
\label{Irescweyl}
The components $W_{ijkl}[\bar{g}] 
= \Lambda^{-1}\,C_{\mu' \nu' \lambda' \rho'}[\bar{g}]
\,v^{\mu'}\,_i\,v^{\nu'}\,_j\,v^{\lambda'}\,_k\,v^{\rho'}\,_l$
of the rescaled conformal Weyl tensor of $\bar{g}$ in the frame $v_k$ extend
as analytic functions to $\bar{{\cal I}}'$.
\end{lemma}

Proof: In the coordinates $x^{\mu'}$ given by (\ref{Bbasemap}) the rescaled
conformal Weyl tensor is obtained as the product of 
\[
- (\Upsilon \circ \Phi)^{-1} = - (\bar{\rho}\,(1 - \bar{\tau}))^{-2},
\]
with the Nomizu-Kulkarni product of 
\[
p' = (h_{ab} \circ \Phi)\,d\,x^a\,d\,x^b 
- ((v\,\Omega) \circ \Phi)^2 \,d\,t^2
= p'_1 + p'_2 +  p'_3 + p'_4,
\]
and $w' = w'_1 + w'_2 + w'_3 + w'_4 + w'_5$,
where
\[
p'_1 = - 2\,((1 - \hat{\tau})\,d\,\bar{\rho} - \bar{\rho}\,d\,\bar{\tau})^2,
\]
\[
p'_2 = ((1 - \bar{\tau})\,d\,\bar{\rho} - \bar{\rho}\,d\,\bar{\tau}) \otimes_s
\left(
\frac{(v\,\Omega)(\bar{\rho}\,(1 - \bar{\tau}))}
{(v\,\Omega)(\bar{\rho})}\,d\,\bar{\rho} +
(v\,\Omega)(\bar{\rho}\,(1 - \bar{\tau}))\,l \right)
\]
\[
p'_3 = - \left(
\frac{(v\,\Omega)(\bar{\rho}\,(1 - \bar{\tau}))}
{(v\,\Omega)(\bar{\rho})}\,d\,\bar{\rho} + 
(v\,\Omega)(\bar{\rho}\,(1 - \bar{\tau}))\,l \right)^2,
\]
\[
p'_4 = \bar{\rho}^2\,(1 - \bar{\tau})^2\,k,
\]
\[
w'_5 = - \left(\left\{\frac{m}{2}\,U\,(1 + \sqrt{\mu \,\sigma})^2
\,r_{ab}[h]\right\} \circ \Phi \right)d\,x^a\,d\,x^b,
\]
and
\[
w'_1 + w'_2 + w'_3 + w_4'= - \left(\left\{\frac{m}{4}\,\frac{U}{\sigma}\,
(2\,s\,h_{ab}
- 3\,\sigma^{-1}\,D_a\,\sigma\,D_b\,\sigma)
\right\} \circ \Phi \right)d\,x^a\,d\,x^b,
\]
with
\[
w'_1 = - \frac{m}{4}\left(\left\{\frac{U}{\sigma}\,
(s\, + 6\,U^{-2})\right\} \circ \Phi \right)p'_1,
\]
\[
w'_2 = - \left(\left\{\frac{m}{2}\,U^3\,s\,
\right\} \circ \Phi \right)k,
\]
\[
w'_3 = - \left(\left\{\frac{3\,m}{U\,\sigma}\,\Upsilon^{1/2}\,
\right\} \circ \Phi \right)
((1 - \bar{\tau})\,d\,\bar{\rho} - \bar{\rho}\,d\,\bar{\tau}) \otimes_s j,
\]
\[
w_4'= - \left(\left\{\frac{m}{U}
\right\} \circ \Phi \right) j^2.
\]
We used above the relation (\ref{ddsigma}) and set
\[
j = (D_a\,U \circ \Phi)\,d\,x^a.
\]

The desired result on the behaviour of the rescaled conformal Weyl tensor
near $\bar{{\cal I}}'$ is obtained now by showing that for arbitrary frame
vector fields
$v_n$ one has
\[
<p' \oslash w';\,v_i,\,v_j,\,v_k,\,v_l,>\,
= O(\bar{\rho}^2\,(1 - \bar{\tau})^2).
\]
From (\ref{NKsymm}) it follows that
\[
p'_1 \oslash w'_1 = p'_1 \oslash w'_3 = p'_2 \oslash w'_1 = 0.
\]
Observing (\ref{Ubehave}) one finds by inspection
\[
< p'_1;\,v_i,\,v_j>\,= O(\bar{\rho}^2),\,\,\,\,\,\,
< p'_M;\,v_i,\,v_j>\,= O(\bar{\rho}^2\,(1 - \bar{\tau})^2),\,\,\,\,\,\,
\mbox{for}\,\,\,M = 2, 3, 4, 
\]
\[
< w'_4;\,v_i,\,v_j>\,= O(\bar{\rho}^2\,(1 - \bar{\tau})^2),\,\,\,\,\,\,
< w'_N;\,v_i,\,v_j>\,= O(1),\,\,\,\,\,\,
\mbox{for}\,\,\,N = 2, 3, 5, 
\]
and thus
\[
<p'_M \oslash w'_N;\,v_i,\,v_j,\,v_k,\,v_l>\, 
= O(\bar{\rho}^2\,(1 - \bar{\tau})^2)\,\,\,\,\,\,
\mbox{for}\,\,\,M = 2, 3, 4,\,\,\,\,\,N = 2, 3, 4, 5,
\]
\[
< p'_1 \oslash w'_4;\,v_i,\,v_j,\,v_k,\,v_l>\, 
= O(\bar{\rho}^2\,(1 - \bar{\tau})^2).
\]
The remaining term is given by
\[
p'_1 \oslash (w'_2 + w'_5) + (p'_3 + p'_4) \oslash w'_1 
= p'_1 \oslash m
\]
with 
\[
m = w'_2 + w'_5 - \frac{m}{4}\left(\left\{\frac{U}{\sigma}\,
(s\, + 6\,U^{-2})\right\} \circ \Phi \right)(p'_3 + p'_4) 
\]
\[
= - \frac{m}{4}\left(\left\{U\,
(3\,s\,U^2 + 6)\right\} \circ \Phi \right)k
- \left(\left\{\frac{m}{2}\,U\,(1 + \sqrt{\mu \,\sigma})^2
\,r_{ab}[h]\right\} \circ \Phi \right)d\,x^a\,d\,x^b
\]
\[
+ \frac{m}{4}\left(\left\{
s\,U^3 + 6\,U\right\} \circ \Phi \right)\,
\left(
\frac{(v\,\Omega)(\bar{\rho}\,(1 - \bar{\tau}))}
{\bar{\rho}\,(1 - \bar{\tau})\,(v\,\Omega)(\bar{\rho})}\,d\,\bar{\rho} + 
\frac{(v\,\Omega)(\bar{\rho}\,
(1 - \bar{\tau}))}{\bar{\rho}\,(1 - \bar{\tau})}\,l
\right)^2
\]
For the three summands to be considered here we get the following. From
$3\,s\,U^2 + 6 = O(\Upsilon)$ it follows that 
\[
< p'_1 \oslash \left(\left\{U\,
(3\,s\,U^2 + 6)\right\} \circ \Phi \right)k;\,v_i,\,v_j,\,v_k,\,v_l>\, 
= O(\bar{\rho}^2\,(1 - \bar{\tau})^2). 
\] 
Because of
\[
d\,x^{(a}\,d\,x^{b)} = 
\bar{\rho}^2\,(1 - \bar{\tau})^2\,d\,e^{(a} \otimes d\,e^{b)}
\]
\[
- \frac{1}{2}\,p'_1\,e^a\,e^b
+ \bar{\rho}\,(1 - \bar{\tau})\,e^{(a}\,\,d\,e^{\,b)} \otimes_s
((1 - \hat{\tau})\,d\,\rho - \rho\,d\,\hat{\tau}),
\]
it follows by (\ref{NKnil}) that
\[
< p'_1 \oslash 
\left(\left\{\frac{m}{2}\,U\,(1 + \sqrt{\mu \,\sigma})^2
\,r_{ab}[h]\right\} \circ \Phi \right)d\,x^a\,d\,x^b
;\,v_i,\,v_j,\,v_k,\,v_l>\, = 
\]
\[
O(\bar{\rho}^2\,(1 - \bar{\tau})^2). 
\] 
It holds that $s\,U^3 + 6\,U = O(1)$ and by inspection it follows 
that
\[
< p'_1 \oslash 
\left(
\frac{(v\,\Omega)(\bar{\rho}\,(1 - \bar{\tau}))}
{\bar{\rho}\,(1 - \bar{\tau})\,(v\,\Omega)(\bar{\rho})}\,d\,\bar{\rho} + 
\frac{(v\,\Omega)(\bar{\rho}\,(1 - \bar{\tau}))}{\bar{\rho}\,(1 - \bar{\tau})}\,l
\right)^2
;\,v_i,\,v_j,\,v_k,\,v_l>\, 
\]
$= O(\bar{\rho}^2\,(1 - \bar{\tau})^2)$. 


\section{Static vacuum solutions near the cylinder at
space-like infinity}
\label{stavacatI}

The conformal extension considered in the previous section relies on
specific features of static fields. We use it to show that the 
construction of the cylinder at space-like infinity in section \ref{Rfivp},
which is based on general concepts and applies to general solutions,
is for static vacuum solutions as smooth as can be expected.
\begin{theorem}
\label{staticIsmooth}
For static vacuum solutions which are asymptotically flat the construction
of section \ref{Rfivp} is analytic in the sense that in the frame
(\ref{01+-frame}) all conformal fields, including the rescaled
conformal Weyl tensor, extend to analytic fields on some neighbourhood
${\cal O}$ of $\bar{{\cal I}}$ in $\bar{{\cal N}}$. This statement does not
depend on a particular choice of (analytic) scaling of the (analytic) free
datum
$h$ on ${\cal S}$.
\end{theorem}
This result will be obtained as a consequence of Lemmas
\ref{Icgsol}, \ref{cggauge smooth}, and \ref{confgaugechange} below. 

The construction of section \ref{Rfivp} will be discussed here for static
solutions in terms of the initial data $h$ and $\Omega$ in the gauge
given by (\ref{mgauge}), and the field $\bar{g}$ given on
$\bar{{\cal M}}'$ in the coordinates defined by (\ref{Bbasemap}).  
The effect of a rescaling of $h$ will be dicussed seperately because it is
of interest in itself.

The conformal factor $\Theta $ is assumed in the form (\ref{ivpconfac}),
(\ref{ivpThetaexpl}) with
\begin{equation}
\label{omegadef}
\kappa = \omega = 2\,\Omega\,|D_{a}\Omega D^{a}\Omega|^{-\frac{1}{2}}
= 2\,\Omega\,(1 + \sqrt{\mu\,\sigma})^{-3}\,\sqrt{2\,|s|\,\sigma}. 
\end{equation}
It follows that $\omega = \Upsilon^{1/2} + O(\Upsilon)$ and  
$\Theta = 1/2\,\,|D_{a}\Omega D^{a}\Omega|^{\frac{1}{2}}
= \Upsilon^{1/2} + O(\Upsilon)$ so that 
\begin{equation}
\label{quotlim} 
\mbox{on}\,\,\,
\bar{{\cal S}}:\,\,\,\,\,\,\,
\lim_{\bar{\rho}\to 0} \Upsilon^{1/2}\,\omega^{-1} =
\lim_{\bar{\rho}\to 0} \omega\,\Lambda^{-1} =
\lim_{\bar{\rho}\to 0} \Theta\,\Lambda^{-1} = 1.
\end{equation}
The metric $\bar{h}$ induced by $g = \Theta^2\,\tilde{g}$ on
$\tilde{{\cal S}}$ is given by $\omega^{-2}\,h$. 

The main ingredient of the gauge for the evolution equations
used in section \ref{evolutiongauge} are the conformal geodesics
generating the conformal Gauss system described in section
\ref{genconffieldequ}. We shall try to control their evolution
on $\bar{{\cal M}}'$ near $\bar{{\cal I}}'$.
Following the prescription in section \ref{genconffieldequ}, 
we assume that the
tangent vectors $\dot{x} = d\,x/d\,\tau$ of the conformal geodesics with
parameter $\tau$ satisfy 
\begin{equation}
\label{xdotonS}
\dot{x} \perp \tilde{{\cal S}},\,\,\,\,\,\,
\Theta^2\,\tilde{g}(\dot{x}, \dot{x}) = 1
\,\,\,\,\,\,\mbox{on}\,\,\,\,\,\,
\tilde{{\cal S}}.
\end{equation}
With the frame (\ref{frames}) and the coordinates (\ref{Bbasemap})
this translates into the initial condition
\begin{equation}
\label{dxindata}
\dot{x} =
\frac{\omega(\bar{\rho})}{\bar{\rho}}\,\partial_{\bar{\tau}}
+ \omega(\bar{\rho})\,\partial_{\bar{\rho}}  
= \frac{\omega(\bar{\rho})}{\bar{\rho}}\,(v_0 + v_1) \equiv X^i\,v_i
\,\,\,\,\mbox{at}\,\,\,\,\bar{\tau} = 0,
\end{equation}
with
\[
X^i = \delta^i\,_0 + \delta^i\,_1 + O(\bar{\rho})
\,\,\,\mbox{as}\,\,\bar{\rho} \rightarrow 0.
\]
For the following we need to observe besides 
$\bar{g} = \Lambda^2\,\tilde{g}$  the relations
\[
g = \Pi^2\,\bar{g} = \Theta^2\,\tilde{g},\,\,\,\,\,\mbox{with}
\,\,\,\,\,
\Pi \equiv \Lambda^{-1}\,\Theta.
\]
For the connections $\tilde{\nabla}$, $\nabla$, and $\bar{\nabla}$
of $\tilde{g}$, $g$, and $\bar{g}$ respectively we have relations
\[
\hat{\nabla} = \tilde{\nabla} + S(\tilde{f}),\,\,\,\,\,\,\,\,\,\,
\hat{\nabla} = \nabla + S(f),\,\,\,\,\,\,\,\,\,\,
\hat{\nabla} = \bar{\nabla} + S(\bar{f}),
\]
\[
\nabla = \tilde{\nabla} + S(\Theta^{-1}\,d\,\Theta),
,\,\,\,\,\,\,\,\,\,\,
\bar{\nabla} = \tilde{\nabla} + S(\Lambda^{-1}\,d\,\Lambda).
\]
The comparison gives
$f = \tilde{f} - \Theta^{-1}\,d\,\Theta$
and
$\bar{f} = \tilde{f} - \Lambda^{-1}\,d\,\Lambda$ 
which imply the relation
\begin{equation}
\label{formcomp}
\bar{f} = f + \Theta^{-1}\,d\,\Theta - \Lambda^{-1}\,d\,\Lambda
= f + \Pi^{-1}\,d\,\Pi,
\end{equation}
between the 1-form $\bar{f}$ which is obtained if the conformal geodesic
equations are written in terms of the metric $\bar{g}$,  
the 1-form $f$ which is supplied by the conformal geodesic
equations written in terms of the metric $g$, and the
conformal factor which relates $\bar{g}$ to $g$.

By the choices of section \ref{evolutiongauge} we have 
$<f, \dot{x}>\, = 0$ everywhere on the space-time and $<d\,\Theta,
\dot{x}>\, = 0$ on $\tilde{{\cal S}}$. 
Since $\Lambda$ has been chosen to be independent of $t$ and 
$\partial_t$ is orthogonal $\tilde{{\cal S}}$, it follows that 
$<d\,\Lambda, \dot{x}>\, = 0$ and thus
$<\bar{f}, \dot{x}>\, = 0$ on $\tilde{S}$. 
Observing the pull back of $f$ to $\tilde{{\cal S}}$ given by
(\ref{ivpctildef}) and the relation
\begin{equation}
\label{Pidata}
\Pi = \Upsilon^{1/2}\,\omega^{-1}
\,\,\,\,\,\mbox{on}\,\,\,\,\tilde{{\cal S}},
\end{equation}
we find that the pull back of $\bar{f}$ to
$\tilde{{\cal S}}$ is given by $1/2\,\,\Upsilon^{-1}\,d\,\Upsilon$. 
From this one gets in the frame (\ref{frames})
and the coordinates (\ref{Bbasemap}) with $\bar{\tau} = 0$
\begin{equation}
\label{vfdata}
\bar{f} = (1/2\,\,\Upsilon^{-1}\,D_a\,\Upsilon) \circ \Phi\,\,d\,x^a
= \bar{f}_i\,\alpha^i
\,\,\,\,\,\mbox{with}\,\,\,\,\bar{f}_i = - \delta^0\,_i + \delta^1\,_i
\,\,\,\,\,\mbox{on}\,\,\,\,\tilde{{\cal S}}.
\end{equation}

The relation $<f, \dot{x}>\, = 0$ and equation (\ref{formcomp}) imply the
ODE
\begin{equation}
\label{Piode}
\dot{\Pi} = \Pi <\bar{f},\dot{x}>,
\end{equation}
along the conformal geodesics, which, together with (\ref{Pidata}),
will allow one to determine $\Pi$ once $<\bar{f},\dot{x}>$ is known.

\subsection{The extended conformal geodesic equation on $I$}
\label{extconfgeodatI}

With respect to the metric (\ref{Xconfext}) a solution to the conformal
geodesic equations is given by a space-time curve 
$x^{\mu}(\tau) = (\bar{\tau}(\tau), \bar{\rho}(\tau), \psi^A(\tau))$ and along
that curve a vector field $X(\tau)$ and a 1-form
$\bar{f}(\tau)$ such that 
\[
\dot{x} = X,
\]
\[
\bar{\nabla}_X\,X = - 2\,<\bar{f}, X> X + \bar{g}(X, X)\,\bar{f}^{\sharp},
\]
\[
\bar{\nabla}_X\,\bar{f} 
= \,<\bar{f}, X> \bar{f} - \frac{1}{2}\,\bar{g}(\bar{f}, \bar{f})\,X^{\flat} +
L(X,\,.\,).
\]
With the expansions $X = X^i\,v_i$, $\bar{f} = \bar{f}_i\,\alpha^i$, 
$\bar{g} = \bar{g}_{jk}\,\alpha^j\,\alpha^k$, 
$\bar{g}^{\sharp} = \bar{g}^{jk}\,v_j\,v_k$, 
$L = L_{jk}\,\alpha^j\,\alpha^k$, 
$e_k = e^i\,_k\,v_i$, 
the equations above take in the domain where $\bar{\rho} > 0$ the form

\[
\frac{d}{d\,\tau}\,\bar{\tau} = X^0,\,\,\,\,\,
\frac{d}{d\,\tau}\,\bar{\rho} = \bar{\rho}\,X^1,\,\,\,\,\,
\frac{d}{d\,\tau}\,\psi^A = X^A,
\]
which is the equation $\dot{x}^{\mu} = X^i\,v^{\mu}\,_i$, relating
the coordinate to the frame expressions, 
\[
\frac{d}{d\,\tau}\, X^i +  \gamma_j\,^i\,_k\,X^j\,X^k =
- 2\,\bar{f}_k\,X^k\,X^i + \bar{g}_{jk}\,X^j\,X^k\,\bar{g}^{il}\,\bar{f}_l,
\]

\[
\frac{d}{d\,\tau}\,\bar{f}_k - \gamma_j\,^i\,_k\,\xi^j\,\bar{f}_i = 
\bar{f}_l\,X^l\,\bar{f}_k
- \frac{1}{2}\,\bar{g}^{lj}\,\bar{f}_l\,\bar{f}_j\,\bar{g}_{lk}\,X^l +
L_{jk}\,X^j. 
\]

Note that the functions $\bar{g}_{jk}$, $\bar{g}^{il}$, $\gamma_j\,^i\,_l$,
$L_{jk}$ entering these equations extend by analyticity
through $\bar{{\cal I}}'$ into a domain where $\bar{\rho} < 0$. Assuming
such an extension, we get the {\it extended conformal geodesic equations}.
Since also the data are analytic on $\bar{{\cal S}}$, it makes sense to
consider these equations in a neighbourhood of $\bar{{\cal I}}'$.

\begin{lemma}
\label{Icgsol}
With the values of $L_{jk}$ on $\bar{{\cal I}}'$ found in Lemma
\ref{LIval}, the initial data $x = (0, 0, \psi^{A'})$ and 
(cf. (\ref{dxindata}), (\ref{vfdata})) 
$X^i = \delta^i\,_0 + \delta^i\,_1$,
$\bar{f}_i = - \delta^0\,_i + \delta^1\,_i$ on ${\cal I}^{0'}$
determine a solution 
$x(\tau)$, $X(\tau)$, $\bar{f}(\tau)$ 
of the extended conformal geodesic equations
with $\tau = 0$ on ${\cal I}^{0'}$ and  
\[
x(\tau) = (\bar{\tau}(\tau), \bar{\rho}(\tau), \psi^A(\tau))
= (\tau, 0, \psi^{A'}).
\] 
By analyticity it extends as a solution into a domain 
$0 \le \tau \le 1 + 2\,\epsilon$ for some $\epsilon > 0$. The
extension to $\bar{{\cal I}}'$ of the conformal factor $\Pi$ which is
determined by (\ref{Pidata}) and (\ref{Piode}) takes the value $\Pi = 1$
on $\bar{{\cal I}}'$.
\end{lemma}

Proof: With the ansatz $x(\tau) = (\bar{\tau}(\tau), 0, \psi^{A'})$, 
$X(\tau) = X^0(\tau)\,v_0 + X^1(\tau)\,v_1$, 
$\bar{f} = \bar{f}_0(\tau)\,\alpha^0 + \bar{f}_1(\tau)\,\alpha^1$ 
those of the extended
conformal geodesic equations which are not identically satisfied because of
(\ref{gapprox}), (\ref{congapprox}), (\ref{Igammarel}) are given by

\[
\frac{d}{d\,\tau}\,\bar{\tau} = X^0,
\]

\[
\frac{d}{d\,\tau}\, X^0 + 
2\,\bar{\tau}\,X^0\,X^1 - \bar{\tau}\,(1 - \bar{\tau}^2)\,X^1\,X^1
\]
\[
=
- 2\,(\bar{f}_0\,X^0 + \bar{f}_1\,X^1)\,X^0 + 
(2\,X^0\,X^1  - (1 - \bar{\tau}^2)X^1\,X^1)\,
((1 - \bar{\tau}^2)\,\bar{f}_0 + \bar{f}_1),
\]

\[
\frac{d}{d\,\tau}\, X^1 - \bar{\tau}\,X^1\,X^1
=
- 2\,(\bar{f}_0\,X^0 + \bar{f}_1\,X^1)\,X^1 + 
(2\,X^0\,X^1  - (1 - \bar{\tau}^2)X^1\,X^1)\,\bar{f}_0,
\]

\[
\frac{d}{d\,\tau}\,\bar{f}_0 - \bar{\tau}\,\bar{f}_0\,X^1
= (\bar{f}_0\,X^0 + \bar{f}_1\,X^1)\,\bar{f}_0
- \frac{1}{2}\,
((1 - \bar{\tau}^2)\,\bar{f}_0\,\bar{f}_0 + 2\,\bar{f}_0\,\bar{f}_1)\,X^1
+ \frac{1}{2}\,X^1, 
\]

\[
\frac{d}{d\,\tau}\,\bar{f}_1
- \bar{\tau}\,\bar{f}_0\,(X^0 - (1 - \bar{\tau}^2)\,X^1) 
+ \bar{\tau}\,\bar{f}_1\,X^1
= (\bar{f}_0\,X^0 + \bar{f}_1\,X^1)\,\bar{f}_1
\]
\[
- \frac{1}{2}\,
((1 - \bar{\tau}^2)\,\bar{f}_0\,\bar{f}_0 + 2\,\bar{f}_0\,\bar{f}_1)\,
(X^0 - (1 - \bar{\tau}^2)\,X^1)
+ \frac{1}{2}\,X^0
- \frac{1 - \bar{\tau}^2}{2}\,X^1. 
\]

A calculation shows that the solution of this system for the prescribed
initial is given by
\begin{equation}
\label{Isol}
\bar{\tau} = \tau,\,\,\,\,\,\,\, 
X^0 = 1,\,\,\,\,\,\,\,
X^1 = \frac{1}{1 + \bar{\tau}},\,\,\,\,\,\,\,
\bar{f}_0 = - \frac{1}{1 + \bar{\tau}},\,\,\,\,\,\,\,
\bar{f}_1 = 1.
\end{equation}
This proves the first assertion. 
With the solution above equation (\ref{Piode}) reads $\dot{\Pi} = 0$ and 
we have $\Pi = 1$ on $I^0$ by (\ref{quotlim}). This proves the second
assertion.

\vspace{.3cm}

\noindent
{\bf Remark}: 
The ODE above is sufficiently complicated so that giving the solution
explicitly deserves an explanation. 
In (\ref{finconfmink}) is given the conformal factor and the conformal
representation of Minkowski space which result from the general procedure
of section \ref{Rfivp}. In (\ref{firstcotrafo}) is given the coordinate
transformation which, together with the conformal factor, 
relates the conformal metric to the standard
representation of Minkowski space in coordinates $t$ and $r$.

If the Minkowski values $m = 0$, $U = 1$,  $h_{ab} = - \delta_{ab}$ are
assumed in section \ref{confextstaticvacuum} the metric $\bar{g}$ reduces
by (\ref{gapprox}) to the metric $g^*_{ik}\,\alpha^i\,\alpha^k$ with 
$m= 0$. One can consider this as the lowest order (in $\bar{\rho}$)
approximation of the general version of $\bar{g}$. Tracing back how the
functions $\bar{\tau}$, $\bar{\rho}$, $\Lambda$ in section
\ref{confextstaticvacuum} are related in the flat case to
$t$ and $r$, one finds \[
r = \frac{1}{\bar{\rho}\,(1 - \bar{\tau})},\,\,\,\,\,\,
t = \frac{\bar{\tau}}{\bar{\rho}\,(1 - \bar{\tau})},\,\,\,\,\,\,
\Lambda = \frac{1}{r}.
\]
The conformal factors in the conformal representations thus agree but the
coordinates are related by the transformation 
\begin{equation}
\label{simptrans}
\bar{\tau} = \tau,\,\,\,\,\,\,\bar{\rho} = \rho\,(1 + \tau).
\end{equation}
This implies  
\begin{equation}
\label{flatmetrics}
2\,\frac{d\,\bar{\rho}}{\bar{\rho}}\,d\,\bar{\tau} - (1 - \bar{\tau}^2)\,
\left(\frac{d\,\bar{\rho}}{\bar{\rho}}\right)^2 - d\sigma^2 =
d\,\tau^2 + 2\,\tau\,\frac{d\,\rho}{\rho}\,
d\,\tau - (1 - \tau^2)
\left(\frac{d\,\rho}{\rho}\right)^2 -  d\sigma^2.
\end{equation}
The left hand side is the conformal Minkowski metric (\ref{gapprox})
with $m= 0$ while the right hand side is the metric $g^{\star}$ given by
(\ref{finconfmink}). The conformal geodesics underlying (\ref{finconfmink})
have tangent vector $X = \partial_{\tau}$ and 1-form 
$f = \frac{d\,\rho}{\rho}$. With (\ref{simptrans}) these
transform into
\[
X = \partial_{\bar{\tau}} + 
\frac{1}{1 + \bar{\tau}}\,\bar{\rho}\,\partial_{\bar{\rho}} 
= v_0 + \frac{1}{1 + \bar{\tau}}\,v_1,
\] 
\[
f
= \frac{d\,\bar{\rho}}{\bar{\rho}}
- \frac{1}{1 + \bar{\tau}}\,d\,\bar{\tau} 
= - \frac{1}{1 + \bar{\tau}}\,\alpha_0 + \alpha_1, 
\]
from which one can read off (\ref{Isol}).

\vspace{.3cm}

The hypersurfaces $\{\bar{\rho} = \bar{\rho}_\# = const. > 0\}$ are in
general time-like for the metric (\ref{Xconfext}). The form of
$\bar{g}^{\sharp}$ suggests that these hypersurfaces approximate null
hypersurfaces in the limit as  
$\bar{\rho}_\# \rightarrow 0$, but the conclusion is delicate because of the
degeneracy of $\bar{g}^{\sharp}$ on $\bar{{\cal I}}'$.
The discussion above shows that they do become null asymptotically
in the sense that for the metric on the left hand side of
(\ref{flatmetrics}) the hypersurfaces $\{\bar{\rho} =
const.\}$ are in fact null. To some extent this explains why the
coordinates  given by (\ref{Bbasemap}) had a chance to extend smoothly to
${\cal J}^+$ and to provide a description of the cylinder at space-like
infinity.

\subsection{The smoothness of the gauge of section \ref{Rfivp} for static
asymptotically flat vacuum solution near ${\cal I}$}
\label{smoothnessnearI}

Let $\bar{{\cal S}}_{ext}$ denote an analytic extension of 
$\tilde{{\cal S}}$ into a range where $\bar{\rho} < 0$ so that
$\bar{\rho}$, $\psi^A$ extend to analytic coordinates. If the set
$\bar{{\cal S}}_{ext} \setminus \bar{{\cal S}}$ is sufficiently small, the
following statements make sense. The initial conditions (\ref{dxindata}),
(\ref{vfdata})  extend analytically to $\bar{{\cal S}}_{ext}$ and 
determine near $\bar{{\cal S}}_{ext}$ an analytic congruence of solutions
to the extended conformal geodesic equations. It therefore follows from
Lemma \ref{Icgsol} and well known results on ODE's that, with the
$\epsilon$ of Lemma \ref{Icgsol}, there exists a $\rho_\# >\,0$ such that
for initial data
$\bar{\tau}(0) = 0$, $\bar{\rho}(0) = \rho'$ with $|\rho'| < \rho_\#$,
$\psi^A(0) = \psi^{A'}$ and those implied at these points by
(\ref{dxindata}), (\ref{vfdata}) 
the solution
\[
\bar{\tau} = \bar{\tau}(\tau, \rho', \psi^{A'}),\,\,\,\,\,
\bar{\rho} = \bar{\rho}(\tau, \rho', \psi^{A'}),\,\,\,\,\,
\psi^A = \psi^A(\tau, \rho', \psi^{A'}),
\]
\[
X^i = X^i(\tau, \rho', \psi^{A'}),\,\,\,\,\,
\bar{f}_k = \bar{f}_k(\tau, \rho', \psi^{A'}),
\]
of the extended conformal geodesic equations exists for the values
$0 \le \tau \le 1 + \epsilon$ of their natural parameter and the function
$\Pi$ is positive in the given range of $\rho'$ and $\tau$.

Taking a derivative of the equation satisfied by $\bar{\rho}$ and observing
(\ref{Isol}) gives 
\[
\frac{d}{d\,\tau}\,
\left(\frac{\partial \bar{\rho}}{\partial{\rho'}}|_{\rho' = 0}\right)
= \left(\frac{\partial \bar{\rho}}{\partial{\rho'}}|_{\rho' = 0}\right)\, 
\frac{1}{1 + \bar{\tau}},
\] 
which implies by (\ref{Isol}) 
\[
\left(\frac{\partial \bar{\rho}}{\partial{\rho'}}|_{\rho' = 0}\right)
= 1 + \tau \ge 1.
\]
It follows that the Jacobian of the analytic map
\[
(\tau, \rho', \psi^{A'}) \rightarrow x^{\mu}(\tau, \rho', \psi^{A'}),
\] 
takes the value $1 + \bar{\tau}$ on $\bar{{\cal I}}'$ and for sufficiently
small $\rho_\# > 0$ the Jacobian does not vanish in the range
$0 \le \tau \le 1 + \epsilon$, $|\bar{\rho}| \le \rho_\#$. The relations
$\Lambda = \Theta\,\Pi^{-1}$, $\Pi > 0$, and 
$\Theta = (\omega^{-1}\,\Omega)_*\left(1 - \tau^2\right)$ imply that the
curves with $\rho' > 0$ cross ${\cal J}^{+'}$ for $\tau = 1$.  
It follows that $\tau$, $\rho'$, and $\psi^{A'}$ define an analytic
coordinate system in a certain neighbourhood ${\cal O}'$ of $\bar{{\cal
I}}'$ in
$\bar{{\cal M}}'$, such that (suppressing again the upper bounds for 
$\rho'$) ${\cal O}' \cap {\cal J}^{+'} = \{\tau = 1, \,\,\rho' > 0\}$,  
${\cal I}' = \{0 \le \tau < 1,\,\, \rho' = 0\}$,   
${\cal I}^{+'} = \{\tau = 1,\,\, \rho' = 0\}$, and ${\cal O}'$ is ruled
by conformal geodesics. 

\vspace{.3cm}

The metric $g = \Pi^2\,\bar{g}$, the connection coefficients of the
connection
$\hat{\nabla}$ and the tensor fields (cf. (\ref{WLtensrel})) 
\[
\hat{L}_{\mu \nu} = L_{\mu \nu}[\bar{g}]
- \nabla_{\mu}\,\bar{f}_{\nu} + \bar{f}_{\mu}\,\bar{f}_{\nu}
- \frac{1}{2}\,\bar{g}_{\mu \nu}\,\bar{f}_{\lambda}\,\bar{f}^{\lambda},
\]
\[
f = \bar{f} - \Pi^{-1}\,d\,\Pi,\,\,\,\,\,
W_{\mu \nu \rho \lambda}[g] =
\Pi\,W_{\mu \nu \rho \lambda}[\bar{g}].
\]
in the frame (\ref{frames}) extend in the new coordinates as analytic
fields to ${\cal O}'$. 

Given these structures and the conformal geodesics
on ${\cal O}'$, the construction of the manifold $\bar{{\cal N}}$ as
decribed in section \ref{Rfivp} poses no problems.
With the given analytic initial data on $\bar{{\cal S}}$ 
it only involves solving linear ODE's corresponding to (\ref{Wccgxequ}),
such as 
\begin{equation}
\label{framecoeffode}
\frac{d}{d\,\tau}\, e^i\,_k +  \gamma_j\,^i\,_l\,X^j\,e^l\,_k =
- \bar{f}_l\,X^l\,e^i\,_k 
- \bar{f}_l\,e^l\,_k\,X^i
+ \bar{g}_{jl}\,X^j\,e^l\,_k\,\bar{g}^{im}\,\bar{f}_m,
\end{equation}
or its spinor analogue, along the conformal geodesics. This allows us to
conclude

\begin{lemma}
\label{cggauge smooth}
Starting with static asymptotically flat initial data in the gauge
(\ref{mgauge}), the construction of section \ref{Rfivp} leads to a
conformal representation of the static vacuum space-time which is real
analytic in a neighbourhood ${\cal O}$ of the set $\bar{{\cal I}}$ in 
$\bar{{\cal N}}$.
\end{lemma}

\subsection{Changing the conformal gauge on the initial slice}
\label{changeofconfgauge}

It will be shown now how the construction described in
section \ref{Rfivp} depends for static vacuum solutions on rescalings 
\[
\omega^{-2}\,h \rightarrow h' = \vartheta^2\,\omega^{-2}\,\bar{h},
\,\,\,\,\,\,\,
\Omega \rightarrow \Omega' = \vartheta\,\Omega\,\,\,\,\,\,
\mbox{on}\,\,\,\,\,{\cal S},
\]
with analytic, positive conformal factors $\vartheta$.

There are harmless consequences such as the change of the normal
coordinates $x^a \rightarrow x^{'a} = x^{'a}(x^c)$ with $x^{'a}(0) = 0$
and a related change $e_a \rightarrow e'_a = \vartheta^{-1}\,s^c\,_a\,e_c$ 
of the frame vector fields tangent to $\tilde{{\cal S}}$. Here $s^c\,_a$
denotes an analytic function on $\tilde{{\cal S}}$ with values in $SO(3)$
such that $s^c\,_a \rightarrow \delta^c\,_a$ as $\bar{\rho} \rightarrow 0$.
These changes will simply be propagated along the new conformal
geodesics.

Critical is the transition from the congruence of conformal geodesics
related to $\Omega$ (the {\it $\Omega$-congruence}) to the new one
related to $\Omega'$ (the {\it $\Omega'$-congruence}). If the curves are
considered as point sets, the two families of curves will be different if
$\Omega^{'-1}\,d\,\Omega' -
\Omega^{-1}\,d\,\Omega = \vartheta^{-1}\,d\,\vartheta \neq 0$ (cf.
\cite{friedrich:cg on vac}). 

The rescaling above implies on $\tilde{S}$ the transitions
\[
||d\,\Omega||_h \rightarrow ||d\,\Omega'||_{h'} = \xi\,\,||d\,\Omega||_h,
\]
\[
\omega = \frac{2\,\Omega}{||d\,\Omega||_h} \rightarrow
\omega' = \frac{2\,\Omega'}{||d\,\Omega'||_h'} 
= \frac{\omega\,\delta}{\xi},
\]
\[
\Theta|_{\tilde{S}} = \omega^{-1}\,\Omega  \rightarrow 
\Theta'|_{\tilde{S}} = \omega^{'-1}\,\Omega'
= \xi\,\,\Theta|_{\tilde{S}},
\]
with the function
\[
\xi = \left|1 
- 3\,\vartheta^{-1}\frac{D_a\Omega\,D^a\vartheta}
{\Delta_h \Omega} 
- \frac{3}{2}\,
\vartheta^{-2}\Omega\,\frac{D_a\vartheta\,D^a\vartheta}
{\Delta_h \Omega}\right|^{\frac{1}{2}},
\]
which extends to $\bar{{\cal S}}$ as a analytic function of $\bar{\rho}$
and $\psi^A$. 

It follows from the initial conditions for the $\Omega$-congruence that
\begin{equation}
\label{xi'data} 
\xi^{-1}\,\dot{x} \perp \tilde{{\cal S}},\,\,\,\,\,\,\,\,
\Theta^{'2}\,\tilde{g}(\xi^{-1}\,\dot{x}, \xi^{-1}\,\dot{x}) = 1,
\end{equation}
and for the transformed 1-form that
\[
<f', \dot{x}>\, = 0,\,\,\,\,\,\,\,\,
f_{\tilde{{\cal S}}} \rightarrow
f'_{\tilde{{\cal S}}} = \omega^{'-1}\,d\,\omega' 
= f_{\tilde{{\cal S}}}
+ \vartheta^{-1}\,d\,\vartheta 
- \xi^{-1}\,d\,\xi,
\]
where the subscripts indicate the pull back to $\tilde{\cal S}$. These two
lines give the initial data for the $\Omega'$-congruence if the conformal
geodesic equations are expressed with respect to the rescaled metric $g'$
and its connection $\nabla'$.

To compare the $\Omega'$-congruence with the $\Omega$-congruence we observe
the conformal invariance of conformal geodesics (cf.
\cite{friedrich:tueb}) and express the equations for the
$\Omega'$-congruence in terms of $g$ and its
connection $\nabla$. The space-time curves, including their
parameter $\tau'$, then remain unchanged. The 1-form is transformed 
because of $g = (\Theta\,\Theta^{'-1})^2\,g'$
according to $f' \rightarrow f^* = f' - 
(\Theta\,\Theta^{'-1})^{-1}\,d\,(\Theta\,\Theta^{'-1})$,
which implies $<f^*, \dot{x}>\, = 0$,
$f^*_{\tilde{{\cal S}}} = \bar{f}_{\tilde{{\cal S}}} +
\vartheta^{-1}\,d\,\vartheta$ on $\tilde{S}$.
If this 1-form is expressed in terms of the $g$-orthonormal frame 
$e_k$ with $e_0 \perp \tilde{{\cal S}}$, one finds  
\begin{equation}
\label{f'gbarekdata} 
{f}^*_0 \equiv \,<{f}^*, e_0>\, = 0,\,\,\,\,\,\,\,\,
{f}^*_a \equiv \,<{f}^*, e_a>\,
= f_a + \vartheta^{-1}<d\,\vartheta, e_a>,\,\,\,a = 1, 2, 3.
\end{equation}
The fields $\xi^{-1}\,\dot{x}$, ${f}^*_k$ are the initial data 
for the $\Omega'$-congruence in terms of $g$, $e_k$, and 
$\nabla$.
Since $\xi \rightarrow 1$ and $<d\,\vartheta, e_a> \,=
O(\bar{\rho})$ as $\bar{\rho} \rightarrow 0$,  it follows
that
\[
\frac{\Theta'}{\Theta} \rightarrow 1,\,\,\,\,\,\,\,\,
\dot{x} - \xi^{-1}\,\dot{x} \rightarrow 0,\,\,\,\,\,\,\,\,
f^*_k - f'_k \rightarrow 0
\,\,\,\,\,\mbox{as}\,\,\,\bar{\rho} \rightarrow 0.
\]
As a consequence, the initial data for the  $\Omega'$- and the
$\Omega$-congruence have coinciding limits on ${\cal I}^{0'}$ and the
corresponding curves are identical on $\bar{{\cal I}}'$. 

Assuming now the conditions of section \ref{smoothnessnearI} and 
using arguments similar to the ones used there, we conclude that in a
certain neighbourhood ${\cal O}'$ of $\bar{{\cal I}}'$ in
$\bar{{\cal M}}'$ the gauge related to the $\Omega'$-congruence is as
smooth and regular as the one related to the $\Omega$-congruence. Thus we
have  

\begin{lemma}
\label{confgaugechange}
In the case of static asymptotically flat space-times the construction of
the set $\bar{{\cal I}}'$ is independent of the choice of $\Omega$ and the
set ${\cal I}'$ introduced in section \ref{confextstaticvacuum} coincides
with the projection $\pi'({\cal I})$ of the cylinder at space-like infinity
as defined in section \ref{Rfivp}.
\end{lemma} 

We note that the comparison of the $\Omega'$- with the $\Omega$-congruence
leads in the case where the solution is not static and thus
not necessarily analytic still to similar results if the solution acquires
a certain smoothness near ${\cal J}^{\pm} \cup {\cal I}^{\pm}$. In the case
of low smoothness, however, the detailed behaviour of the different
congruences needs to be analysed in the context of an existence theorem.

\section{Concluding remarks}
\label{conclusion}

Concerning the regularity conditions we have now the following situation.
For {\it static} asymptotically flat solutions with $m \neq 0$ the conformal
extensions to ${\bar{\cal I}}$ are smooth (in the sense discussed above) and
their data satisfy the regularity condition (\ref{regcond}) with $p_* =
\infty$.  In the {\it massless case} condition (\ref{regcond}) with $p_* =
\infty$ is necessary and sufficient for space-like infinity to be
represented by a regular point in a smooth conformal extension. In the {\it
general time reflection case with} $m \neq 0$ conditions (\ref{regcond}) are
necessary but not sufficient for the s-jets $J^p_{\cal I}(u)$, $p \in
\mathbb{N}$,  to be regular at the critical sets ${\cal I}^{\pm}$.
Thus, the mass $m = 2\,W(i)$ and also the derivatives of 
$\partial^{\alpha}_{x^a}W(i)$, $\alpha \in \mathbb{N}^3$, play a crucial
role for the behaviour of the $J^p_{\cal I}(u)$ at ${\cal I}^{\pm}$. 
The mechanism which decides on the smoothness remains to be
understood. 

Only the d-jet $J^{p}_{{\cal I}^0}(u)$ and the 
s-jets $J^{p-1}_{\cal I}(u)$ are needed to obtain $J^p_{\cal I}(u)$
by integrating the transport equations on ${\cal I}$.
Since the left hand sides of the transport equations are universal in the
sense that they do not depend on the data, it follows that 
$J^p_{\cal I}(u)$ is uniquely determined by $J^{p}_{{\cal I}^0}(u)$ for 
$p \in \mathbb{N}$. In the static case the s-jets $J^p_{\cal I}(u)$ are
regular. In \cite{chrusciel:delay:2003} has been exhibited a class of data
which are asymptotically static of order $p$ for given 
$p \in \mathbb{N} \cup \{\infty\}$ and which are essentially arbitrary on
given compact sets. It follows that for prescribed
differentiability order
$p$ there exists a large class of data for which the s-jet $J^{p}_{\cal
I}(u)$ is regular on ${\cal I}$. 

We expect there to be a threshold in $p$ beyond which the regularity
of $J^{p}_{\cal I}(u)$ ensures peeling resp. asymptotic smoothness of a
given order of differentiability and below which the singularity of
$J^{p}_{\cal I}(u)$ implies a failure of peeling. This order is likely
to be low enough such that the behaviour of $J^{q}_{\cal I}(u)$ with
$q \le p$ can be controlled by a direct, though tedious, calculation.
However, if asymptotic staticity does play a role here, one should try to
understand the underlying mechanism. It would be
quite a remarkable feature of Einstein's equations if asymptotic
staticity could be {\it deduced} from asymptotic regularity at null
infinity.



\begin{thebibliography}{11}



\bibitem{andersson:chrusciel:as}
L. Andersson, P.T. Chru\'sciel.
\newblock On hyperboloidal Cauchy data for the vacuum Einstein
equations and obstructions to the smoothness of scri.
\newblock { Comm. Math. Phys.} 161 (1994) 533--568.

\bibitem{andersson:chrusciel:ph}
L. Andersson, P.T. Chru\'sciel.
\newblock Solutions of the constraint equations in general
relativity satisfying hyperboloidal boundary conditions.
\newblock { Dissertationes Mathematicae Polska Akademia Nauk, Inst.
Matem.}, Warszawa, 1996.

\bibitem{friedrich:ACF}
L. Andersson, P.T. Chru\'sciel, H. Friedrich.
\newblock On the regularity of solutions to the Yamabe equation
and the  existence  of smooth hyperboloidal initial data for
Einstein's field equations.
\newblock { Comm. Math. Phys.} 149 (1992) 587--612.

\bibitem{ashtekar}
A. Ashtekar.
\newblock Asymptotic properties of isolated systems: Recent developments.
\newblock In: B. Bertotti et. al (eds.) {\em General Relativity and
Gravitation}
\newblock Dordrecht, Reidel, 1984. 

\bibitem{bartnik:norton}
R. A. Bartnik, A. H. Norton.
\newblock Numerical experiments at null infinity.
\newblock In: J. Frauendiener, H. Friedrich (eds.),
{\em The Conformal Structure of Spacetime: Geometry, Analysis, Numerics}.
\newblock Springer, Berlin, 2002.

\bibitem{beig:simon}
R. Beig, W. Simon.
\newblock Proof of a multipole conjecture due to Geroch.
\newblock { Comm. Math. Phys.} 79 (1981) 581--589.

\bibitem{beig:schmidt:2000}
R. Beig, B. Schmidt.
Time-independent gravitational fields.
In: B. G. Schmidt (ed.): Einstein's field equations and their physical
implications. Springer, Berlin, 2000.

\bibitem{besse}
A. L. Besse.
\newblock Einstein Manifolds.
\newblock Springer, Berlin, 1987.

\bibitem{blanchet}
L. Blanchet.
\newblock Post-Newtonian Gravitational Radiation.
\newblock In: B. Schmidt (ed.), {\em Einstein's Field Equations and 
Their Physical Implications}.
\newblock Berlin, Springer, 2000. 

\bibitem{bondi:et.al}
H. Bondi, M. G. J. van der Burg, A. W. K. Metzner.
\newblock Gravitational waves in general relativity VII. Waves from
axi-symmetric isolated systems.
\newblock { Proc. Roy. Soc} A 269 (1962) 21--52.

\bibitem{choquet-bruhat}
Y. Choquet-Bruhat.
\newblock Th\'eor\`emes d'existence pour certains syst\`emes 
d'\'equations aux de\'riv\'ees
partielles non lin\'eaires.
\newblock {\em Acta. Math.} 88 (1952) 141 - 225.

\bibitem{choptuik}
M. W. Choptuik.
\newblock Universality and scaling in gravitational collapse of a
massless scalar field.
\newblock {\em Phys. Rev. Lett.} 70 (1993) 9.

\bibitem{christ}
D. Christodoulou.
\newblock The Global Initial Value Problem in General Relativity.
\newblock In: V. G. Gurzadyan et al. (eds.) {Proceedings of the 9th
Marcel Grossmann Meeting}
\newblock World Scientific, New Jersey, 2002.

\bibitem{christ:klain}
D. Christodoulou, S. Klainerman.
The Global Nonlinear Stability of the Minkowski Space.
Princeton University Press, Princeton, 1993.

\bibitem{chrusciel:delay:2002}
P. T. Chru\'sciel, E. Delay.
\newblock Existence of non-trivial, vacuum, asymptotically simple
spacetimes.
\newblock {\em Class. Quantum Grav.}, 19 (2002) L 71 - L 79.
\newblock Erratum
\newblock {\em Class. Quantum Grav.}, 19 (2002) 3389.

\bibitem{chrusciel:delay:2003}
P. T. Chru\'sciel, E. Delay.
\newblock On mapping properties of the general relativistic constraints
operator in weighted function spaces, with application.
\newblock {\it M\'{e}m. Soc. Math. France} submitted.
\newblock http://xxx.lanl.gov/abs/gr-qc/0301073

\bibitem{cms}
P. T. Chru\'sciel, M.A. H. MacCallum, D. B. Singleton.
\newblock Gravitational Waves in General Relativity. XIV: Bondi Expansions
and the ``Polyhomogeneity'' of $Scri$.
\newblock {\it Phil. Trans. Royal Soc., London} A 350 (1995) 113 - 141.

\bibitem{chrusciel:lengard}
P.T. Chru\'sciel, O. Lengard.
\newblock Solutions of wave equations in the radiation regime.
\newblock { Preprint} (2002).
\newblock {http://xxx.lanl.gov/archive/math.AP/0202015}

\bibitem{chrusciel:jezierski:kijowski}
P. T. Chru\'sciel, J. Jezierski, J. Kijowski.
\newblock Hamiltonian Field Theory in the Radiating Regime.
\newblock Springer, Berlin, 2002.

\bibitem{corvino}
J. Corvino.
\newblock Scalar curvature deformation and a gluing
construction for the Einstein constraint equations.
\newblock { Comm. Math. Phys.} 214 (2000) 137--189. 

\bibitem{corvino:schoen}
J. Corvino, R. Schoen
\newblock On the Asymptotics for the Vacuum Einstein Constraint
Equations.
\newblock  http://xxx.lanl.gov/abs/gr-qc/0301071

\bibitem{courant:hilbert:II}
R. Courant, D. Hilbert.
\newblock Methods of mathematical physics, Vol II.
\newblock J. Wiley, New York, 1962.

\bibitem{dain:stationary}
S. Dain.
\newblock Initial data for stationary space-times near space-like
infinity.
\newblock {\em Class. Quantum Grav.}, 18 (2001) 4329 - 4338.

\bibitem{dain:in prepr}
S. Dain.
\newblock Asymptotically Flat Initial Data with
Prescribed Regularity II.
\newblock In preparation 

\bibitem{dain:friedrich}
S. Dain, H. Friedrich.
\newblock Asymptotically Flat Initial Data with
Prescribed Regularity.
\newblock { Comm. Math. Phys.}  222 (2001) 569--609. 

\bibitem{ellis:1984}
Ellis, G. F. R. (1984) 
Relativistic Cosmology: Its Nature, Aims, and Problems,
{\em General Relativity and Gravitation},
{B. Bertotti et. al (eds.)}, Reidel, Dordrecht.

\bibitem{frauendiener:tueb}
J. Frauendiener.
\newblock Some aspects of the numerical treatment of the
conformal field equations.
\newblock In: J. Frauendiener, H. Friedrich (eds.),
{\em The Conformal Structure of Spacetime: Geometry, Analysis, Numerics}.
\newblock Springer, Berlin, 2002.

\bibitem{friedrich:1981a}
H. Friedrich.
\newblock On the regular and the asymptotic characteristic initial value
problem for Einstein's vacuum field equations. 
\newblock Proceedings of the 3rd Gregynog Relativity
Workshop on  Gravitational Radiation Theory
\newblock MPI-PEA/Astro 204 (1979) 137--160 and
\newblock { Proc. Roy. Soc.}, 375 (1981) 169--184.

\bibitem{friedrich:1981b}
H. Friedrich.
\newblock The asymptotic characteristic initial value problem for Einstein's
vacuum field equations as an initial value problem for a first-order
quasilinear symmetric hyperbolic system. 
\newblock { Proc. Roy. Soc.}, A 378 (1981) 401--421.

\bibitem{friedrich:1982}
H. Friedrich.
\newblock On the Existence of Analytic Null Asymptotically Flat
Solutions of Einstein's Vacuum Field Equations.
\newblock { Proc. Roy. Soc.\ Lond.\ A} 381 (1982) 361--371.

\bibitem{friedrich:hypivp}
H. Friedrich.
\newblock Cauchy Problems for the Conformal Vacuum Field
Equations in  General Relativity.
\newblock{ Comm. Math. Phys.} 91 (1983) 445--472.

\bibitem{friedrich:1hyp red}
H. Friedrich.
\newblock On the hyperbolicity of Einstein's and other gauge field 
equations.
\newblock { Comm.\ Math.\ Phys.} 100 (1985) 525--543.

\bibitem{friedrich:n-geod}
H. Friedrich.
\newblock On the existence of n-geodesically complete or future 
complete solutions of Einstein's equations with smooth 
asymptotic structure. 
\newblock { Comm. Math. Phys.} 107 (1986) 587--609.

\bibitem{friedrich:static}
H. Friedrich.
\newblock On static and radiative space-times.
\newblock { Comm. Math. Phys.}, 119 (1988) 51--73.

\bibitem{friedrich:global}
H. Friedrich.
\newblock On the global existence and the asymptotic behaviour
of solutions to the Einstein-Maxwell-Yang-Mills equations.
\newblock {\em J. Diff. Geom.}, 34 (1991) 275 - 345. 

\bibitem{friedrich:AdS}
H. Friedrich.
\newblock Einstein equations and conformal structure: existence of
anti-de Sitter-type space-times.
\newblock { J. Geom. Phys.}, 17 (1995) 125--184.

\bibitem{friedrich:i-null}
H. Friedrich.
\newblock Gravitational fields near space-like and null
infinity.
\newblock { J. Geom. Phys.}  24 (1998)  83--163.

\bibitem{friedrich:tueb}
H. Friedrich.
\newblock Conformal Einstein Evolution
\newblock In: J. Frauendiener, H. Friedrich (eds.),
{\em The Conformal Structure of Spacetime: Geometry, Analysis, Numerics}.
\newblock Springer, Berlin, 2002.

\bibitem{friedrich:spin-2}
H. Friedrich.
\newblock Spin-2 fields on Minkowski space near spacelike and
null infinity.
\newblock {\em Class. Quantum. Grav.} 20 (2003) 101 - 117.

\bibitem{friedrich:cg on vac}
H. Friedrich.
\newblock Conformal geodesics on vacuum space-times.
\newblock {\em Commun. Math. Phys.} 235 (2003) 513 - 543.

\bibitem{friedrich:kannar1}
H. Friedrich, J. K\'ann\'ar
\newblock Bondi systems near space-like infinity and the
calculation  of the NP-constants.
\newblock { J. Math. Phys.} 41, (2000), 2195--2232.

\bibitem{friedrich:nagy}
H. Friedrich, G. Nagy.
\newblock The initial boundary value problem for Einstein's vacuum 
field equations.
\newblock { Comm. Math. Phys.} 201 (1999) 619--655.

\bibitem{friedrich:rendall}
H. Friedrich, A. Rendall.
\newblock The Cauchy Problem for the Einstein Equations.
\newblock In: B. Schmidt (ed.), {\em Einstein's Field Equations and 
Their Physical Implications}.
\newblock Berlin, Springer, 2000. 

\bibitem{friedrichs}
K. O. Friedrichs.
\newblock Symmetric hyperbolic linear differential equations.
\newblock { Comm. Pure Appl. Math.} 7 (1954) 345--392.

\bibitem{garabedian}
P. Garabedian.
\newblock {\em Partial Differential Equations}.
\newblock J. Wiley, New York, 1964.

\bibitem{geroch}
R. Geroch.
\newblock Asymptotic structure of space-time.
\newblock In: F. P. Esposito, L. Witten (eds.) {\em Asymptotic Structure
of Space-Time.}
\newblock New York, Plenum, 1977.

\bibitem{geroch:horowitz}
R. P. Geroch, G. Horowitz.
\newblock Asymptotically simple does not imply asymptotically
Minkowskian.
\newblock {Phys. Rev. Lett.} 40 (1978) 203--206.

\bibitem{gues}
O. Gu\`es. 
\newblock Probl\`eme mixte hyperbolique quasi-lin\'eaire charact\'eristique. 
\newblock {\em Commun. Part. Diff. Equ.}, 15 (1990) 595 - 645.

\bibitem{gundlach}
C. Gundlach.
\newblock Critical Phenomena in gravitational collapse.
\newblock {\em Physica Reports} 367 (2003) 339 - 405.

\bibitem{helgason}
S. Helgason.
\newblock {\em Differential Geometry and Symmetric Spaces}.
\newblock Academic Press, New York, 1962

\bibitem{hoermander:III}
L. H\"ormander.
\newblock The Analysis of Linear Partial Differential Operators III.
\newblock Springer, Berlin, 1985.

\bibitem{huebner:2001}
P. H\"ubner.
\newblock From now to timelike infinity on a finite grid.
\newblock { Class. Quantum Grav.} 18 (2001) 1871--1884.

\bibitem{husa:tueb}
S. Husa.
\newblock Problems and successes in the numerical approach to
the conformal field equations.
\newblock In: J. Frauendiener, H. Friedrich (eds.),
{\em The Conformal Structure of Spacetime: Geometry, Analysis, Numerics}.
\newblock Springer, Berlin, 2002.

\bibitem{kennefick:o'murchadha}
D. Kennefick, N. O'Murchadha. 
\newblock Weakly decaying asymptotically flat static and
stationary solutions to the Einstein equations. 
\newblock { Class. Quantum. Grav.} 12 (1995) 149--158.

\bibitem{klainerman:nicolo}
S. Klainerman, F. Nicol\`o.
\newblock The Evolution Problem in General Relativity.
\newblock Birkh\"auser, Basel, 2003. 

\bibitem{klainerman:nicolo:II}
S. Klainerman, F. Nicol\`o.
\newblock Peeling properties of asymptotically flat solutions to
the Einstein vacuum equations.
\newblock {\em Class. Quantum Grav.} 20 (2003) 3215 - 3257. 

\bibitem{lengard}
O. Lengard.
\newblock Solution of the Einstein equation, wave maps, and
semilinear waves in the radiation regime. 
\newblock Ph. D. thesis, Universit\'e de Tours, 2001.
\newblock {http://www/phys.uni-tours.fr/~piotr/papers/batz}

\bibitem{newman:penrose}
E. T. Newman, R. Penrose.
\newblock {\em An approach to gravitational radiation by a methods of spin
coefficients}.
\newblock{ J. Math. Phys.} 3 (1962) 566--578.

\bibitem{penrose:scri:let}
R. Penrose.
\newblock Asymptotic properties of fields and space-time.
\newblock { Phys. Rev. Lett.}, 10 (1963) 66--68.

\bibitem{penrose:scri}
R. Penrose.
\newblock Zero rest-mass fields including gravitation:
asymptotic behaviour.
\newblock { Proc. Roy. Soc. Lond.}, A 284 (1965) 159--203.

\bibitem{penrose:rindler:I}
R. Penrose, W. Rindler.
\newblock {\it Spinors and space-time}, Vol. 1 and 2.
\newblock Cambridge University Press, 1984.

\bibitem{pirani}
F. A. E. Pirani.
\newblock Invariant Formulation of Gravitational Radiation Theory.
\newblock {\em Phys. Rev.} 105 (1957) 1089 - 1099. 

\bibitem{sachs:waves VI}
R. K. Sachs.
\newblock Gravitational waves in general relativity VI. 
The outgoing radiation condition.
\newblock { Proc. Roy. Soc} A 264 (1961) 309--338.

\bibitem{sachs:waves VIII}
R. K. Sachs.
\newblock Gravitational waves in general relativity VIII. Waves in
asymptotically flat space-time.
\newblock { Proc. Roy. Soc} A 270 (1962) 103--126.

\bibitem{schoen:yau}
R. Schoen, S.-T. Yau.
\newblock Proof of the positive mass theorem II.
\newblock {\em Commun. Math. Phys.}, 79 (1981) 231 - 260.

\bibitem{schutz:2002}
B. F. Schutz.
\newblock Mathematical and Physical Perspectives on Gravitational
Radiation.
\newblock Talk given at the summer school ``50 years of the
Cauchy problem in general relativity'', Carg\`ese July 29 - August
10, 2002. 
\newblock http://fanfreluche.math.univ-tours.fr

\bibitem{secchi:II}
P. Secchi.
\newblock Well-Posedness of Characteristic Symmetric Hyperbolic Systems. 
\newblock {\em Arch. Rational Mech. Anal.}, 134 (1996) 155 - 197.

\bibitem{szego}
G. Szeg\"o.
\newblock Orthogonal Polynomials. 4th Edition.
\newblock A.M.S. Colloqu. Publ. Vol. 23, Providence, 1978.

\bibitem{trautman}
A. Trautman.
\newblock Radiation and boundary conditions in the theory of gravitation.
\newblock {\em Bull. Acad. Pol. Sci., S\'erie sci. math., astr. et phys.} VI
(1958) 407 - 412.

\bibitem{valiente kroon:2001}
J. A. Valiente Kroon.
\newblock Can one detect a non-smooth null infinity ?
\newblock {\em Class. Quantum Grav.} 18 (2001) 4311 - 4316.

\bibitem{valiente kroon:2003}
J. A. Valiente Kroon.
\newblock A new class of obstructions to the smoothness of null
infinity.
\newblock {\em Commun. Math. Phys.} 244 (2004) 133 - 156.

\bibitem{valiente kroon:2003B}
J. A. Valiente Kroon.
\newblock Does asymptotic simplicity allow for radiation near spatial
infinity ?  
\newblock {\em Commun. Math. Phys.} (2004) to appear.

\bibitem{winicour}
J. Winicour.
\newblock Logarithmic Asymptotic Flatness.
\newblock {\em Foundations of Physics} 15 (1985) 605 - 616.


\end{thebibliography}
\end{document}